\documentclass[preprints,review,moreauthors,accept,pdftex]{Definitions/mdpi}

\firstpage{1} 
\pubvolume{xx}
\issuenum{1}
\articlenumber{5}
\pubyear{2021}
\copyrightyear{2021}
\history{Received: date; Accepted: date; Published: date}
\pdfoutput=1

\usepackage{bm, amsmath, amssymb}

\def\cE{\mathcal{E}}

\def\cN{\mathcal{N}}
\def\cO{\mathcal{O}}

\def\mint{\int_{-\infty}^\infty\!\cdots\!\int_{-\infty}^\infty}

\newcommand{\be}{\begin{equation}}
\newcommand{\ee}{\end{equation}}
\newcommand{\ba}{\begin{aligned}}
\newcommand{\ea}{\end{aligned}}

\DeclareMathOperator{\Ai}{Ai}
\DeclareMathOperator{\Bi}{Bi}

\def\({\left(}
\def\){\right)}

\newcommand{\del}{\partial}

\DeclareMathOperator{\diag}{diag}

\Title{Spectral Problems for Quasinormal Modes of Black Holes}

\Author{Yasuyuki Hatsuda and Masashi Kimura}

\AuthorNames{Yasuyuki Hatsuda}
\AuthorNames{Masashi Kimura}

\address{%
Department of Physics, Rikkyo University, Toshima, Tokyo 171-8501, Japan%\\
}

\abstract{This is an unconventional review article on spectral problems in black hole perturbation theory. Our purpose is to explain how to apply various known techniques in quantum mechanics to such spectral problems. The article includes analytical/numerical treatments, semiclassical perturbation theory, the (uniform) WKB method and useful mathematical tools: Borel summations, Pad\'e approximants, etc. The article is not comprehensive, but rather looks into a few examples from various points of view. The techniques in this article are widely applicable to many other examples. }

\begin{document}
%%%%%%%%%%%%%%%%%%%%%%%%%%%%%%%%%%%%%%%%%%

%%%%%%%%%%%%%%%%%%%%%%%%%%%%%%%%%%%%%%%%%%
\tableofcontents

%\newpage

\section{Introduction}
The main motivation to write up this article is the following.
We collect various traditional approaches to one-dimensional eigenvalue problems in quantum mechanics. Some of them are well explained in usual textbooks, but some are not. 
We would like to demonstrate that these approaches are widely applicable to spectral problems in black hole perturbation theory.  We also present a unified manner to treat bound states and resonant states together. The latter problem is particularly important in analysis of quasinormal modes (QNMs) of black holes.

We guess that some of the readers are familiar with general relativity but may not be so familiar with quantum mechanics.
Conversely, some may be familiar with quantum mechanics but not with general relativity.
The article is pedagogical and designed for both of these people.

There have already been many excellent review articles \cite{nakamura1987, kokkotas1999, nollert1999, ferrari2008, berti2009, konoplya2011} on black hole perturbation theory.
For differentiation from these articles, the contents of the current article are intentionally quite biased. 
%For instance, contrary to its title, any explicit metrics of black holes do not appear, nor discuss we black hole perturbation theory itself. 
Although we review the linear perturbation of the four-dimensional Schwarzschild spacetime in Appendix~\ref{app:BH-pert},
the reader who wants to learn black hole perturbation theory in more general cases should see other good reviews \cite{nakamura1987, kokkotas1999, nollert1999, ferrari2008, berti2009, konoplya2011} or textbooks \cite{chandrasekhar1998, maggiore2018, andersson2020, ferrari2020} and references therein. In the main text, we rather concentrate our attention on practical aspects of computational schemes on eigenvalue problems associated with black hole perturbation theory.

Let us briefly comment on physical significances of QNMs.
They are defined as poles of Green's function in initial value problems of the linear perturbation around black hole solutions,
and describe exponentially damped oscillation as the late time behavior of the perturbation\footnote{We note that there is also a branch cut contribution which corresponds to the power law tails. We do not discuss this contribution in the present article.}
~\cite{leaver1986b, nollert1992, andersson1995, andersson1997, berti2006c}.
Interestingly, they also describe the last stage of a gravitational waveform
in a black hole merger, which was firstly observed by LIGO--Virgo~\cite{abbott2016b},
while the black hole merger process should be treated as a fully non-linear problem.
A recent argument~\cite{giesler2019} shows that the gravitational waveform of the binary black hole merger can be fit by the superposition of QNMs even before the peak amplitude. This suggests that the linear perturbation treatment around the final state black hole is a good approximation even just after the black hole merger.
In general relativity, because of the uniqueness of the Kerr black holes, the QNM spectra are completely 
characterized by their mass and spin.
If we can check it by observation, it will be a good test for general relativity.
In other words, we have a possibility to test theories beyond general relativity through the observation of QNMs~\cite{barack2019}.
For this purpose, the QNM spectra are particularly important.
The computational technique introduced in the present article can be also applied for such problems.

Mathematically, the spectral problems in this article are connection problems of local solutions to ordinary differential equations at two different spatial points, i.e., two-point boundary value problems. Since these problems require global information on the solutions, they are not solved analytically in general.
This fact makes spectral problems rich and interesting. We would like to investigate the spectral problems as analytic as possible.

The organization of the article is as follows. In the next section, we review several classic approaches to spectral problems in quantum mechanics. Some of them are less familiar.
It includes analytical/numerical treatments, perturbation theory and the WKB method. %and a few less familiar approaches.
All of them have direct applications to black hole physics.
In most cases, spectral problems are not exactly solvable.
It is desirable to present widely applicable ways to various problems.
To understand this section, the reader needs basics on second order ordinary differential equations, asymptotic analysis and Pad\'e approximants, which are summarized in Appendices~\ref{app:ODE} and \ref{app:Pade}.
In Section~\ref{sec:BH1}, we discuss how to apply these techniques to spectral problems in black hole physics.
We consider two spectral problems: quasinormal modes and mode (in)stability of black holes by seeing concrete examples.
More basics on black hole perturbation theory are explained in Appendix~\ref{app:BH-pert}.

%%%%%%%%%%%%%%%%%%%%%%%%%%%%%%%%%%%%%%%%%%
\section{A quantum mechanical approach to spectral problems}\label{sec:traditional}
In this section, we review various ways for spectral problems in quantum mechanics.
In the next section, we will discuss actual applications to black hole physics.

\subsection{Bound states and resonant states}\label{subsec:bound}
We begin by reviewing bound states and resonant states in quantum mechanics.
Let us consider the one-dimensional time-independent Schr\"odinger equation of the form:
\begin{equation}
\begin{aligned}
\left( -\hbar^2 \frac{d^2}{dx^2}+V(x) \right) \psi(x)=E \psi(x).
\end{aligned}
\label{eq:eigen}
\end{equation}
This type of the eigen-equations appears in many places in theoretical physics.
Different contexts often provide us different perspectives to the same problem.
In many cases, $\hbar$ just appears as a parameter of the equation.
A typical problem in quantum mechanics is to compute energy eigenvalues of bound states for a given potential.
Another interesting problem is a potential scattering.
As we see just below, this problem is related to resonant states.
These two problems turn out to be interrelated to each other.
The former problem is related to mode stability of black holes, and
the latter has a direct connection with quasinormal modes.

As is well-known, the bound states require the normalizability of the wave function.
Therefore the potential needs to have a global minimum in a considered region. However this does not mean that a potential with a minimum always has a bound state.
Unless a well is deep enough, there exist no bound states. This point is crucially important in analysis of (in)stability problems of black holes.

\begin{figure}[tbp]
\begin{center}
\includegraphics[width=0.45\linewidth]{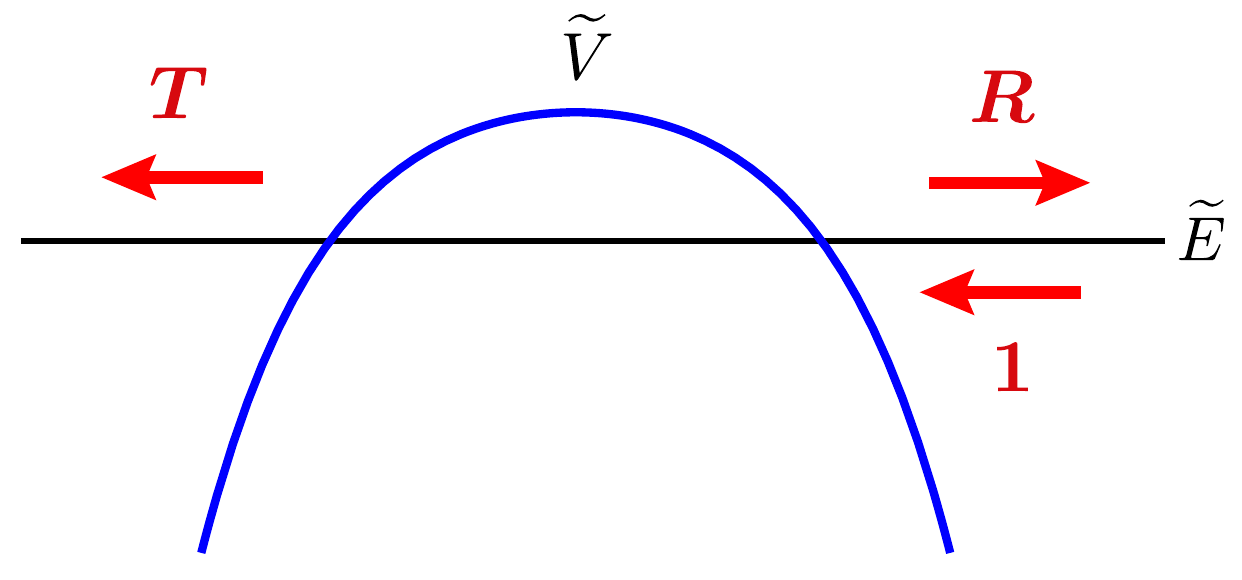}
\caption{Scattering process by a potential wall. Resonant states are defined by the absence of the incoming wave. This occurs when $1/R \to 0$ (or roughly $|T| \sim |R| \gg 1$).}
\label{fig:scatteing}
\end{center}
\end{figure}

The resonant states are less familiar. Let us consider a scattering problem for a potential with a wall, shown in Figure~\ref{fig:scatteing}.\footnote{In this section, we use $\,\widetilde{~}\,$ for resonant state problems to distinguish them from bound state problems.}
The Schr\"odinger equation in this setup is
\begin{equation}
\begin{aligned}
\left( -\widetilde{\hbar}^2 \frac{d^2}{dx^2}+\widetilde{V}(x) \right) \widetilde{\psi}(x)=\widetilde{E} \widetilde{\psi}(x),
\end{aligned}
\label{eq:eigen-2}
\end{equation}
There is an incoming wave from $x=+\infty$, and it scatters with the potential wall. As a result, there are a reflected and a transmitted waves.
In the region $x\sim +\infty$, the wave function is written as
\begin{equation}
\begin{aligned}
\widetilde{\psi}(x) \sim \widetilde{\psi}^\text{in}(x)+R \widetilde{\psi}^\text{r}(x), \qquad x \to +\infty ,\\
\end{aligned}
\end{equation}
where $R$ is the reflection coefficient.
In the region $x\sim -\infty$, only the transmitted wave exists:
\begin{equation}
\begin{aligned}
\widetilde{\psi}(x) \sim T \widetilde{\psi}^\text{t}(x), \qquad x \to -\infty,
\end{aligned}
\end{equation}
where $T$ is the transmission coefficient.
It is more convenient to change the overall normalization as follows:
\begin{equation}
\begin{aligned}
B\widetilde{\psi}^\text{t}(x) \stackrel{-\infty \leftarrow x}{\longleftarrow} \widetilde{\psi}(x) \stackrel{x\rightarrow +\infty}{\longrightarrow} A\widetilde{\psi}^\text{in}(x)+\widetilde{\psi}^\text{r}(x),
\end{aligned}
\end{equation}
where $A=1/R$ and $B=T/R$.
The resonant states are defined by the absence of the incoming wave:
\begin{equation}
\begin{aligned}
A=\frac{1}{R}=0.
\end{aligned}
\end{equation}
Clearly it happens for characteristic energies that are located on singularities of the reflection coefficient (or equivalently the transmission coefficient).
Hence this phenomenon is called a resonance.
In general, these characteristic eigenvalues are discrete but complex-valued \cite{konishi2009}.
Note that the boundary condition for the resonant states is the same as that for the quasinormal modes of black holes.
Computing the resonant energies is the main issue in this article.

\begin{figure}[tb]
\begin{center}
  \begin{minipage}[b]{0.45\linewidth}
    \centering
    \includegraphics[width=0.95\linewidth]{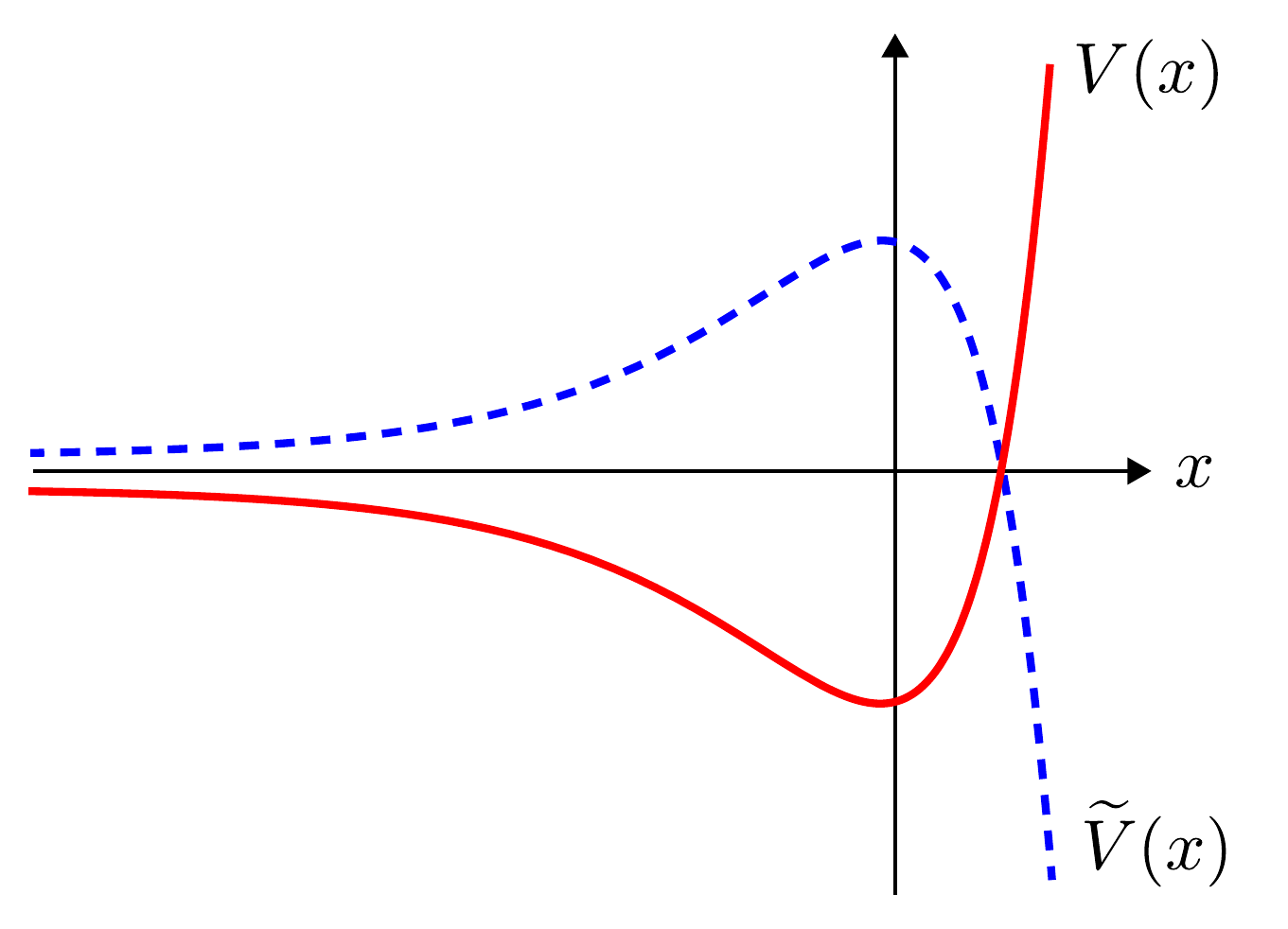}
  \end{minipage}
    \begin{minipage}[b]{0.45\linewidth}
    \centering
    \includegraphics[width=0.95\linewidth]{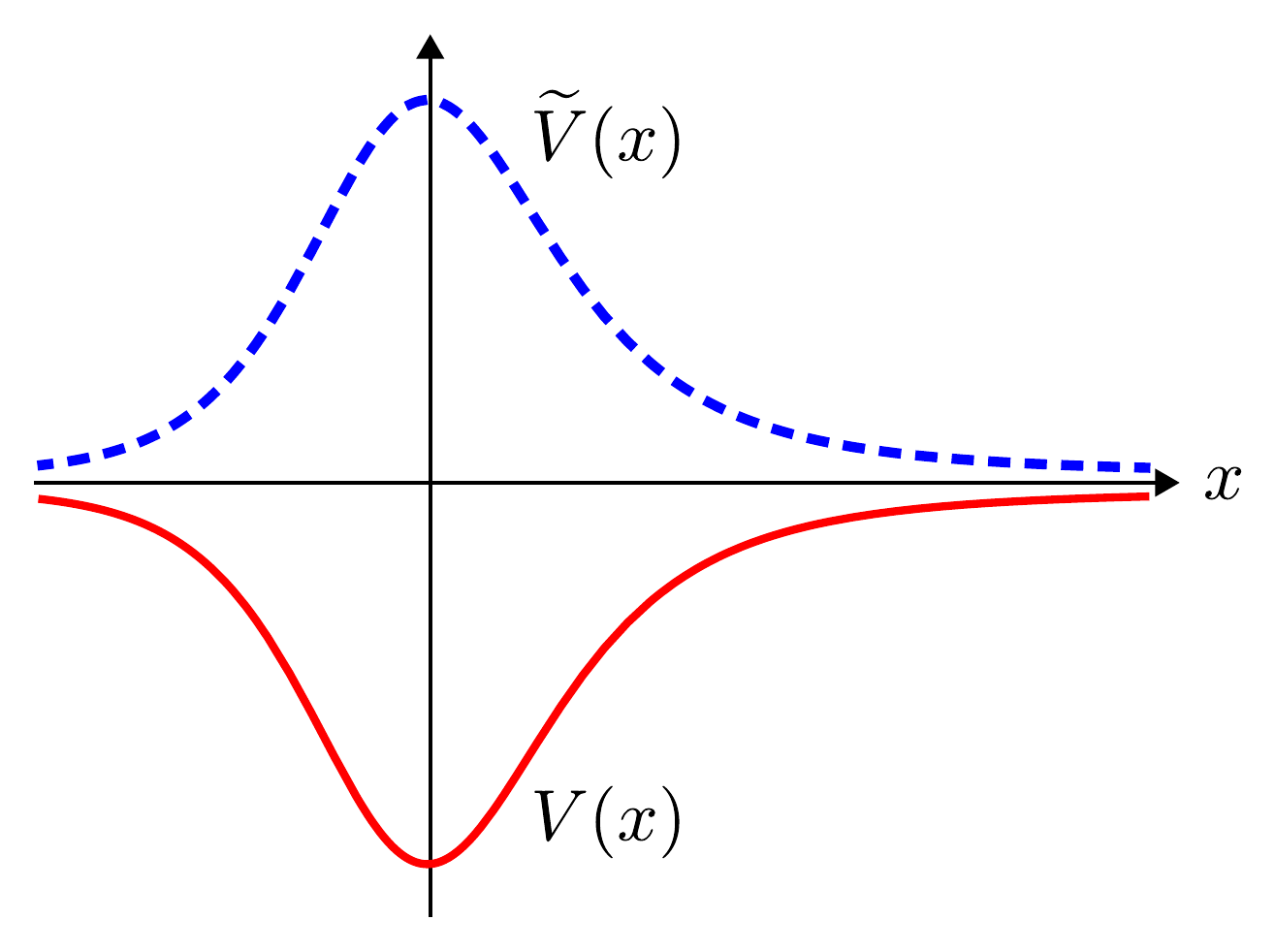}
  \end{minipage}
%    \begin{minipage}[b]{0.32\linewidth}
%    \centering
%    \includegraphics[width=0.95\linewidth]{pot-2-mMathieu.pdf}
%  \end{minipage}
\end{center}
  \caption{In this article, we mainly consider two types of potentials. These appear in the Morse potential \eqref{eq:Morse} (left) and in an effective potential \eqref{eq:RW-potential} in the linear perturbation of the Schwarzschild spacetime (right). Both cases can be treated in a unified manner.}
  \label{fig:potentials}
\end{figure}

Next we see a relation between bound states and resonant states.
In this article, we particularly focus on two types of potentials $V(x)$ and their sign-flipped ones $\widetilde{V}(x)=-V(x)$, as shown in Figure~\ref{fig:potentials}.
Since the potential $V(x)$ has a well in these cases, it admits bound states. Then its inverted potential $\widetilde{V}(x)$ has resonant states.
The energy eigenvalues of these two states are simply related by an analytic continuation of the quantum parameter $\hbar$.
If setting $\hbar=i\widetilde{\hbar}$,\footnote{We have a freedom to set $\hbar=-i\widetilde{\hbar}$, and it leads to another branch of the resonance. Since these two branches are symmetric, it is sufficient to consider either of them.}
the eigen-equation \eqref{eq:eigen} is related to \eqref{eq:eigen-2} with $\widetilde{E}=-E$ and $\widetilde{\psi}(x)=\psi(x)|_{\hbar \to \widetilde{\hbar}}$.\footnote{Concerning boundary conditions, one can see the following observation. If the wave function behaves as $e^{-\sqrt{-E}x/\hbar}$ in $x \to \infty$, then it satisfies the bound state condition in $x \to \infty$. After the analytic continuation, the wave function behaves as $e^{i\sqrt{\widetilde{E}}x/\widetilde{\hbar}}$, and satisfies the resonant boundary condition.}
Therefore physics in the system \eqref{eq:eigen-2} may be described by that in the system \eqref{eq:eigen}.
In particular, we expect the following exact spectral relation:%
\footnote{Strictly, there is a subtlety on the number of the allowed bound states. This point is discussed in the next section. Also this relation is based on an assumption that the potential $V(x)$ has bound states. In the case of the cubic potential for example, both $V(x)$ and $\widetilde{V}(x)$ have only the resonant states, and the relation \eqref{eq:B-R} should be modified for these eigenvalues.}
\begin{equation}
\begin{aligned}
\widetilde{E}_n^\text{resonance}(\widetilde{\hbar})=-E_n^\text{bound}(\hbar=i\widetilde{\hbar}),\qquad n=0,1,2,\dots,
\end{aligned}
\label{eq:B-R}
\end{equation}
where $E_n^\text{bound}(\hbar)$ is the bound state energy for the Schr\"odinger equation \eqref{eq:eigen}, and $\widetilde{E}_n^\text{resonance}(\widetilde{\hbar})$ is the resonant energy for its sign-flipped one \eqref{eq:eigen-2}.
We will test this equality in a few examples in detail.
We can extract the resonant energy from the bound state energy.
This idea has been applied to the computation of the QNM frequencies in \cite{hatsuda2020, eniceicu2020} (see \cite{ferrari1984, Ferrari:1984zz, zaslavskii1991} for earlier works).

\subsection{Lessons from an exactly solvable model}
In the context of black hole perturbation theory, the P\"oschl--Teller potential $\widetilde{V}_\text{PT}(x)=V_0/(2\cosh^2 \beta x)$ is usually taken as an exactly solvable example \cite{ferrari1984}.
This is reasonable because its shape is similar to effective potentials of master equations appearing in black hole perturbations, as in the right panel of Figure~\ref{fig:potentials}.
For differentiation, we take another less-familiar example: the Morse potential.
The Morse potential is also exactly solvable. 
Its asymptotic behavior is quite different from the potentials in black hole perturbations. See the left panel of Figure~\ref{fig:potentials}.
However, from the viewpoint of the singularity structure of ordinary differential equations (see Appendix~\ref{app:ODE}),
the Morse potential is more similar to the spectral problem for asymptotically flat black holes.
The P\"oschl--Teller potential is rather similar to the spectral problem for asymptotically (anti-)de Sitter black holes.
Since the P\"oschl--Teller potential has been reviewed in detail in \cite{berti2009}, we do not discuss it here. 

Let us consider the following very special potential:
\begin{equation}
\begin{aligned}
V_\text{Morse}(x)=V_0(e^{2\beta x}-2e^{\beta x}),\qquad -\infty<x<\infty,
\end{aligned}
\label{eq:Morse}
\end{equation}
where we assume $V_0>0$ and $\beta>0$. Its shape is shown in the left panel in Figure~\ref{fig:potentials}.
It has the global minimum at $x=0$.
A simple reason of the solvability of the Morse potential is a shape invariance \cite{cooper1995}.
Another explanation is that the Schr\"odinger equation is mapped to the well-known equation as we will see below.
To make expressions simpler, we introduce the following rescaled parameters:
\begin{equation}
\begin{aligned}
q:=\beta x,\qquad \varepsilon:=\frac{E}{V_0},\qquad g:=\frac{\beta \hbar}{\sqrt{V_0}}.
\end{aligned}
\end{equation}
Then the Schr\"odinger equation is rewritten as
\begin{equation}
\begin{aligned}
\(-g^2 \frac{d^2}{dq^2}+e^{2q}-2e^q \) \psi(q)=\varepsilon \psi(q).
\end{aligned}
\label{eq:reduced}
\end{equation}
We change the variable $z=(2/g) e^{q}$.
The differential equation \eqref{eq:reduced} becomes
\begin{equation}
\begin{aligned}
\left[ -z \frac{d}{dz} z\frac{d}{dz} +\frac{z}{4}\left( z-\frac{4}{g} \right) \right]\psi(z)=\frac{\varepsilon}{g^2} \psi(z).
\end{aligned}
\end{equation}
We further perform the transform
\begin{equation}
\begin{aligned}
\psi(z)=e^{-z/2}z^{\frac{\kappa}{g}} y(z),
\end{aligned}
\end{equation}
where $\kappa:=\sqrt{-\varepsilon}$. Since we are interested in the bound states, we implicitly assume $-1<\varepsilon<0$ and $\kappa>0$.
However, in the construction of the solutions, we do not need this assumption.
Then the differential equation reduces to the generalized (or associated) Laguerre equation:
\begin{equation}
\begin{aligned}
zy''(z)+(\alpha+1-z)y'(z)+\nu y(z)=0,
\end{aligned}
\label{eq:Laguerre}
\end{equation}
where
\begin{equation}
\begin{aligned}
\alpha=\frac{2\kappa}{g},\qquad \nu=\frac{1-\kappa}{g}-\frac{1}{2}.
\end{aligned}
\label{eq:para-Morse}
\end{equation}
The generalized Laguerre equation has the regular singular point at $z=0$ and the irregular singular point at $z=\infty$.
It is essentially equivalent to the confluent hypergeometric equation.
The (characteristic) exponents at $z=0$ are $0$ and $-\alpha$.
The former solution corresponds to the generalized Laguerre function $L_\nu^\alpha(z)$.
Some of basics on ordinary differential equations are explained in Appendix~\ref{app:ODE}.

Let us see the eigenvalues of the bound states. In the limit $z \to 0$ (i.e., $q \to -\infty$), since the wave function behaves as $\psi(z) \sim z^{\frac{\kappa}{g}}$ or $\psi(z) \sim z^{-\frac{\kappa}{g}}$, the normalizability requires the regularity of $y(z)$ at $z=0$.
This means that we have to choose $y(z)=L_\nu^\alpha(z)$.
Therefore the analytic solution satisfying the boundary condition at $z=0$ is given by
\begin{equation}
\begin{aligned}
\psi(z)=e^{-z/2}z^{\frac{\kappa}{g}} L_\nu^\alpha (z).
\end{aligned}
\label{eq:ef-Morse}
\end{equation}
This solution of course does not satisfy the boundary condition at $z =\infty$ for arbitrary $E$ because $L_\nu^\alpha(z)$ exponentially grows in $z \to \infty$.
The boundary condition at infinity is satisfied if and only if $L_\nu^\alpha (z)$ is a polynomial.
This requires $\nu$ to be a non-negative integer. 
Therefore we obtain an exact quantization condition for the bound states:
\begin{equation}
\begin{aligned}
\frac{1-\sqrt{-\varepsilon_n}}{g}-\frac{1}{2}=n,\qquad n=0,1,2,\dots
\end{aligned}
\label{eq:eQC-Morse}
\end{equation}
This is easily solved, and we finally obtain
\begin{equation}
\begin{aligned}
\varepsilon_n=-\left[ 1-g\( n+\frac{1}{2} \) \right]^2,\qquad n=0,1,2\dots.
\end{aligned}
\label{eq:spec-Morse}
\end{equation}
The same result is obtained by the asymptotic expansion of $L_\nu^\alpha(z)$.
It is well-known that the generalized Laguerre function has the following asymptotic expansion:
\begin{equation}
\begin{aligned}
&L_\nu^\alpha(z)\simeq \frac{(-z)^\nu}{\Gamma(\nu+1)} \sum_{k=0}^\infty \frac{(-1)^k (-\nu)_k (-\alpha-\nu)_k}{k! z^k}\\
&\quad-\frac{\sin(\pi \nu) \Gamma(\alpha+\nu+1)z^{-\alpha-\nu-1}e^z}{\pi} \sum_{k=0}^\infty \frac{(\nu+1)_k(\alpha+\nu+1)_k}{k! z^k}\qquad (z \to \infty),
\end{aligned}
\label{eq:L-asympt}
\end{equation}
where $(a)_k$ is the Pochhammer symbol. We have used $\simeq$ instead of $=$ because the right hand side contains formal divergent series, i.e., the radius of convergence is just zero.
We have to treat it in a careful manner.
We return to this issue in the next subsection and in Appendix~\ref{app:Borel}.
Since the second line in this asymptotic expansion is a source of the exponential grow,
the boundary condition at infinity requires the absence of it: $\sin(\pi \nu)=0$. Then, the right hand side in the first line has finite terms, and it reduces to a polynomial.

\begin{figure}[tb]
\begin{center}
  \begin{minipage}[b]{0.45\linewidth}
    \centering
    \includegraphics[width=0.95\linewidth]{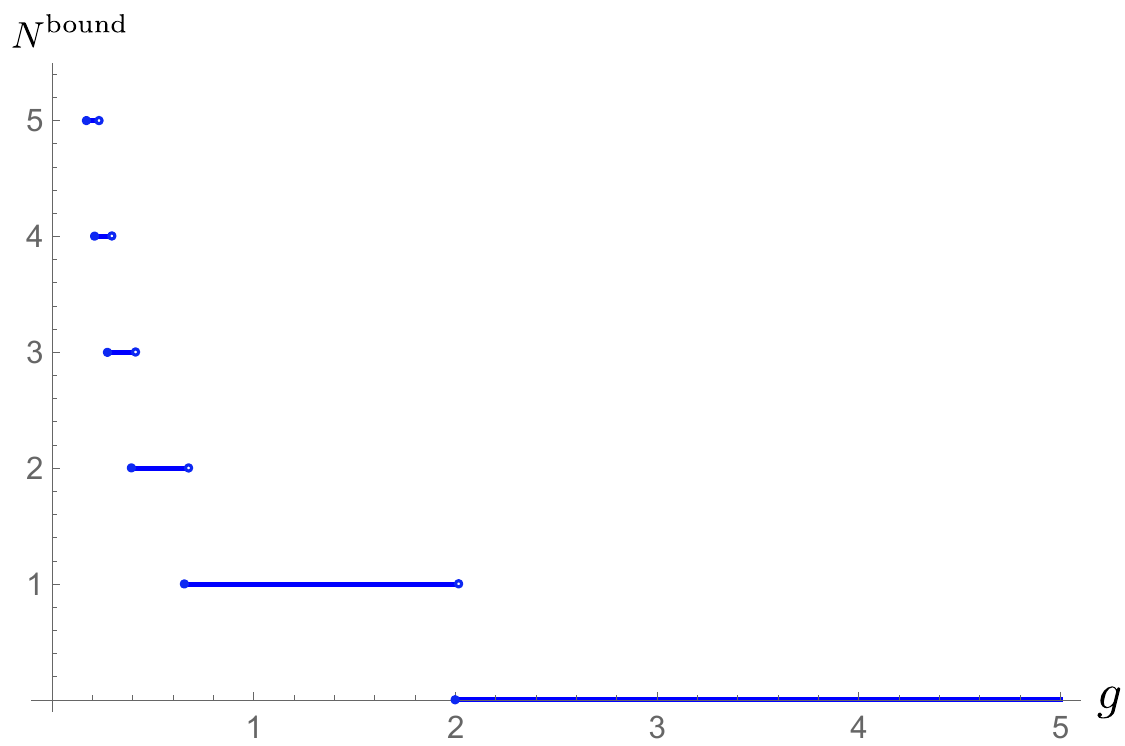}
  \end{minipage}
  \begin{minipage}[b]{0.45\linewidth}
    \centering
    \includegraphics[width=0.95\linewidth]{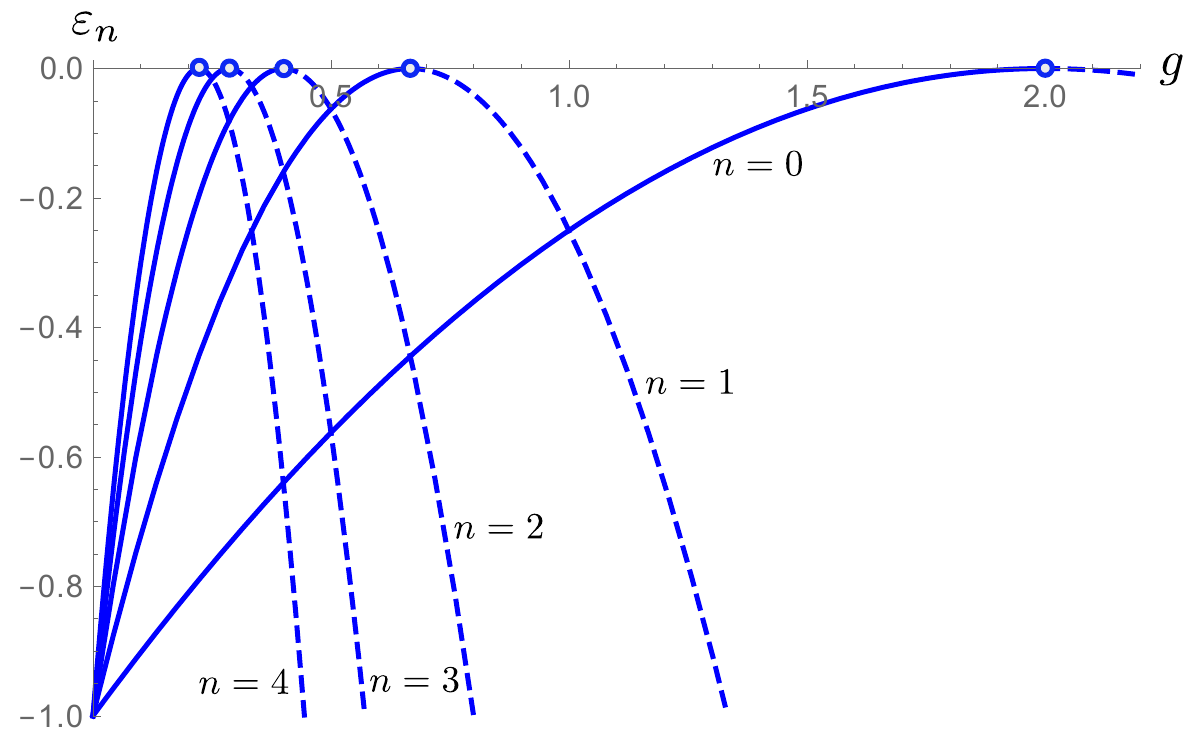}
  \end{minipage} 
\end{center}
  \caption{Left: The number of the allowed bound states in the Morse potential. Right: The exact spectrum $\varepsilon_n$ against the parameter $g$.
  The solid lines are the physical bound state energies while the dashed lines are on the unphysical branch. For $2/3<g<2$, the branch for $n=0$ is physical but those for $n \geq 1$ are all unphysical. The total number of the bound states is $N^\text{bound}=1$, which is consistent with the left panel.}
  \label{fig:spectrum}
\end{figure}

Let us count the number of the bound states.
The positivity $\kappa_n=\sqrt{-\varepsilon_n}>0$ for the bound states leads to the upper bound to the quantum number $n$:
\begin{equation}
\begin{aligned}
n<\frac{1}{g}-\frac{1}{2}\,.
\end{aligned}
\end{equation}
Therefore the number of the allowed bound states, $N^\text{bound}$, is given by
\begin{equation}
\begin{aligned}
N^\text{bound}=\left\lceil \frac{1}{g}-\frac{1}{2} \right\rceil,
\end{aligned}
\label{eq:number}
\end{equation}
where $\lceil p \rceil $ means the least integer greater than or equal to $p$.
For instance, if $g\geq 2$, there are no bound states although the potential still has the well.
We show $N^\text{bound}$ against the parameter $g$ in the left panel of Figure~\ref{fig:spectrum}.

Let us recall the exact spectrum \eqref{eq:spec-Morse}. At first glance, it seems to give an infinite number of the eigenvalues for $n=0,1,2,\dots$.
This is not the case. As seen above, for relatively large $n$, $\kappa_n$ becomes imaginary.
How do we exclude such $n$'s by looking at only the spectrum \eqref{eq:spec-Morse}?
In the representation \eqref{eq:spec-Morse}, this is understood as two branches of the square root of $-\varepsilon_n$.
We show in in the right panel of Figure~\ref{fig:spectrum} the spectrum $\varepsilon_n$ against the parameter $g$.
For a given negative energy ranging $-1<\varepsilon<0$, we have two $\varepsilon_n$ with quantum number $n$ corresponding to different $g$. The left branch (solid line) gives the physical bound state energy because it is continuously connected with the classical point $g=0$. The right branch (dashed line) is an unphysical mode that does not satisfy the bound state boundary condition.\footnote{In fact, this state satisfies the boundary condition such that there is no contamination of the decaying mode in $|q|\to \infty$. That is, the solution purely grows in $|q| \to \infty$.} For a fixed value of $g$, highly excited states are mostly on the unphysical branch.

Things are different for the resonant states. As mentioned in the introductory section, the resonant states are related to the bound states by the analytic continuation: $g=i\widetilde{g}$ (or $g=-i\widetilde{g}$).
In this continuation, the parameters become
\begin{equation}
\begin{aligned}
z=\frac{2}{g} e^{q} = \frac{2}{i\tilde{g}} e^{q},\qquad
\alpha=\frac{2\kappa}{i\widetilde{g}},\qquad
\nu=\frac{1-\kappa}{i\widetilde{g}}-\frac{1}{2}.
\end{aligned}
\end{equation}
The eigenvalues and the eigenfunctions for the inverted potential $\widetilde{V}$ are given by
\begin{equation}
\begin{aligned}
\widetilde{\varepsilon}_n&:=-\varepsilon_n=\left[ 1-i\widetilde{g}\( n+\frac{1}{2} \) \right]^2,\\
\widetilde{\psi}_n(q)&=e^{ie^{q}/\tilde{g}}\( \frac{2}{i\tilde{g}} e^{q} \)^{-i\kappa_n/\widetilde{g}} L_n^\alpha\( \frac{2}{i\tilde{g}} e^{q} \).
\end{aligned}
\end{equation}
One can check that this eigenfunction always satisfies the boundary condition of the resonant states for all $n=0,1,2,\dots$.
We conclude that there are always an infinite number of the resonant states\footnote{Moreover the infinite number of the resonant states is ``doubled'' by the other analytic continuation $\hbar=-i\widetilde{\hbar}$. This corresponds to including $n=-1,-2,-3, \dots$.} while the number of the bound states is finite.
To construct highly excited resonant states, we need unphysical ``false bound states.''

\subsection{Numerical methods}\label{subsec:numerical}
There are many ways to evaluate the eigen-energies numerically. Since our purpose is not to cover all of these methods, we present only two particular methods that are useful in our later analysis.

\subsubsection{Milne's method}
Here we revisit a very old result by Milne \cite{milne1930} because it is useful to numerically count the number of bound states.
We first rewrite the Schr\"odinger equation \eqref{eq:eigen} as
\begin{equation}
\begin{aligned}
\( \hbar^2 \frac{d^2}{dx^2}+Q(x) \) \psi(x)=0,
\end{aligned}
\label{eq:Milne}
\end{equation}
where
\begin{equation}
\begin{aligned}
Q(x)=E-V(x).
\end{aligned}
\end{equation}
We assume that $Q(x)$ does not have any singular points for $x \in \mathbb{R}$.
We consider two independent solutions satisfying conditions:
\begin{equation}
\begin{aligned}
\psi_1(x_0)&=1, \qquad \psi_1'(x_0)=0, \\
\psi_2(x_0)&=0, \qquad \psi_2'(x_0)=1,
\end{aligned}
\end{equation}
where $x_0$ is a certain point on the real axis.
It is well-known that the Wronskian of these two solutions does not depend on $x$, and it always equals to unity.
Let us define
\begin{equation}
\begin{aligned}
w_E(x):=\sqrt{\psi_1(x)^2+\psi_2(x)^2},
\end{aligned}
\end{equation}
where we use the subscript $E$ to emphasize the energy dependence of $w$.
The fact that the Wronskian of $\psi_1(x)$ and $\psi_2(x)$ is the unity guarantees that $w_E(x)$ never vanishes.
It satisfies
\begin{equation}
\begin{aligned}
w_E(x_0)=1,\qquad w_E'(x_0)=0.
\end{aligned}
\label{eq:ini-w}
\end{equation}
One can easily check that $w_E(x)$ satisfies the following non-linear differential equation:
\begin{equation}
\begin{aligned}
\( \hbar^2 \frac{d^2}{dx^2}+Q(x)\) w_E(x)=\frac{\hbar^2}{w_E(x)^3}.
\end{aligned}
\label{eq:deq-w}
\end{equation}
The two solutions are conversely reconstructed by
\begin{equation}
\begin{aligned}
\psi_1(x)&=w_E(x) \cos \left[ \int_{x_0}^x \frac{dx'}{w_E(x')^2} \right], \\
\psi_2(x)&=w_E(x) \sin \left[ \int_{x_0}^x \frac{dx'}{w_E(x')^2} \right].
\end{aligned}
\end{equation}
Now, we would like to construct a solution that satisfies the decaying boundary condition at $x=-\infty$.
Let us consider the following function:
\begin{equation}
\begin{aligned}
\psi(x)=w_E(x) \sin \left[ \int_{-\infty}^x \frac{dx'}{w_E(x')^2} \right].
\end{aligned}
\label{eq:Milne-sol}
\end{equation}
This is a solution to \eqref{eq:Milne} because the function $\psi(x)$ is written as a superposition of the two solutions $\psi_1(x)$ and $\psi_2(x)$. 
\begin{equation}
\begin{aligned}
\psi(x)=\psi_1(x) \sin\left[ \int_{-\infty}^{x_0} \frac{dx'}{w_E(x')^2} \right]+\psi_2(x) \cos \left[ \int_{-\infty}^{x_0} \frac{dx'}{w_E(x')^2} \right].
\end{aligned}
\end{equation}
Also, this function behaves as $\psi(x) \sim C/(x w_E(x))$ in $x \to -\infty$, and actually satisfies the boundary condition at $x=-\infty$.
Therefore, this is what we want. 
Then, the decaying boundary condition at $x=\infty$ requires
\begin{equation}
\begin{aligned}
 \sin \left[ \int_{-\infty}^\infty \frac{dx}{w_E(x)^2} \right]=0.
\end{aligned}
\end{equation}
Let us define
\begin{equation}
\begin{aligned}
N^\text{Milne}(E):=\frac{1}{\pi}\int_{-\infty}^\infty \frac{dx}{w_E(x)^2}-1.
\end{aligned}
\label{eq:N-Milne}
\end{equation}
The important fact found in \cite{milne1930} is that this function is a non-decreasing function of $E$.%
\footnote{We are not sure whether this is strictly proved or not. As far as we checked for many models, it indeed holds.}
It turns out that for the bound state energy $E_n$, the function satisfies\footnote{
The function $N^\text{Milne}(E)$ in Eq.~\eqref{eq:N-Milne} 
counts the number of nodes, i.e., the number of zeros, of the wave function Eq.~\eqref{eq:Milne-sol}.
According to the nodal theorem \cite{hartman1982, courant2008, messiah2014}, it corresponds to the number of the bound states with energy less than or equal to $E$.}
\begin{equation}
\begin{aligned}
N^\text{Milne}(E_n)=n,\qquad n=0,1,2,\dots.
\end{aligned}
\end{equation}
This is a kind of quantization conditions.
Therefore $N^\text{Milne}(E)$ can be regarded as a counting function of the bound states.
If $m-1\leq N^\text{Milne}(E)<m$, then there are $m$ bound states below the energy $E$.

Now we apply this simple method to the Morse potential \eqref{eq:Morse}.
In the actual computation, we use \eqref{eq:reduced}.
To fit the notation in this subsection, we equivalently fix the parameters $V_0=\beta=1$.
There is no difficulty to solve \eqref{eq:deq-w} with \eqref{eq:ini-w} numerically for given $E$.
This is just an initial value problem of the second order ordinary differential equation.
We can easily evaluate $N^\text{Milne}(E)$. Solving $N^\text{Milne}(E)=n$, we obtain the bound state energy at level $n$.
In the computation, we set $x_0=0$.
It is interesting to notice that we have another exact quantization condition \eqref{eq:eQC-Morse}.
Since the left hand side in \eqref{eq:eQC-Morse} is obviously an increasing function,
\begin{equation}
\begin{aligned}
N^\text{exact}(E):=\frac{1-\sqrt{-E}}{\hbar}-\frac{1}{2}
\end{aligned}
\end{equation}
is also a counting function.
\begin{figure}[tbp]
\begin{center}
\includegraphics[width=0.4\linewidth]{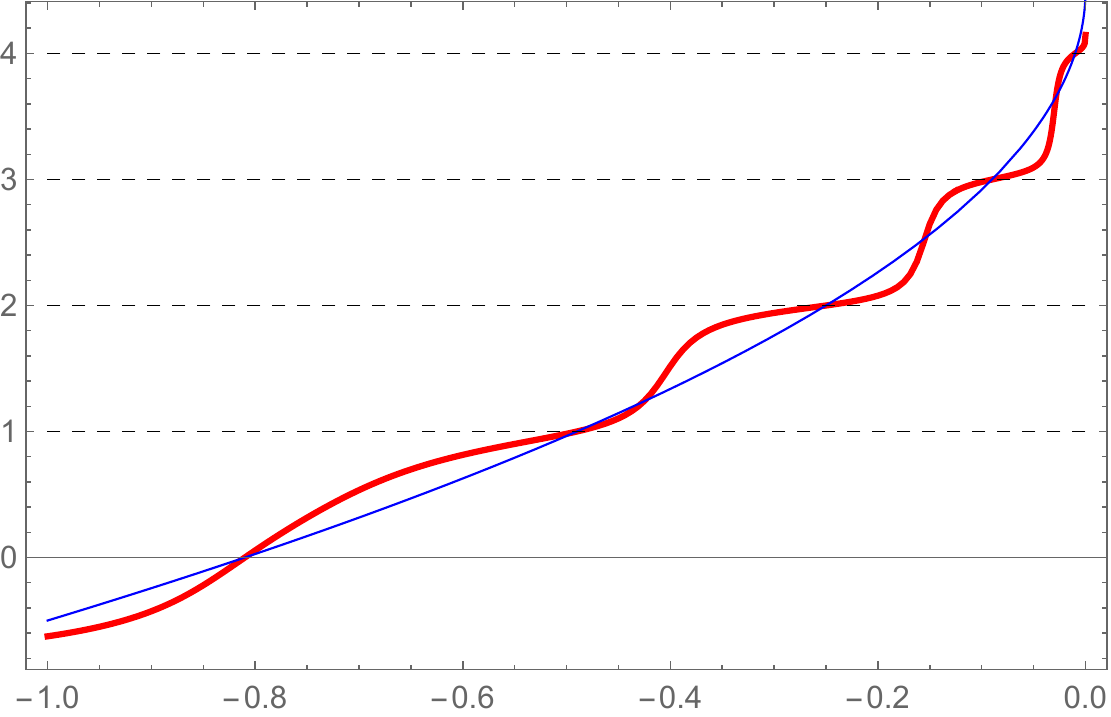}
\caption{Two counting functions of the Morse potential. The (red) thick line is Milne's one, and the (blue) thin line is $N^\text{exact}(E)$.}
\label{fig:counting}
\end{center}
\end{figure}
We show in Figure~\ref{fig:counting} the behaviors of $N^\text{Milne}(E)$ and $N^\text{exact}(E)$ for $\hbar=1/5$.
Interestingly, these two functions are quite different, but $N^\text{Milne}(E)$ indeed gives the correct bound state spectra.
The total number of the bound states is evaluated by the values of the counting functions at $E=0$ because $V_\text{Morse} \to 0$ in $x \to -\infty$.
In the case of $\hbar=1/5$, since $N^\text{Milne}(0) \approx 4.1$ or $N^\text{exact}(0)=4.5$, we have the five bound states.
Of course it is consistent with the exact result \eqref{eq:number}.
In other words, we always have
\begin{equation}
\begin{aligned}
N^\text{bound}=\lceil N^\text{exact}(0) \rceil=\lceil N^\text{Milne}(0) \rceil.
\end{aligned}
\end{equation}
We can use this relation in the mode stability analysis of black holes.
On the other hand, it seems hard to apply this method to resonant state problems.
To compute their eigenvalues numerically, one needs other approaches.

\subsubsection{Wronskian method}
We also review the well-known Wronskian method.
The Wronskian is useful to evaluate connection coefficients among local solutions.
Let $y_{p1}(x)$ and $y_{p2}(x)$ be two independent solutions around $x=p$.
Let us consider a two-point boundary value problem between $x=a$ and $x=b$.
The four solutions should be related by the connection coefficients:
\begin{equation}
\begin{aligned}
y_{a1}(x)&=C_{11} y_{b1}(x)+C_{12} y_{b2}(x), \\
y_{a2}(x)&=C_{21} y_{b1}(x)+C_{22} y_{b2}(x).
\end{aligned}
\end{equation}
The connection coefficients $C_{ij}$ are evaluated by Wronskians.
Let us assume that $y_{a1}(x)$ and $y_{b1}(x)$ satisfy boundary conditions we are interested in.
Then these two boundary conditions require
\begin{equation}
\begin{aligned}
C_{12}=\frac{W[y_{a1}, y_{b1}]}{W[y_{b2}, y_{b1}]}=0 \quad \Longrightarrow \quad W[y_{a1}, y_{b1}]=0.
\end{aligned}
\end{equation}
This determines the eigenvalues of the boundary value problem.
In other words, $C_{12}=0$ means that the two solutions $y_{a1}(x)$ and $y_{b1}(x)$ are linearly dependent,
and their Wronskian must vanish.
This method works not only for bound state problems but also for resonant state problems.

\begin{figure}[tb]
\begin{center}
  \begin{minipage}[b]{0.45\linewidth}
    \centering
    \includegraphics[width=0.95\linewidth]{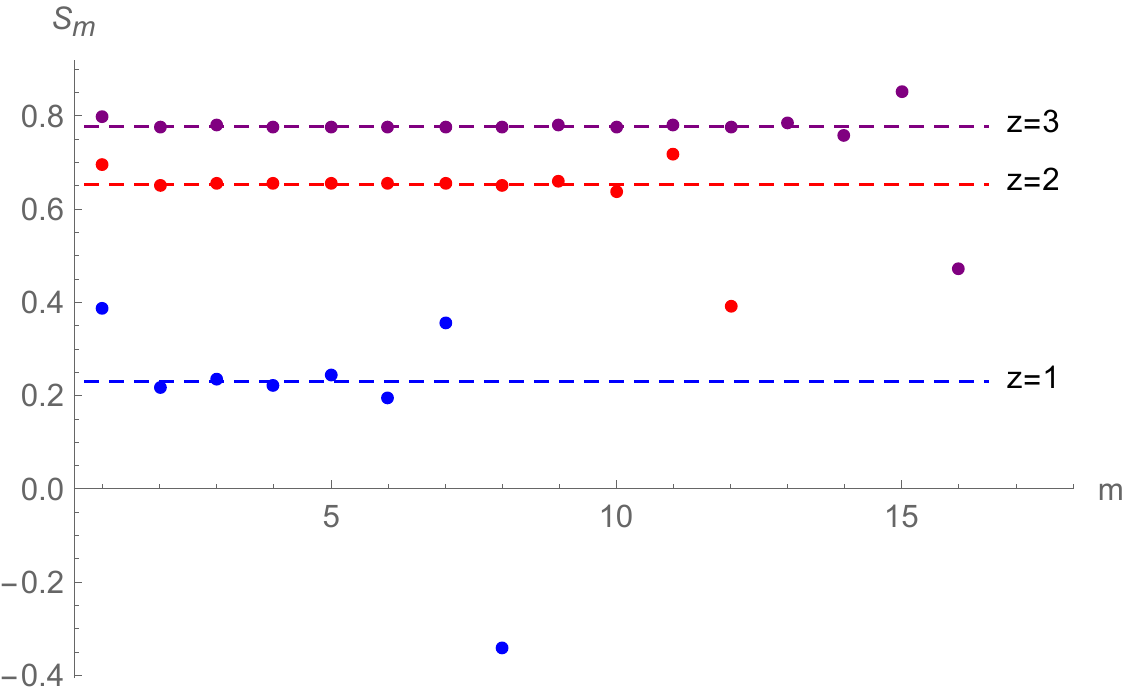}
  \end{minipage}
  \begin{minipage}[b]{0.45\linewidth}
    \centering
    \includegraphics[width=0.95\linewidth]{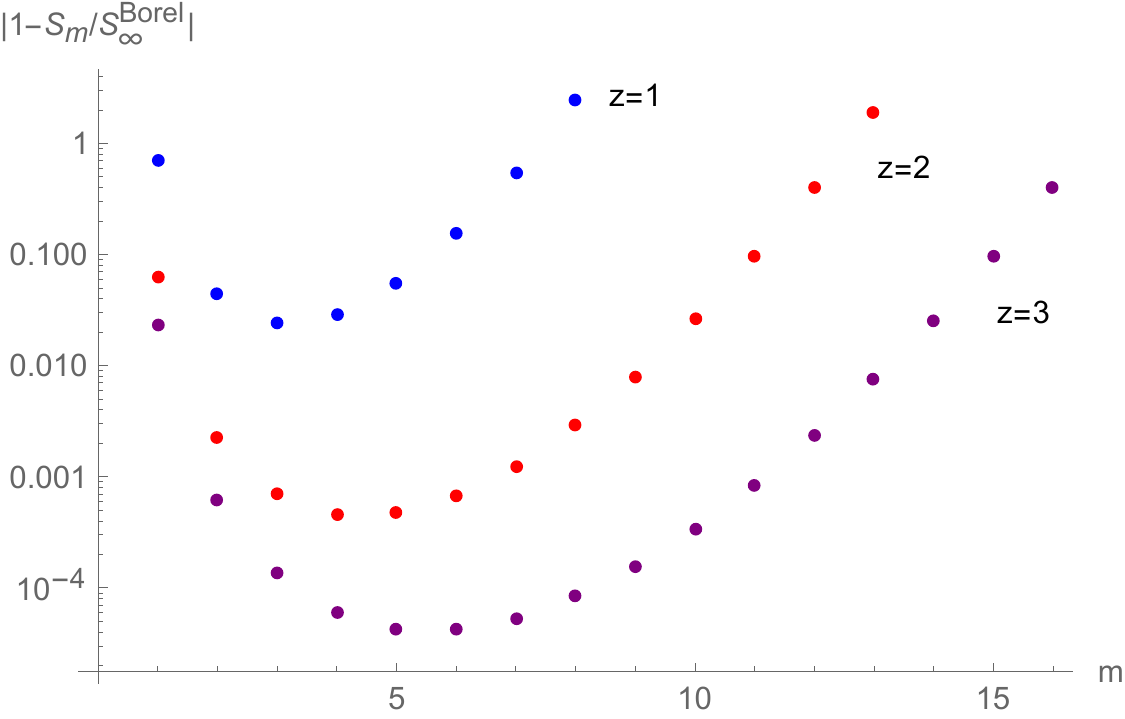}
  \end{minipage} 
\end{center}
\caption{The left panel shows the divergence of the asymptotic sum $S_m(z)$. We set $\nu=1/3$, $\alpha=3/2$ and $z=1,2,3$. The dashed lines represent the values of the Borel sum $S_\infty^\text{Borel}(z)$. The Borel sum gives the ``true value'' of the formal divergent sum $S_\infty(z)$.
The right panel shows the optimal order of $S_m(z)$. Beyond this order, $S_m(z)$ gets away from the ``true value'' $S_\infty^\text{Borel}(z)$.}
\label{fig:asymp-sum}
\end{figure}

Let us apply it to the Morse potential.
It is more convenient to use the $z$-variable.
We consider the boundary value problem for the differential equation \eqref{eq:Laguerre}.
As already seen, the regular solution at $z=0$ is given by $y_{01}(z)=L_{\nu}^\alpha(z)$.
The construction of the local solutions at infinity is more involved because $z=\infty$ is the irregular singular point.
See Appendix~\ref{app:irregular} on formal series solutions at an irregular singular point. Here we take a shortcut. We use the asymptotic expansion \eqref{eq:L-asympt}.
This tells us the all-order expressions of the two formal series solutions at $z=\infty$:
\begin{equation}
\begin{aligned}
y_{\infty1}^\text{formal}(z) &= z^\nu \sum_{k=0}^\infty \frac{(-1)^k (-\nu)_k (-\alpha-\nu)_k}{k! z^k},\\
y_{\infty2}^\text{formal}(z) &=  z^{-\alpha-\nu-1}e^z\sum_{k=0}^\infty \frac{(\nu+1)_k(\alpha+\nu+1)_k}{k! z^k}.
\end{aligned}
\label{eq:local-sols-inf}
\end{equation}
One can check that these are actually solutions to \eqref{eq:Laguerre}.
The former solution satisfies the boundary condition at infinity.
An important caution is that the sums on the right hand side are divergent series.
They do not converge for any values of $z$ and generic values of $\alpha$ and $\nu$.
This is why we call these solutions formal.
To see it in detail, let us consider the truncated sum in the first equation in \eqref{eq:local-sols-inf}:
\begin{equation}
\begin{aligned}
S_m(z) := \sum_{k=0}^m \frac{(-1)^k (-\nu)_k (-\alpha-\nu)_k}{k! z^k}.
\end{aligned}
\end{equation}
We plot some values of $S_m(z)$ for $\nu=1/3$, $\alpha=3/2$ and $z=1,2,3$ in Figure~\ref{fig:asymp-sum}. 
Obviously $S_m$ diverges when $m$ gets large. 
The dashed lines are the values of the Borel summation (see Appendix~\ref{app:irregular}) of the infinite sum $S_\infty(z)$.
By setting $p=1$ in \eqref{eq:Borel}, the Bore sum of $S_\infty(z)$ is analytically given by
\begin{equation}
\begin{aligned}
S_\infty^\text{Borel}(z)=\int_0^\infty d\zeta\, e^{-\zeta} {}_2F_1(-\nu,-\alpha-\nu;1;-\zeta/z),
\end{aligned}
\end{equation}
where ${}_2F_1(a,b;c;z)$ is the Gauss hypergeometric function.
This integral is convergent for any $z\in (0,\infty) $, and reproduces the asymptotic series $S_{\infty}(z)$.
Up to a certain order, $S_m(z)$ gets close to the Borel sum $S_\infty^\text{Borel}(z)$. This maximal order is called an \textit{optimal order}.
The optimal order $m^\text{opt}(z)$ depends on $z$. Its simple estimation method is explained in \cite{marino2014}. In the present case, $m^\text{opt}(1)=3$, $m^\text{opt}(2)=4$ and $m^\text{opt}(3)=6$.
See the right panel in Figure~\ref{fig:asymp-sum}.
Beyond the optimal order, the approximation of $S_\infty^\text{Borel}(z)$ by the finite sum $S_m(z)$ gets worse.
This fact sometimes causes a confusion in numerics. If a series expansion is a divergent series, one has to choose a truncation order very carefully.
A facile calculation makes an approximation worse.\footnote{However, up to the optimal order, the truncated sum gives a very good approximate value. This is why perturbation theory in physics is a successful approximation method.}
The Borel summation avoids this problem.

Now we apply the Borel summation to the formal power series \eqref{eq:local-sols-inf}.
The Borel summation of the first equation in \eqref{eq:local-sols-inf} is given by
\begin{equation}
\begin{aligned}
y_{\infty 1}^\text{Borel}(z)=z^\nu S_\infty^\text{Borel}(z)= z^{\nu} \int_0^\infty d\zeta\, e^{-\zeta} {}_2F_1(-\nu,-\alpha-\nu;1;-\zeta/z).
\end{aligned}
\end{equation}
By construction, it has the correct asymptotic expansion in $1/z$.
This solution is analytic in a Stokes sector $-\pi< \arg z< \pi$, but the Stokes phenomenon happens along $\arg z=\pm \pi$.
See Appendix~\ref{app:Borel} for the Stokes phenomenon.
In our analysis here, the Stokes phenomenon is not important because we are interested in $\arg z=0$.
We look for the zeros of the Wronskian for the two analytic solutions $y_{01}(z)$ and $y_{\infty1}^\text{Borel}(z)$.
One can check that the Wronskian for $z>0$ vanishes when $\nu$ is a non-negative integer. 

Of course, in most examples, we cannot construct analytic solutions at boundary points.
We have to use truncated series solutions to evaluate their Wronskians.
If a boundary point is an ordinary point or a regular singular point of a differential equation, we can immediately apply Pad\'e approximants to the truncated series solutions.
It provides us an approximate solution of the analytical one in the complex domain.
For irregular singular points, formal series solutions are usually divergent. For divergent series, Pad\'e approximants usually still works (see Appendix~\ref{app:Pade-div}), but sometimes, due to the Stokes phenomenon, we need the Borel--Pad\'e summation method or numerical solutions.

\subsection{Perturbation theory}\label{subsec:perturbation}
In this subsection, we consider perturbative expansions of eigenvalues in the quantum parameter~$\hbar$.
Since the Schr\"odinger equation \eqref{eq:eigen} is the form of singular perturbation in $\hbar$,
we cannot na\"ively apply the ordinary method of Rayleigh--Schr\"odinger perturbation theory.
Normally singular perturbations are treated by the WKB method that is reviewed in the next subsection.
Here we use a more unconventional treatment.
We will show that results from this method agree with those from the uniform WKB method in the next subsection. 

We first rescale the variable by $x=\sqrt{\hbar}q$. The Schr\"odinger equation \eqref{eq:eigen} is then written as
\begin{equation}
\begin{aligned}
\biggl( -\frac{1}{2}\frac{d^2}{dq^2}+\frac{V(\sqrt{\hbar}q)}{2\hbar} \biggr) \psi(q)=\frac{E}{2\hbar} \psi(q).
\end{aligned}
\label{eq:eigen-pert}
\end{equation}
It is more convenient to define new variables by
\begin{equation}
\begin{aligned}
g=\sqrt{\hbar},\qquad v(x)=\frac{V(x)-V(0)}{2},\qquad \epsilon=\frac{E-V(0)}{2\hbar}.
\end{aligned}
\end{equation}
The Schr\"odinger equation is now written as
\begin{equation}
\begin{aligned}
-\frac{1}{2} \psi''(q)+\frac{v(gq)}{g^2} \psi(q)=\epsilon \psi(q).
\end{aligned}
\label{eq:BW-eq}
\end{equation}
In our analysis, the potential generically has the following form:
\begin{equation}
\begin{aligned}
\frac{v(gq)}{g^2}=\frac{1}{g^2} \sum_{m=0}^M g^{m} v_{m}(gq).
\end{aligned}
\end{equation}
Taking the limit $g \to 0$ with $q$ kept finite, we can expand the potential functions as
\begin{equation}
\begin{aligned}
\frac{v_0(gq)}{g^2}&=\frac{1}{2}\Omega^2 q^2+\sum_{j=3}^\infty g^{j-2}  v_{0,j} q^j, \\
g^{m-2}v_{m}(gq)&=\sum_{j=1}^\infty g^{j+m-2}  v_{m,j} q^j,\qquad m \geq 1,
\end{aligned}
\label{eq:v-pert}
\end{equation}
where $\Omega=\sqrt{v_0''(0)}$ and $v_{m,j}=v_{m}^{(j)}(0)/j!$. We have used $v(0)=0$.
Therefore \eqref{eq:BW-eq} can be regarded as a perturbation of the harmonic oscillator.
In this picture, we zoom in the potential minimum. Near the minimum the potential is a deformation of the harmonic potential.

Now we can apply perturbation theory.
The eigenvalue at the $n$-th energy level is perturbatively expanded as
\begin{equation}
\begin{aligned}
\epsilon_n=\sum_{l=0}^\infty g^{l} \epsilon_{n,l},\qquad \epsilon_{n,0}=\Omega\( n+\frac{1}{2} \).
\end{aligned}
\label{eq:epsilon-pert}
\end{equation}
The usual textbook method is quite complicated to compute high-order corrections for the potential \eqref{eq:v-pert}. Instead we use a smart way by Bender and Wu \cite{bender1969} and a generalization by Sulejmanpasic and \"Unsal \cite{sulejmanpasic2018}.
The idea is very simple. We put the following ansatz of the normalizable eigenfunction:%\footnote{We do not need to normalize it.}
\begin{equation}
\begin{aligned}
\psi_n(q)=\cN_n(g) e^{-\Omega q^2/2}\sum_{l=0}^\infty g^{l} F_{n,l}(q).
\end{aligned}
\label{eq:psi-pert}
\end{equation}
In the perturbative computation, we can fix the overall factor $\cN_n(g)$ as we want.
The zeroth order term $F_{n,0}(q)$ is of course the $n$-th Hermite polynomial $H_n(\sqrt{\Omega} q)$ for the unperturbed harmonic oscillator.
The $l$-th order correction $F_{n,l}(q)$ is a polynomial of at most degree $n+3l$,
\begin{equation}
\begin{aligned}
F_{n,l}(q)=\sum_{k=0}^{n+3l} A_{n,l}^{k} q^k,
\end{aligned}
\end{equation}
where we use a freedom of $\cN_n(g)$ so that
\begin{equation}
\begin{aligned}
A_{n,l}^n=\delta_{l0}.
\end{aligned}
\end{equation}
This ansatz is very powerful. In fact, the Schr\"odinger equation determines all the coefficients in the polynomial $F_{n,k}(q)$ as well as $\epsilon_{n,k}$ recursively.
This fact was first observed in the quartic anharmonic oscillator \cite{bender1969}, and it was recently generalized in \cite{sulejmanpasic2018} to arbitrary potentials with a locally harmonic minimum. 

We borrow the result in \cite{sulejmanpasic2018}.
First, we have relations
\begin{equation}
\begin{aligned}
A_{n,l}^k=\frac{1}{2\Omega(k-n)}\biggl[(k+2)(k+1)A_{n,l}^{k+2}+\sum_{j=1}^{l-1}2\epsilon_{n,j} A_{n,l-j}^k
-2\sum_{j=1}^l v_{0,j}A_{n,l-j}^{k-j-2}\\
-2\sum_{j=1}^l \sum_{m=1}^M v_{m,j}A_{n,l-j+2-m}^{k-j}\biggr]\qquad
(n+1 \leq k \leq n+3k).
\end{aligned}
\label{eq:A-recur-1}
\end{equation}
where we set $A_{n,l}^k=0$ for $k>n+3k$. This determines $A_{n,l}^k$ for $k=n+3k, n+3k-1, \dots, n+1$ uniquely.
Next, the energy correction is obtained by
\begin{equation}
\begin{aligned}
\epsilon_{n,l}=-\frac{(n+2)(n+1)}{2}A_{n,l}^{n+2}-\sum_{j=1}^{l-1}\epsilon_{n,j} A_{n,l-j}^n+\sum_{j=1}^l v_{0,j}A_{n,l-j}^{n-j-2}+\sum_{j=1}^l \sum_{m=1}^M v_{m,j}A_{n,l-j+2-m}^{n-j}.
\end{aligned}
\end{equation}
Finally, we fix the remaining coefficients by
\begin{equation}
\begin{aligned}
A_{n,l}^k=\frac{1}{2\Omega(k-n)}\biggl[(k+2)(k+1)A_{n,l}^{k+2}+\sum_{j=1}^{l}2\epsilon_{n,j} A_{n,l-j}^k
-2\sum_{j=1}^l v_{0,j}A_{n,l-j}^{k-j-2}\\
-2\sum_{j=1}^l \sum_{m=1}^M v_{m,j}A_{n,l-j+2-m}^{k-j}\biggr]\qquad
(0 \leq k \leq n-1).
\end{aligned}
\end{equation}
These relations are easily solved by symbolic computation systems.
In general, there are no odd order corrections in the energy: $\epsilon_{n,l}=0$ for all odd $l$.
The perturbative series of the original energy is then given by
\begin{equation}
\begin{aligned}
E_n=V(0)+2\hbar \epsilon_n=V(0)+2\hbar \Omega\biggl(n+\frac{1}{2}\biggr)+2\hbar \sum_{l=1}^\infty \hbar^{l}\epsilon_{n,2l}.
\end{aligned}
\end{equation}

Let us apply it to the Morse potential. We start with the rescaled eigen-equation \eqref{eq:reduced}.
After changing variables $q \to \sqrt{g}q$ and $\cE=(\varepsilon+1)/(2g)$, this is rewritten as
\begin{equation}
\begin{aligned}
\biggl( -\frac{1}{2} \frac{d^2}{dq^2}+\frac{1}{2}q^2+\sum_{k=1}^\infty g^{k/2} \frac{2^{k+2}-2}{2(k+2)!} q^{k+2} \biggr) \psi(q)=\cE \psi(q).
\end{aligned}
\end{equation}
It turns out that the energy eigenvalue has only the first order perturbative correction:
\begin{equation}
\begin{aligned}
\cE_n=n+\frac{1}{2}-\frac{g}{2}\(n+\frac{1}{2} \)^2+0g^2+0g^3+0g^4+\cdots.
\end{aligned}
\end{equation}
Of course, this is a reflection of the hidden supersymmetric structure in the Morse potential.
This cancellation is directly confirmed by the \textit{Mathematica} package in \cite{sulejmanpasic2018} up to any desired orders in $g$.
The result is consistent with the exact result \eqref{eq:spec-Morse}.
However, we should note that the wave function receives an infinite number of corrections.
For instance, the ground state eigenfunction is given by
\begin{equation}
\begin{aligned}
\psi_0(q)=e^{-q^2/2}\biggl( 1-\frac{g^{1/2}}{6}q(q ^2+3)+\frac{g}{72} q^2 (q ^4+3 q ^2+9)
-\frac{g^{3/2}}{6480}q^3(5 q ^6+54 q ^2+135) \\
+\frac{g ^2}{155520}q ^4 (5 q ^8-30 q ^6+81 q ^4+162 q ^2+405)+\cdots \biggr).
\end{aligned}
\end{equation}
It is a good exercise to compare it with the exact eigenfunction.

A merit of the perturbative series is that one can easily analytically continue the Planck parameter to the complex domain.
Therefore the resonant eigenvalues are obtained in this way.
In the case of the Morse potential, the perturbative series accidentally stops at the first order, but this is not the case for most models.
Moreover the perturbative series \eqref{eq:epsilon-pert} is not convergent in general. Therefore we need to resum it by the Borel summation method (or Pad\'e approximants).
This point is important to obtain correct numerical values.

\subsection{The WKB method}\label{subsec:WKB}
The WKB method is a very powerful tool to investigate global properties of the wave function.
Typically it is used to analyze quantum tunneling effects.
We review two WKB methods.

\subsubsection{Standard WKB}
Let us start with the original WKB method. We use \eqref{eq:Milne}.
Since the equation takes the form of the singular perturbation in $\hbar$, we put the ansatz:\footnote{If one puts the ansatz $\psi(x)=\exp [\frac{i}{\hbar} \int^x \widehat{p}(x') dx' ]$, one obtains the Riccati equation for $\widehat{p}(x)$. It is solved order by order in $\hbar$. The odd order part in the perturbative solution can be expressed by a derivative of the even order part, and it finally leads to \eqref{eq:WKB-sol}. Hence $p(x)$ has only the even order corrections as in \eqref{eq:p-pert}. See \cite{froman1996, kawai2005} for a rigorous proof.}
\begin{equation}
\begin{aligned}
\psi(x)=\frac{1}{\sqrt{p(x)}} \exp \biggl[ \frac{i}{\hbar} \int^x p(x') dx' \biggr],
\end{aligned}
\label{eq:WKB-sol}
\end{equation}
where $p(x)$ satisfies the non-linear differential equation:\footnote{We notice that this equation is formally equivalent to Milne's non-linear equation \eqref{eq:deq-w} if identifying $p(x)=-i\hbar/w_E(x)^2$.}
\begin{equation}
\begin{aligned}
p(x)^2-Q(x)+\hbar^2 \biggl( \frac{p''(x)}{2p(x)}-\frac{3p'(x)^2}{4p(x)^2} \biggr)=0.
\end{aligned}
\label{eq:eq-p}
\end{equation}
We can solve \eqref{eq:eq-p} perturbatively in $\hbar$:
\begin{equation}
\begin{aligned}
p(x)=\sum_{k=0}^\infty \hbar^{2k} p_k(x).
\end{aligned}
\label{eq:p-pert}
\end{equation}
At the lowest order, we have two branches of the solution:
\begin{equation}
\begin{aligned}
p_{0}^{\pm}(x)=\pm \sqrt{Q(x)}=\pm \sqrt{E-V(x)}.
\end{aligned}
\end{equation}
For each branch, the quantum corrections are uniquely fixed.
Hence these two branches leads to the two solutions of the Schr\"odinger equation. 
In general, these two are independent.
Once one of them is constructed, the other is easily obtained.
Therefore, we abbreviate the subscripts $\pm$. 
The first two corrections in $p(x)$ are given by
\begin{equation}
\begin{aligned}
p_1&=-\frac{p_0''}{4p_0^2}+\frac{3p_0'^2}{8p_0^3}, \\
p_2&=\frac{p_0''''}{16p_0^4}-\frac{5p_0' p_0'''}{8p_0^5}-\frac{13p_0''^2}{32p_0^5}+\frac{99p_0'^2 p_0''}{32p_0^6}-\frac{297p_0'^4}{128p_0^7}.
\end{aligned}
\end{equation}
The higher order corrections rapidly get so lengthy.

For practical computations, a sophisticated way in \cite{froman1996} is very useful.%
\footnote{In the phase-integral formalism in \cite{froman1996}, there is a freedom to choose a ``base function'' that depends on situations. %We use this freedom in subsection~\ref{sec:BH2}. In this subsection, 
We simply choose it as $Q(x)$ itself. A change of the base function might improve the approximation. This technical issue is not a purpose of this review. See \cite{froman1996}.}
Let us introduce the following notations:
\begin{equation}
\begin{aligned}
q(x):=\frac{p(x)}{p_0(x)},\qquad \hat{D}:= \frac{1}{p_0(x)}\frac{d}{dx}.
\end{aligned}
\end{equation}
The non-linear equation \eqref{eq:eq-p} now becomes
\begin{equation}
\begin{aligned}
1-q(x)^2+\hbar^2 \Big( \epsilon_0(x)+q(x)^{1/2}\hat{D}^2 q(x)^{-1/2} \Big)=0,
\end{aligned}
\label{eq:eq-q}
\end{equation}
where
\begin{equation}
\begin{aligned}
\epsilon_0(x):=p_0(x)^{-3/2} \frac{d^2}{dx^2} p_0(x)^{-1/2}
=\frac{5Q'(x)^2}{16Q(x)^3}-\frac{Q''(x)}{4Q(x)^2}.
\end{aligned}
\end{equation}
We solve \eqref{eq:eq-q} perturbatively,
\begin{equation}
\begin{aligned}
q(x)=1+\sum_{k=1}^\infty \hbar^{2k} Y_k(x).
\end{aligned}
\end{equation}
By construction we have $p_k(x)=Y_k(x)p_0(x)$.
The non-linear equation \eqref{eq:eq-q} leads to a recurrence relation for $Y_k(x)$ \cite{froman1996}.
This function can be separated as
\begin{equation}
\begin{aligned}
Y_k(x)=Z_k(x)+\hat{D} U_k(x).
\end{aligned}
\label{eq:separation}
\end{equation}
Note that this separation is not unique. We require that $Z_k(x)$ gets as ``simple'' as possible, as shown below. 
It has the following advantage.
In the WKB approach, a contour integral around two turning points\footnote{A turning point $a$ is defined by $Q(a)=E-V(a)=0$.} 
\begin{equation}
\begin{aligned}
\oint_{C(a,b)} p_k(x) dx,
\end{aligned}
\end{equation}
is particularly important. Here the contour $C(a,b)$ is a closed curve encircling only the two turning points $x=a$ and $x=b$.
Using the above notation, this integral is written as
\begin{equation}
\begin{aligned}
\oint_{C(a,b)} p_k(x) dx &=\oint_{C(a,b)} Y_k(x)p_0(x) dx=\oint_{C(a,b)} \( Z_k(x)p_0(x)+\frac{d}{dx}U_k(x)\) dx \\
&=\oint_{C(a,b)} Z_k(x)p_0(x) dx.
\end{aligned}
\end{equation}
Therefore the derivative term $\hat{D} U_k(x)$ does not contribute to the contour integral.
This fact drastically reduces the computational cost.
For our purpose in spectral problems, it is sufficient to compute $Z_k$. These are much simpler than $p_k$.
Up to order $\hbar^{12}$ ($k=6$), we have 
\begin{equation}
\begin{aligned}
Z_1&=\frac{1}{2}\epsilon_0, \quad Z_2=-\frac{1}{8}\epsilon_0^2,\quad
Z_3=\frac{1}{32}( 2\epsilon_0^3-\epsilon_1^2 ), \\
Z_4&=-\frac{1}{128}(5\epsilon_0^4-10\epsilon_0 \epsilon_1^2+\epsilon_2^2),\quad
Z_5=\frac{1}{512}(14\epsilon_0^5-70\epsilon_0^2\epsilon_1^2+14\epsilon_0 \epsilon_2^2-\epsilon_3^2),\\
Z_6&=-\frac{1}{2048}(42\epsilon_0^6-420\epsilon_0^3\epsilon_1^2-35\epsilon_1^4+126\epsilon_0^2\epsilon_2^2
+20\epsilon_2^3-18\epsilon_0\epsilon_3^2+\epsilon_4^2),
\end{aligned}
\label{eq:Z_k}
\end{equation}
where $\epsilon_k:=\hat{D}^k \epsilon_0$. We perform the separation \eqref{eq:separation} so that $Z_k$ does not contain higher derivative functions $\epsilon_k$. This requirement uniquely fixes the separation.
The explicit forms of $U_k$ are found in \cite{froman1996, froman1992}. They are not important in our analysis.

Sometimes, the function $Q(x)$ depends on $\hbar$ explicitly.
For example, in the next section, we encounter the situation such as
\begin{equation}
\begin{aligned}
Q(x)=Q_0(x)+\hbar^2 Q_1(x).
\end{aligned}
\end{equation}
Even in this case, we can still use the above formulae. The equation \eqref{eq:eq-p} is now written as
\begin{equation}
\begin{aligned}
p(x)^2-Q_0(x)+\hbar^2 \biggl(-Q_1(x) +\frac{p''(x)}{2p(x)}-\frac{3p'(x)^2}{4p(x)^2} \biggr)=0.
\end{aligned}
\label{eq:eq-p-2}
\end{equation}
Then the equation \eqref{eq:eq-q} becomes
\begin{equation}
\begin{aligned}
1-q(x)^2+\hbar^2 \Big( \epsilon_0(x)+\frac{Q_1(x)}{p_0(x)^2}+q(x)^{1/2}\hat{D}^2 q(x)^{-1/2} \Big)=0,
\end{aligned}
\label{eq:eq-q-2}
\end{equation}
where $p_0(x)=\sqrt{Q_0(x)}$. If we define
\begin{equation}
\begin{aligned}
\bar{\epsilon}_0(x):=\epsilon_0(x)+\frac{Q_1(x)}{p_0(x)^2},
\end{aligned}
\label{eq:bar-epsilon_0}
\end{equation}
then the formulae \eqref{eq:Z_k} still hold by replacing $\epsilon_k \to \bar{\epsilon}_k=\hat{D}^k \bar{\epsilon}_0$.

In general, the turning points in the contour integral are complex, and we have to consider the WKB solutions in the complex domain.
The WKB method still works in the complex domain \cite{landau1981, voros1983}.
In the resonant spectral problem, such a complex analysis is particularly important.
Since the lowest function $p_0(x)$ is a multivalued function, we have to choose branch cuts carefully in numerical calculations.

\begin{figure}[tb]
\begin{center}
  \begin{minipage}[b]{0.4\linewidth}
    \centering
    \includegraphics[width=0.95\linewidth]{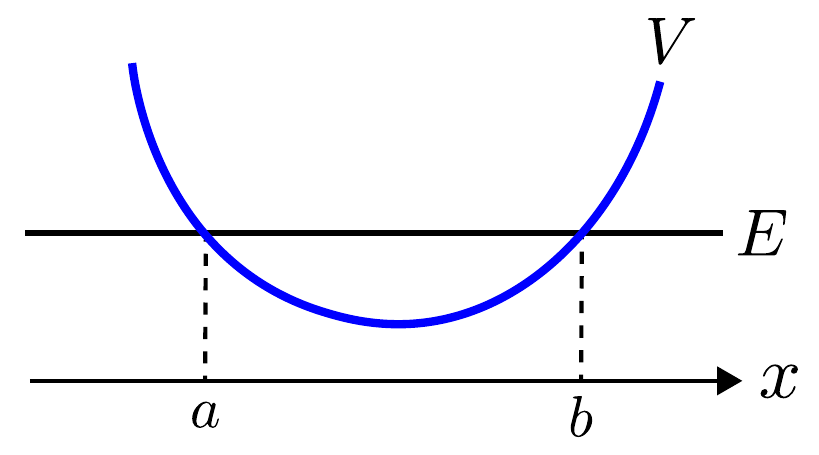}
  \end{minipage}\hspace{1truecm}
  \begin{minipage}[b]{0.4\linewidth}
    \centering
    \includegraphics[width=0.95\linewidth]{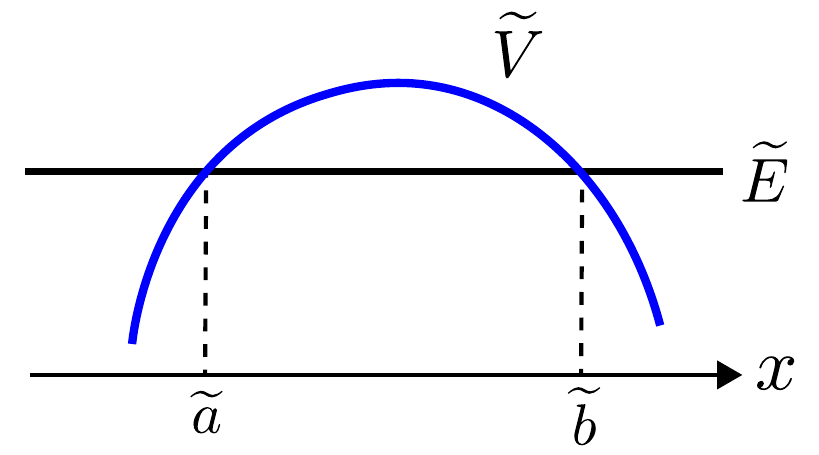}
  \end{minipage} 
\end{center}
  \caption{Two turning points in the WKB method.}
  \label{fig:WKB}
\end{figure}

Let us see the computations of the bound state energy and the resonant energy in the WKB method.
To find these eigenvalues, we need to solve the connection problems at the semiclassical level.
We start with the connection problem for the bound states shown in the left panel in Figure~\ref{fig:WKB}.
There are two real turning points $a<b$.
We use the lowest order WKB solution:
\begin{equation}
\begin{aligned}
\psi(x)\sim \frac{1}{\sqrt{p_0(x)}} \exp \biggl[ \frac{i}{\hbar} \int^x p_0(x') dx' \biggr]. 
\end{aligned}
\label{eq:WKB-0}
\end{equation}
Let us define
\begin{equation}
\begin{aligned}
w(x_1, x_2):=\frac{1}{\hbar} \int_{x_1}^{x_2} \sqrt{E-V(x)}dx,\qquad
\sigma(x_1, x_2):=\frac{1}{\hbar} \int_{x_1}^{x_2} \sqrt{V(x)-E}\, dx.
\end{aligned}
\end{equation}
The solution satisfying the boundary condition at $x \to -\infty$ is given by
\begin{equation}
\begin{aligned}
\psi(x) \sim \frac{1}{\sqrt{p_0(x)}} e^{-\sigma(x,a)},
\end{aligned}
\end{equation}
where we ignore irrelevant constants.
Using the connection formula in \cite{konishi2009}, we can continue this solution to the region in $x \to +\infty$.
The result is
\begin{equation}
\begin{aligned}
\frac{1}{\sqrt{p_0(x)}} e^{-\sigma(x,a)}
\stackrel{-\infty \leftarrow x}{\longleftarrow} \psi(x) \stackrel{x\rightarrow +\infty}{\longrightarrow}
\frac{\cos \phi}{\sqrt{p_0(x)}} e^{-\sigma(b,x)}+\frac{2\sin \phi}{\sqrt{p_0(x)}} e^{\sigma(b,x)},
\end{aligned}
\end{equation}
where
\begin{equation}
\begin{aligned}
\phi=\frac{\pi}{2}-w(a,b).
\end{aligned}
\end{equation}
In the region $x \to +\infty$, the first term satisfies the boundary condition.
We conclude that the bound state boundary condition requires
\begin{equation}
\begin{aligned}
\sin \phi=0 \quad \Longleftrightarrow \quad w(a,b)=\pi \( n+\frac{1}{2} \), \quad n=0,1,2,\dots.
\end{aligned}
\end{equation}
where we have used $w(a,b)>0$ for $a<b$.
This equation is well-known as the Bohr--Sommerfeld quantization condition.

It seems hard to extend this argument when the quantum corrections are taken into account.
Instead we rewrite the quantization condition as follows:
\begin{equation}
\begin{aligned}
\oint_{C(a,b)} p_0(x) dx=2\pi \hbar \( n+\frac{1}{2} \),
\end{aligned}
\label{eq:BS}
\end{equation}
where $p_0(x)=\sqrt{Q(x)}$ should be understood as a multi-valued complex function that defines a Riemann surface.
Geometrically the left hand side gives an area surrounded by the curve $p^2+V(x)=E$ in the phase space $(x,p)$. 
This quantization condition is simply derived by the following argument. We assume that there are no singular points of the Schr\"odinger equation inside the closed curve $C(a,b)$.
Then the wave function should be single-valued when $x$ goes around $C(a,b)$.
After encircling it, the WKB solution \eqref{eq:WKB-0} apparently becomes
\begin{equation}
\begin{aligned}
\frac{1}{\sqrt{p_0(x)}} \exp\biggl[ \frac{i}{\hbar} \int^x p_0(x') dx' \biggr] \to -\frac{1}{\sqrt{p_0(x)}}\exp\biggl[ \frac{i}{\hbar} \oint_{C(a,b)} p_0(x) dx +\frac{i}{\hbar} \int^x p_0(x') dx' \biggr],
\end{aligned}
\label{eq:WKB-monodromy}
\end{equation}
where the minus sign comes from the multi-valuedness of $\sqrt{p_0(x)}$.\footnote{Since $p_0(x)$ has a square root branch cut, it comes back to the same value when going around the two turning points. However $\sqrt{p_0(x)}$ changes its sign.} 
The single-valued condition requires the quantization condition \eqref{eq:BS}.
This argument is easily extended to the formal WKB solution \eqref{eq:WKB-sol} \cite{dunham1932}.
The result is given by
\begin{equation}
\begin{aligned}
\sum_{k=0}^\infty \hbar^{2k} \oint_{C(a,b)} p_k(x) dx \simeq 2\pi \hbar \( n+\frac{1}{2} \),
\end{aligned}
\label{eq:qBS}
\end{equation}
or equivalently
\begin{equation}
\begin{aligned}
\sum_{k=0}^\infty \hbar^{2k} \oint_{C(a,b)} Z_k(x)p_0(x) dx \simeq 2\pi \hbar \( n+\frac{1}{2} \).
\end{aligned}
\label{eq:qBS-2}
\end{equation}

Two remarks should be mentioned in order.
As for usual perturbative series, the left hand side is a formal divergent series in $\hbar$.
Therefore the equality in \eqref{eq:qBS} and \eqref{eq:qBS-2} should be understood in the asymptotic sense.
The Borel resummation of the left hand side causes the Stokes phenomenon in the complex $\hbar$-plane.
Therefore, in general, the left hand side may have non-perturbative corrections in $\hbar$.
Its analysis is quite complicated. This is beyond the scope of this article.
We refer to \cite{balian1978, voros1983, delabaere1999, kawai2005} on this issue.
The other remark is that if there is a singular point of the Schr\"odinger equation inside $C(a,b)$, one has to consider a monodromy (see Appendix~\ref{app:regular}).
Then the required condition is a matching condition of two monodromies of the analytic solution and of the WKB solution.
The monodromy of the WKB solution is given by \eqref{eq:WKB-monodromy}, and our task is to know the monodromy of the true solution.

The argument of the resonant states is in parallel. See the right panel in Figure~\ref{fig:WKB}.
Let us define
\begin{equation}
\begin{aligned}
\widetilde{w}(x_1, x_2):=\frac{1}{\widetilde{\hbar}} \int_{x_1}^{x_2} \sqrt{\widetilde{E}-\widetilde{V}(x)}dx,\qquad
\widetilde{\sigma}(x_1, x_2):=\frac{1}{\widetilde{\hbar}} \int_{x_1}^{x_2} \sqrt{\widetilde{V}(x)-\widetilde{E}}\, dx.
\end{aligned}
\end{equation}
In the case of the resonance, we start with the transmitted wave in $x \to -\infty$, and connect it to the waves in $x \to +\infty$.
The connection formula leads to
\begin{equation}
\begin{aligned}
T\frac{1}{\sqrt{\widetilde{p}_0(x)}} e^{-i\widetilde{w}(x,\widetilde{a})}
\stackrel{-\infty \leftarrow x}{\longleftarrow} \widetilde{\psi}(x) \stackrel{x\rightarrow +\infty}{\longrightarrow}
\frac{1}{\sqrt{\widetilde{p}_0(x)}} e^{-i\widetilde{w}(\widetilde{b},x)}+R\frac{1}{\sqrt{\widetilde{p}_0(x)}} e^{i\widetilde{w}(\widetilde{b},x)},
\end{aligned}
\end{equation}
where $\widetilde{a}$ and $\widetilde{b}$ are two turning points, and
\begin{equation}
\begin{aligned}
R=\frac{e^{-i\delta}}{\sqrt{1+e^{-2\widetilde{\sigma}(\widetilde{a},\widetilde{b})}}},\qquad
T=\frac{e^{-\widetilde{\sigma}(\widetilde{a},\widetilde{b})-i\delta}}{\sqrt{1+e^{-2\widetilde{\sigma}(\widetilde{a},\widetilde{b})}}}.
\end{aligned}
\end{equation}
with some phase shift $\delta$ that is not relevant in our analysis. See \cite{konishi2009}.
As seen in subsection~\ref{subsec:bound}, the resonant states require $1/R=0$, and semiclassically it yields
\begin{equation}
\begin{aligned}
1+e^{-2\widetilde{\sigma}(\widetilde{a},\widetilde{b})}=0.
\end{aligned}
\end{equation}
Therefore we obtain the Bohr--Sommerfeld condition for the resonant energy:
\begin{equation}
\begin{aligned}
\widetilde{\sigma}(\widetilde{a},\widetilde{b})=\pi i \( n+\frac{1}{2} \),\qquad n=0,\pm 1, \pm 2, \dots.
\end{aligned}
\label{eq:BS-resonance}
\end{equation}
For the resonant states, $\widetilde{\sigma}(\widetilde{a},\widetilde{b})$ must be complex-valued. There are no reasons to exclude $n<0$.
However, the spectra for $n=m$ and for $n=-m-1$ ($m=0,1,2,\dots$) are symmetric.\footnote{The case of $n<0$ corresponds to another analytic continuation $\hbar=-i\widetilde{\hbar}$ in our framework.} This is checked by taking the complex conjugate to \eqref{eq:BS-resonance}. We can restrict on $n \geq 0$ without loss of generality.

If two turning points $\widetilde{a}$ and $\widetilde{b}$ are both real, $\widetilde{\sigma}(\widetilde{a},\widetilde{b})$ is necessarily real.
Therefore the resonance condition is never satisfied. This implies that the resonant energy should be complex-valued.
We have to carefully choose the correct complex turning points when we solve the quantization condition \eqref{eq:BS-resonance}.
A simple criteria is to trace continuous variations from real turning points to complex ones.
To extend to the quantum corrected result, it is useful to rewrite the quantization condition as
\begin{equation}
\begin{aligned}
\oint_{C(\widetilde{a},\widetilde{b})} \widetilde{p}_0(x) dx=2\pi\widetilde{\hbar} \( n+\frac{1}{2} \),
\end{aligned}
\label{eq:BS-resonance-contour}
\end{equation}
It is straightforward to derive the quantum corrected condition:
\begin{equation}
\begin{aligned}
\sum_{k=0}^\infty \widetilde{\hbar}^{2k} \oint_{C(\widetilde{a},\widetilde{b})} \widetilde{p}_k(x) dx\simeq 2\pi \widetilde{\hbar}\( n+\frac{1}{2} \),
\end{aligned}
\end{equation}
or equivalently
\begin{equation}
\begin{aligned}
\sum_{k=0}^\infty \widetilde{\hbar}^{2k} \oint_{C(\widetilde{a},\widetilde{b})} \widetilde{Z}_k(x) \widetilde{p}_0(x) dx\simeq 2\pi \widetilde{\hbar}\( n+\frac{1}{2} \).
\end{aligned}
\label{eq:qBS-res}
\end{equation}
We have seen that the (all-order) Bohr--Sommerfeld quantization conditions for the bound states and the resonant states are almost same.
These are related by the analytic continuation. This is a reason why we can expect the spectral relation \eqref{eq:B-R}.

For the Morse model \eqref{eq:reduced}, we compute the Bohr--Sommerfeld integral.
The two turning points are given by
\begin{equation}
\begin{aligned}
a=\log(1-\sqrt{1+\varepsilon}),\qquad b=\log(1+\sqrt{1+\varepsilon}).
\end{aligned}
\end{equation}
These are real as long as $-1\leq \varepsilon<0$.
The classically allowed integral is exactly given by
\begin{equation}
\begin{aligned}
w(a,b)=\frac{1}{g}\int_a^b \sqrt{\varepsilon-e^{2q}+2e^q}\, dq=\frac{\pi}{g} (1-\sqrt{-\varepsilon}).
\end{aligned}
\end{equation}
The Bohr--Sommerfeld condition turns out to give the exact result in this case:
\begin{equation}
\begin{aligned}
\frac{\pi}{g} (1-\sqrt{-\varepsilon_n})=\pi \( n+\frac{1}{2}\) \quad \Longrightarrow \quad
\varepsilon_n=-\biggl[ 1-g\biggl(n+\frac{1}{2} \biggr) \biggr]^2.
\end{aligned}
\end{equation}
The computation of the resonant states for the inverted potential is almost the same:
\begin{equation}
\begin{aligned}
\widetilde{\sigma}(\widetilde{a},\widetilde{b})=\frac{1}{\widetilde{g}}\int_{\widetilde{a}}^{\widetilde{b}} \sqrt{-e^{2q}+2e^{q}-\widetilde{\varepsilon}}\; dq=
\frac{\pi}{\widetilde{g}}(1-\sqrt{\widetilde{\varepsilon}}),
\end{aligned}
\end{equation}
where\footnote{At this moment, we assume that $\widetilde{a}$ and $\widetilde{b}$ are real numbers such that $\widetilde{a}<\widetilde{b}$, but for the resonant eigenvalues these are actually complex. We easily perform an analytic continuation in this case. If we cannot evaluate the integral analytically, things are more delicate.} $\widetilde{a}=\log(1-\sqrt{1-\widetilde{\varepsilon}})$, $\widetilde{b}=\log(1+\sqrt{1-\widetilde{\varepsilon}})$ and
\begin{equation}
\begin{aligned}
\frac{\pi}{\widetilde{g}} (1-\sqrt{\widetilde{\varepsilon}_n})=\pi i\( n+\frac{1}{2}\) \quad \Longrightarrow \quad
\widetilde{\varepsilon}_n=\biggl[ 1-i\tilde{g}\biggl(n+\frac{1}{2} \biggr) \biggr]^2.
\end{aligned}
\end{equation}
It would be interesting to check that the quantum corrections really do not contribute to the all-order Bohr--Sommerfeld quantization condition.

\subsubsection{Uniform WKB}
The uniform WKB method has a little bit different flavor.
In the standard WKB method, it is not so easy (but possible) to derive the semiclassical perturbative expansion of the energy eigenvalue itself.
In perturbation theory, this is easy, but it is rather hard (but possible) to see the global structure of the solutions because we zoom in the potential well/wall.
The uniform WKB method has both advantages of these two approaches in a sense.
This is particularly powerful in the analysis of non-perturbative instanton corrections to the spectrum. See \cite{alvarez2004, dunne2014} for instance.
In this article, we just show a perturbative aspect to mention a relation to previous works in black hole physics.

The starting point is the following uniform WKB ansatz of the solution:
\begin{equation}
\begin{aligned}
\psi(x)=\frac{1}{\sqrt{u'(x)}} D_\nu \biggl( \frac{u(x)}{\sqrt{\hbar}} \biggr),
\end{aligned}
\end{equation}
where $D_\nu(z)$ is the parabolic cylinder function that satisfies the Weber equation:
\begin{equation}
\begin{aligned}
y''(z)+\( \nu+\frac{1}{2}-\frac{z^2}{4} \)y(z)=0.
\end{aligned}
\end{equation}
This differential equation is essentially same as the Sch\"odinger equation for the harmonic oscillator.
The other solution is given by $D_{-\nu-1}(iz)$. Plugging the uniform ansatz into the Schr\"odinger equation \eqref{eq:eigen}, one obtains
\begin{equation}
\begin{aligned}
\frac{1}{4} u^2 u'^2+E-V-\hbar\( \nu+\frac{1}{2} \) u'^2+\hbar^2\( \frac{3u''^2}{4u'^2}-\frac{u'''}{2u'} \)=0.
\end{aligned}
\end{equation}
For simplicity, we assume that the potential has the (global) minimum $V_\text{min}=0$ at $x=0$.
This is easily realized by constant shifts of $x$ and $E$.
The equation is solved perturbatively
\begin{equation}
\begin{aligned}
u(x)=\sum_{l=0}^\infty \hbar^l u_l(x),\qquad
E=\sum_{l=1}^\infty \hbar^l E_l.
\end{aligned}
\end{equation}
At the lowest order, we have
\begin{equation}
\begin{aligned}
u_0(x) u_0'(x)=\pm 2\sqrt{V(x)}.
\end{aligned}
\label{eq:u0-u0'}
\end{equation}
It is formally solved by
\begin{equation}
\begin{aligned}
u_0(x)^2= 4\int_{0}^x \sqrt{V(x')} dx',
\end{aligned}
\label{eq:u0}
\end{equation}
where we have chosen the positive branch. We have also required the condition $u_0(0)=0$ to fix the integration constant.
The reason is as follows.
If $u_0(0)\ne 0$, then $u_0'(0)=0$ because of $V(0)=0$ and \eqref{eq:u0-u0'}. In this case, the wave function is divergent at $x=0$.
We exclude this unphysical case.
As in the standard WKB, two branches of the square root in \eqref{eq:u0} lead to two independent solutions.
We can choose the plus sign without loss of generality.

Let us expand the potential around $x=0$:
\begin{equation}
\begin{aligned}
V(x)=\sum_{k=2}^\infty V_k x^k.
\end{aligned}
\end{equation}
Then the solution $u_0(x)$ has the series representation
\begin{equation}
\begin{aligned}
u_0(x)=\sqrt{2}V_2^{1/4}x+\frac{V_3}{3\sqrt{2}V_2^{3/4}}x^2+\frac{-13V_3^2+36V_2V_4}{144\sqrt{2}V_2^{7/4}}x^3+\cO(x^4),\quad x \to 0.
\end{aligned}
\end{equation}

At the first order, we have
\begin{equation}
\begin{aligned}
E_1-\( \nu+\frac{1}{2} \)u_0'^2+\frac{1}{2} u_0 u_0' (u_0 u_1)'=0.
\end{aligned}
\end{equation}
It is not easy to solve this equation for $u_1$ analytically. However our goal is to get the spectrum. This is done by the regularity condition of $u_1$ at $x=0$.
The result is
\begin{equation}
\begin{aligned}
E_1=2\sqrt{V_2} \( \nu+\frac{1}{2} \).
\end{aligned}
\end{equation}
Similarly at the second order, the regularity of $u_2$ at $x=0$ as well as $u_0$ and $u_1$ determines
\begin{equation}
\begin{aligned}
E_2=-\frac{7+60\alpha^2}{32} \frac{V_3^2}{V_2^2}+\frac{3(1+4\alpha^2)}{8}\frac{V_4}{V_2},
\end{aligned}
\end{equation}
where $\alpha:=\nu+1/2$. In this way, we can fix the perturbative coefficients order by order.
There is no difficulty to push the computation for higher orders.

So far, we do not require the normalizability of the wave function.
It is realized by the special condition $\nu=n$ ($n=0,1,2,\dots$) because in this case the parabolic cylinder function $D_\nu(z)$ becomes normalizable.
The perturbative coefficients of the energy eigenvalue should be compared with the result from the perturbative method in subsection~\ref{subsec:perturbation}.
It turns out that both are in agreement as expected. See Eq. (2.7) in \cite{hatsuda2020}.

A very similar approach is found in the context of black hole perturbation theory \cite{mashhoon1983, schutz1985, iyer1987}.
There, the connection problem near the potential peak was solved by using the parabolic cylinder function.
In black hole perturbation theory, the method in \cite{mashhoon1983, schutz1985, iyer1987} is conventionally called the WKB approach.
In this article, we refer to it as the uniform WKB approach to distinguish it from the standard WKB reviewed in the previous subsection.
At the perturbative level, the uniform WKB method leads to the same result in perturbation theory.
Its true value is revealed in non-perturbative analysis, which is beyond the scope of this article.

\section{Applications to black hole perturbation theory}\label{sec:BH1}
Now we proceed to applications to two kinds of problems in black hole perturbation theory.
First, we review computations of QNM frequencies of Schwarzschild black holes.
They are well-studied and a very good playground for the applications.
Next, we look at a mode stability problem of black strings in five-dimension. 
It is well-known that they show the Gregory--Laflamme instability.
We show that the method in subsection~\ref{subsec:numerical} allows us to evaluate a critical value of a parameter in this phase transition numerically.
This method also judges a mode (in)stability of a given black hole easily.

\subsection{Quasinormal modes of Schwarzschild black holes} 
For notational simplicity, we abbreviate tildes in this subsection though we consider a resonant state problem. It will cause no confusions.
Let us see the QNM problem of the four-dimensional Schwarzschild spacetime. We will present basics on linear perturbation theory of this geometry in Appendix~\ref{app:BH-pert}. There are also several excellent review articles on the QNMs \cite{kokkotas1999, nollert1999, ferrari2008, berti2009, konoplya2011}. 
The odd parity gravitational perturbation finally leads to the master equation with the effective potential \eqref{eq:V-Regge--Wheeler}.
As we will discuss in Appendix~\ref{app:BH-pert}, it is well-known that the odd parity and the even parity perturbations have the same QNM spectra. Since the former master equation is simpler than the latter one, we can concentrate our attention to the odd parity sector.

For later convenience, we use a dimensionless variable $z=r/r_H=r/(2M)$ rather than the radial variable $r$ itself.
The master equation then takes the form
\begin{equation}
\begin{aligned}
\( f(z)\frac{d}{dz} f(z)\frac{d}{dz} +(2M\omega)^2-V_s(z) \) \phi(z)=0,
\end{aligned}
\label{eq:RW}
\end{equation}
where
\begin{equation}
\begin{aligned}
f(z)=1-\frac{1}{z},\qquad
V_s(z)=f(z)\( \frac{\ell(\ell+1)}{z^2}+(1-s^2) \frac{1}{z^3} \).
\end{aligned}
\label{eq:RW-potential}
\end{equation}
Here $s=0,1,2$ is the spin weight of the perturbing field.
To see a direct connection to the Schr\"odinger equation, it is useful to introduce the so-called tortoise coordinate by
\begin{equation}
\begin{aligned}
\frac{dx}{dz}=\frac{1}{f(z)} \quad \Longleftrightarrow \quad x=z+\log\( z-1\)+c,
\end{aligned}
\label{eq:tortoise}
\end{equation}
where we fix the integration constant $c$ so that the potential $V_s$ has a peak at $x=0$. This choice is not essential in the following analysis.
In the tortoise coordinate $x$, the potential $V_s$ has the shape shown as the dashed line in the right panel of Figure~\ref{fig:potentials}. The analytic form in terms of $x$ is complicated.
The resonant boundary condition is then given by
\begin{equation}
\begin{aligned}
\phi \to \begin{cases}
e^{-2iM \omega x}\sim (z-1)^{-2iM \omega} \qquad &(x \to -\infty) ,\\
e^{+2iM \omega x}\sim e^{+2iM \omega z} \qquad &(x \to +\infty).
\end{cases}
\end{aligned}
\end{equation}
As we will see below, sometimes it is more convenient to use the original variable $z$ rather than the tortoise variable $x$.

We apply previous methods to compute quasinormal mode frequencies (i.e., resonant energies) satisfying this boundary condition. 
Our purpose is to present widely applicable ways. The computation here is just an example. 
For instance, one can easily extend it to other spherically symmetric geometries such as the Reissner-N\"ordstrom solution.
Although the Kerr solution is not spherically symmetric, the approach in this article is probably applicable even in this case.
To do so, we note that an alternative master equation, which is isospectral to Teukolsky's master equation \cite{teukolsky1972}, for the Kerr spacetime was recently conjectured in \cite{hatsuda2021}. This master equation is expected to be useful for studying the QNM spectra.

\subsubsection{Wronskian method}

We first apply the Wronskian method to this problem. For this purpose, we use the $z$-variable.
Transforming the wave function by\footnote{This transformation is not always necessary. We do it in order to map the problem into a known differential equation.} 
\begin{equation}
\begin{aligned}
\phi(z)=z^{1+s}(z-1)^{-2iM\omega}e^{2iM\omega z} y(z),
\end{aligned}
\label{eq:transform-Heun}
\end{equation}
then the new function $y(z)$ satisfies the differential equation
\begin{equation}
\begin{aligned}
y''(z)+\( \frac{\gamma}{z}+\frac{\delta}{z-1}+\epsilon\)y'(z)+\frac{\alpha z-q}{z(z-1)}y(z)=0.
\end{aligned}
\label{eq:cHeun}
\end{equation}
where
\begin{equation}
\begin{aligned}
q&=\ell(\ell+1)-s(s+1)+4iM\omega(1+2s),\qquad \alpha=(4M \omega)^2+4iM\omega(1+s) ,\\
\gamma&=1+2s,\qquad \delta=1-4iM\omega,\qquad \epsilon=4iM\omega.
\end{aligned}
\end{equation}
This differential equation is well-known as the confluent Heun equation \cite{ronveaux1995}.
The relation between the master equation \eqref{eq:RW} and the confluent Heun equation was discussed in great detail in \cite{fiziev2006}.
We construct formal series solutions at $z=1$ (event horizon) and $z=\infty$ (spacial infinity).
The local solutions at $z=1$ can be expressed by the local confluent Heun solution in \cite{ronveaux1995}, but we rather construct the series solutions directly for extensions to other cases.

The point $z=1$ is a regular singular point of \eqref{eq:cHeun}.
Its exponents are $\rho_1=0$ and $\rho_2=1-\delta$. Recalling the transformation \eqref{eq:transform-Heun}, the QNM boundary condition at $z=1$ is satisfied by the solution with $\rho_1=0$. Therefore the Frobenius series solution we want is given by
\begin{equation}
\begin{aligned}
y_{11}(z)=\sum_{k=0}^\infty a_k (z-1)^k.
\end{aligned}
\label{eq:y11-RW}
\end{equation}
Plugging it into the confluent Heun equation, we obtain the following three-term recursion:
\begin{equation}
\begin{aligned}
\alpha_k^1 a_{k+1}+\beta_k^1 a_k+\gamma_k^1 a_{k-1}=0,
\end{aligned}
\end{equation}
where
\begin{equation}
\begin{aligned}
\alpha_k^1&=(k+1)(k+\delta), \\
\beta_k^1&=k^2+k(\gamma+\delta+\epsilon-1)+\alpha-q,\\
\gamma_k^1&=(k-1)\epsilon+\alpha.
\end{aligned}
\end{equation}
Using this recursion, we can easily evaluate the coefficients $a_k$ to very high orders.
The series solution is convergent for $|z-1|<1$ in general.
In the practical computation, we have to truncate the sum at a certain finite order.
The standard method to reconstruct an (approximate) analytic solution from a truncated convergent series solution is Pad\'e approximants.
We review the Pad\'e approximants in Appendix~\ref{app:Pade}.

Let us proceed to the solutions at infinity. The confluent Heun equation has the irregular singular point at $z=\infty$.
The Poincar\'e rank of this singular point is just $r=1$, and the asymptotic series solutions take the form (see Appendix~\ref{app:irregular})
\begin{equation}
\begin{aligned}
y_{\infty}^\text{formal}(z)=e^{az}z^b \sum_{k=0}^\infty \frac{c_k}{z^k}.
\end{aligned}
\end{equation}
The leading asymptotic behavior determines $a$ and $b$ as
\begin{equation}
\begin{aligned}
(a,b)=\(0,-\frac{\alpha}{\epsilon}\), \(-\epsilon,-\gamma-\delta+\frac{\alpha}{\epsilon} \).
\end{aligned}
\end{equation}
The QNM boundary condition at infinity is satisfied by the former case. Therefore we seek the formal solution with the form
\begin{equation}
\begin{aligned}
y_{\infty1}^\text{formal}(z)=z^{-\alpha/\epsilon} \sum_{k=0}^\infty \frac{c_k}{z^k}.
\end{aligned}
\label{eq:yinf1-RW}
\end{equation}
The coefficients $c_k$ again satisfy the following three-term recursion:
\begin{equation}
\begin{aligned}
\alpha_k^\infty c_{k+1}+\beta_k^\infty c_k+\gamma_k^\infty c_{k-1}=0,
\end{aligned}
\end{equation}
where
\begin{equation}
\begin{aligned}
\alpha_k^\infty&=(k+1)\epsilon, \\
\beta_k^\infty&=-k^2+k\(\gamma+\delta-\epsilon-1-\frac{2\alpha}{\epsilon} \)+q-\alpha+\frac{\alpha(\gamma+\delta-1)}{\epsilon}-\frac{\alpha^2}{\epsilon^2},\\
\gamma_k^\infty&=\(k-1+\frac{\alpha}{\epsilon} \) \(k-\gamma+\frac{\alpha}{\epsilon} \).
\end{aligned}
\end{equation}
A big difference from the solution near the horizon $z=1$ is that this formal series is not convergent for any $z$.
We have to truncate or resum it properly as in the Morse potential. As shown  in subsection~\ref{subsec:numerical}, one way is the Borel summation.
In practice, however, the computational cost is much saved by using the Pad\'e approximants.
We discuss an application of the Pad\'e approximants to divergent series in Appendix~\ref{app:Pade-div}.

In summary, to evaluate the Wronskian practically, we use the Pad\'e approximants $y_{11}^{[M_1/N_1]}(z)$ and $y_{\infty1}^{[M_\infty/N_\infty]}(z)$ of the formal series solutions \eqref{eq:y11-RW} and \eqref{eq:yinf1-RW} by regarding $z-1$ and $1/z$ as expansion parameters, respectively.
We have a freedom to choose $z_0$, at which we search zeros of the Wronskian.
We fix it so that both the approximants $y_{11}^{[M_1/N_1]}(z_0)$ and $y_{\infty1}^{[M_\infty/N_\infty]}(z_0)$ have the same numerical accuracy.\footnote{Such an accuracy is roughly estimated by $|1-y^{[(M-1)/(N+1)]}(z_0)/y^{[M/N]}(z_0)|$. Therefore our requirement for $z_0$ is
\[
 |1-y_{11}^{[(M_1-1)/(N_1+1)]}(z_0)/y_{11}^{[M_1/N_1]}(z_0)| \sim  |1-y_{\infty 1}^{[(M_{\infty}-1)/(N_{\infty}+1)]}(z_0)/y_{\infty 1}^{[M_{\infty}/N_{\infty}]}(z_0)|.
\]
}

Let us show an explicit result. We set $(s,\ell)=(2,2)$, and computed the series solutions \eqref{eq:y11-RW} and \eqref{eq:yinf1-RW} up to $(z-1)^{120}$ and $1/z^{120}$ respectively for a given numerical value of $M\omega$. By using the recurrence relations, they are very quickly done. We used the Pad\'e approximants $y_{11}^{[60/60]}(z)$ and $y_{\infty1}^{[60/60]}(z)$. Setting $z_0=8$, we searched zeros of the Wronskian $W[y_{11}^{[60/60]},y_{\infty1}^{[60/60]}]$ by Newton's method for an initial value $M\omega=0.4-0.07i$.
Then we finally get
\begin{equation}
\begin{aligned}
M\omega_{s=2,\ell=2}^\text{QNM}\approx 
0.373671684418041835793492-0.0889623156889356982804609i.
\end{aligned}
\label{eq:QNM-22}
\end{equation}
Comparing it with a zero of $W[y_{11}^{[59/61]},y_{\infty1}^{[59/61]}]$, the accuracy estimation is about $\cO(10^{-30})$.
Therefore we can get quite good numerical values of the QNM eigenvalues. However, in this approach, it is not easy to identify the overtone number of the obtained QNMs. This identification is usually fixed by comparing with other approaches like the perturbative method or the (uniform) WKB method. In these methods, we can easily know the overtone number of the QNM, as will be seen below.

The Wronskian method presented here is similar to the well-known continued fraction method by Leaver \cite{leaver1985}. We do not explain Leaver's method in detail because it is discussed in many places.
An advantage of the Wronskian method is that it still works even when we do not have an explicit recurrence relation. We just need two series (or numerical) solutions satisfying proper boundary conditions. This can be done without information on the recurrence relation in principle. Moreover, to use Leaver's continued fraction method, one has to treat a recurrence relation carefully if it is not a three-term relation \cite{leaver1990, onozawa1996}.
In this sense, the application range of the Wronskian method is wider than that of Leaver's method. 
However, since Leaver's method seems to be the best numerical method to obtain the precise QNM frequencies for the Schwarzschild spacetime, we can use it as a reference way for comparison. For instance, the numerical result \eqref{eq:QNM-22} shows about 30-digit agreement with Leaver's result as expected.

\subsubsection{WKB method}
We proceed to another method. In the WKB method, we have two options which variables $x$ or $z$ we use.
It turns out that the $z$-variable looks more useful technically.
The similar approach is found in \cite{froman1992}.

To see a quantum mechanical aspect more clearly, we rewrite the master equation \eqref{eq:RW} as follows.
First, we divide it by $\ell(\ell+1)/2$,
\begin{equation}
\begin{aligned}
\biggl[ \frac{2}{\ell(\ell+1)} f(z)\frac{d}{dz}f(z)\frac{d}{dz}+\frac{2}{\ell(\ell+1)}(2M \omega)^2-f(z)\( \frac{2}{z^2}+\frac{2}{\ell(\ell+1)} \frac{1-s^2}{z^3} \) \biggr]\phi(z)=0,
\end{aligned}
\end{equation}
We regard the coefficient in front of the derivative term as the square of a Planck parameter,
\begin{equation}
\begin{aligned}
\hbar&:=\biggl( \frac{2}{\ell(\ell+1)} \biggr)^{1/2}.
\end{aligned}
\end{equation}
Of course, this is not the unique way to introduce $\hbar$.
Ultimately, one can freely introduce it by hand, and set $\hbar=1$ at last. Such a freedom are related to a choice of ``base functions'' in \cite{froman1996}.
Here we choose a very particular base function, which naturally relates $\hbar$ to the multipole index $\ell$.
It has an advantage that we do not have to introduce an additional freedom of parameters.

Then, we obtain 
\begin{equation}
\begin{aligned}
\biggl[ \hbar^2 f(z)\frac{d}{dz}f(z)\frac{d}{dz}+E-f(z)\( \frac{2}{z^2}+\hbar^2\frac{1-s^2}{z^3} \) \biggr]\phi(z)=0,
\end{aligned}
\label{eq:RW-2}
\end{equation}
where
\begin{equation}
\begin{aligned}
E:=\frac{2}{\ell(\ell+1)}(2M \omega)^2.
\end{aligned}
\end{equation}
In terms of the tortoise variable $x$, this is nothing but the Schr\"odinger equation. The potential receives the quantum correction in general. In our picture, the master equation for general spins $s$ is regarded as a quantum deformation from that for the electromagnetic perturbation $s=1$. Note that this picture is just a computational technique.

The role of the multipole number $\ell$ is now mapped to (the inverse of) the Planck parameter.
The eikonal limit $\ell \to \infty$ corresponds to the classical limit $\hbar \to 0$.
We should note that $\ell=0$ is singular in our picture.%
\footnote{Note that the $\ell=0$ mode exists only for $s=0$.} 
We have to consider this case separately.
The treatment in this particular case is however straightforward by introducing a formal Planck parameter by hand.

Next, we rewrite the differential equation as the so-called normal form. To do so, we transform the dependent variable by
\begin{equation}
\begin{aligned}
\psi(z)=\sqrt{f(z)} \phi(z).
\end{aligned}
\end{equation}
We finally obtain the Schr\"odinger-like equation for the $z$-variable:
\begin{equation}
\begin{aligned}
\( \hbar^2 \frac{d^2}{dz^2}+Q_0(z)+\hbar^2 Q_1(z) \) \psi(z) =0,
\end{aligned}
\label{eq:Sch-RW}
\end{equation}
where
\begin{equation}
\begin{aligned}
Q_{0}(z)=\frac{2-2z+Ez^3}{z(z-1)^2},\qquad
Q_{1}(z)=\frac{1-4s^2+4s^2 z}{4z^2(z-1)^2}.
\end{aligned}
\end{equation}
Note that the domain we are interested in is $z>1$, not the whole real line.

We apply the WKB method to the differential equation \eqref{eq:Sch-RW}.
This is essentially the same form as \eqref{eq:Milne}, but $Q$ depends on $\hbar$ explicitly.
Even in this case, we can apply the WKB method, as was discussed in the previous section. 
The all-order Bohr--Sommerfeld quantization condition is given by \eqref{eq:qBS-res}.
We start with
\begin{equation}
\begin{aligned}
p_0(z)=\sqrt{Q_0(z)}=\frac{\sqrt{2-2z+Ez^3}}{(z-1)\sqrt{z}}.
\end{aligned}
\end{equation}
Three zeros of $Q_0(z)$ are turning points. For $0<E<8/27$, two of them are in the regime $z>1$.
Let us denote these two turning points by $a(E)$ and $b(E)$ with $a(E)<b(E)$, as shown in Figure~\ref{fig:WKB}.
In the computation of the quasinormal mode frequencies, we need to analytically continue these turning points to the complex domain. 

We want to compute the contour integral
\begin{equation}
\begin{aligned}
\Omega_0(E):=\oint_{C(a,b)} p_0(z) dz=\oint_{C(a(E),b(E))} \frac{\sqrt{2-2z+Ez^3}}{(z-1)\sqrt{z}} dz.
\end{aligned}
\label{eq:Omega_0}
\end{equation}
At first glance, it looks hard to perform the integral analytically.
In principle, it is evaluated numerically by analytic continuations of $a(E)$ and $b(E)$ for complex values of $E$.
In the numerical computation, we have to choose directions of branch cuts carefully.%
\footnote{For instance, the branch cut of $\sqrt{z}$ in \textit{Mathematica} is along the negative real axis. This is not useful in evaluating the contour complex integral \eqref{eq:Omega_0} numerically. }
In our case, however, it is evaluated analytically. We need no integration formulae.

A basic strategy to obtain an analytic result is to look for a differential equation that $\Omega_0(E)$ satisfies.
To do so, we notice that the known function $p_0(z)$ satisfies the following partial differential equation:
\begin{equation}
\begin{aligned}
\( 8E^2(8-27E) \frac{\del^3}{\del E^3}+4E(40-189E) \frac{\del^2}{\del E^2}+2(16-123E)\frac{\del}{\del E}+15 \) p_0\\
=\frac{\del}{\del z} \( \frac{8\sqrt{z}(-15+20z-3z^2-(2+3E)z^3+Ez^4)}{(2-2z+Ez^3)^{3/2}}\).
\end{aligned}
\label{eq:PDE-p0}
\end{equation}
Since the integration contour $C(a,b)$ is a closed path that does not encircle the branch point $z=0$ nor the other turning point,
the contour integral of the right hand side trivially vanishes.  As the result, we immediately find an ODE for $\Omega_0(E)$,
\begin{equation}
\begin{aligned}
\( 8E^2(8-27E) \frac{d^3}{d E^3}+4E(40-189E) \frac{d^2}{d E^2}+2(16-123E)\frac{d}{d E}+15 \)\Omega_0(E)=0.
\end{aligned}
\label{eq:PF}
\end{equation}
Such ODEs for contour integrals are known as \textit{Picard--Fuchs equations} \cite{clemens2002}.
To find the Picard--Fuchs equation \eqref{eq:PF}, the partial differential equation \eqref{eq:PDE-p0} for $p_0$ is crucial.
This happens very non-trivially, and we do not have a systematic algorithm to find it.
An approach in \cite{clemens2002} is a hint to do this for general cases.
Here we have put ansatz of differential operators appropriately.

Once we find the Picard--Fuchs equation \eqref{eq:PF}, it is not difficult to solve it.
The standard technique is to construct series solutions at singular points, as reviewed in Appendix~\ref{app:ODE}.
In the current case, we find analytic solutions in terms of generalized hypergeometric functions by using a new variable,\footnote{In the original variable $E$, the differential equation is also hypergeometric-type, but two of the solutions near $E=0$ are expressed in terms of so-called Meijer's $G$-functions.}
\begin{equation}
\begin{aligned}
\xi=\frac{8}{27E}.
\end{aligned}
\end{equation} 
The general solution is given by
\begin{equation}
\begin{aligned}
\Omega_0(E)=\frac{C_1}{\sqrt{\xi}}+C_2 \xi^{1/6}{}_3F_2\( \frac{1}{6},\frac{1}{6},\frac{2}{3};\frac{1}{3},\frac{5}{3};\xi \)
+C_3\xi^{5/6}{}_3F_2\( \frac{5}{6},\frac{5}{6},\frac{4}{3};\frac{5}{3},\frac{7}{3};\xi \),
\end{aligned}
\label{eq:Omega0-sol}
\end{equation}
There are three integration constants, which should be fixed by conditions of $\Omega_0(E)$ with the integral form \eqref{eq:Omega_0}.
This is obtained by considering the limit $E \to 8/27$ or $\xi \to 1$.
In this limit, the two turning points $a$ and $b$ meet together at $z=3/2$.
The integrand $p_0(z)$ has the expansion around $\xi=1$,
\begin{equation}
\begin{aligned}
p_0(z)=\frac{2z-3}{3(z-1)}\sqrt{\frac{2(z+3)}{3z}}-\frac{2z^2}{3(z-1)(2z-3)}\sqrt{\frac{2z}{3(z+3)}}(\xi-1)\\
+\frac{2z^2(z^3-9z+9)}{(z-1)(z+3)(2z-3)^3}\sqrt{\frac{2z}{3(z+3)}} (\xi-1)^2+\cO((\xi-1)^3).
\end{aligned}
\end{equation}
The turning points $a$ and $b$ are now shrinking in all orders in $\xi -1$.
Then, the integral around the turning points $a$ and $b$ is simply evaluated by the residue theorem,
\begin{equation}
\begin{aligned}
\Omega_0(E)=-\sqrt{2}\pi i(\xi-1)+\frac{49\pi i}{36\sqrt{2}}(\xi-1)^2+\cO((\xi-1)^3),\qquad
\xi \to 1.
\end{aligned}
\end{equation}
Our requirement for the integration constants is to reproduce this expansion from \eqref{eq:Omega0-sol}.
With the help of \textit{Mathematica}, we find the following exact values:
\begin{equation}
\begin{aligned}
C_1=\frac{8\pi i}{3} \sqrt{\frac{2}{3}},\qquad
C_2=-\frac{i}{2} \sqrt{\frac{3}{2}} \frac{\Gamma^2(1/6)}{\Gamma(1/3)},\qquad
C_3=\frac{3i}{8}\sqrt{\frac{3}{2}} \frac{\Gamma^2(5/6)}{\Gamma(2/3)}.
\end{aligned}
\end{equation}

Once we find the exact form of $\Omega_0(E)$, it is a easy task to continue it to the complex $E$-plane.
Our leading Bohr--Sommerfeld quantization condition is
\begin{equation}
\begin{aligned}
\Omega_0(E)=2\pi \hbar \( n+\frac{1}{2} \),\qquad E=\hbar^2 (2M \omega)^2,\qquad
\hbar=\biggl(\frac{2}{\ell(\ell+1)}\biggr)^{1/2}.
\end{aligned}
\label{eq:BS-0}
\end{equation}
Note that there is no spin dependence at the leading order while the multipole dependence is included through the Planck parameter. This is expected in the eikonal limit.
The spin dependence appears as ``quantum corrections'' to the potential.
As mentioned above, at the leading order, the quantization condition should give an approximate value for $s=1$.
The integer $n$ characterizes quasinormal modes. It corresponds to the overtone number.

By solving the quantization condition \eqref{eq:BS-0} for $\ell=1$ and $n=0$, we find the following values of the frequencies:
\begin{equation}
\begin{aligned}
M\omega_{\ell=1, n=0}^{(0)} \approx 0.2675 - 0.09695i.
\end{aligned}
\end{equation}
where the superscript $(0)$ means the leading Bohr--Sommerfeld approximation.
This value is compared to the QNM frequency for the $s=1$ perturbation,
\begin{equation}
\begin{aligned}
M\omega_{s=1, \ell=1, n=0}^\text{QNM}\approx 0.248263264178-0.0924877179529i.
\end{aligned}
\end{equation}
The leading approximation already gives a relatively good numerical value.
Of course, the semiclassical approximation should get better for the eikonal limit $\ell \gg 1$.
The $\ell=1$ case is the most ``severe'' situation.

The approximation is expected to be improved by including the quantum corrections as in \eqref{eq:qBS-res}.
To check it, we need
\begin{equation}
\begin{aligned}
\Omega_k(E)=\oint_{C(a,b)} Z_k(z)p_0(z) dz,
\end{aligned}
\end{equation}
where $Z_k$ are given by \eqref{eq:Z_k} but we have to replace $\epsilon_0$ by $\bar{\epsilon}_0$ in \eqref{eq:bar-epsilon_0} due to the quantum correction $Q_1(z)$ to the potential.

It turns out that these integrals are also evaluated analytically in the Schwarzschild case.
The idea is similar to the Picard--Fuchs equation. We look for certain differential operators acting on $\Omega_0(E)$.
Let us see the computation for $k=1$ in detail. We want to evaluate
\begin{equation}
\begin{aligned}
\Omega_1(E)=\oint_{C(a,b)} Z_1(z) p_0 (z) dz=\frac{1}{2}\oint_{C(a,b)} \bar{\epsilon}_0(z) p_0 (z) dz,
\end{aligned}
\end{equation}
where
\begin{equation}
\begin{aligned}
\bar{\epsilon}_0(z)=p_0(z)^{-3/2} \frac{d^2}{dz^2} p_0(z)^{-1/2}+\frac{Q_1(z)}{p_0(z)^2}.
\end{aligned}
\end{equation}
We seek a good differential relation for the integrand $Z_1(z) p_0 (z)$.
We find the following one
\begin{equation}
\begin{aligned}
Z_1(z) p_0 (z)=D_1 p_0(z)+\frac{\del}{\del z} \frac{a_{1,0}+a_{1,1} z+\cdots+a_{1,5} z^5}{\sqrt{z}(2-2z+Ez^3)^{3/2}},
\end{aligned}
\label{eq:Z1p0}
\end{equation}
where $D_1$ is a second order differential operator,
\begin{equation}
\begin{aligned}
D_1&=\( \frac{1}{12}(5-24s^2)-\frac{9}{8}(1-6s^2)E \) \frac{\del^2}{\del E^2}\\
&\quad+\( \frac{1}{12}(1-12s^2)-\frac{3}{4}(1-6s^2)E \) \frac{\del}{\del E}+\(\frac{1}{16}+\frac{3}{32}(1-6s^2)E \),
\end{aligned}
\label{eq:D1}
\end{equation}
and $a_{1,i}$ are some coefficients. We do not know a general algorithm to find \eqref{eq:Z1p0}, but once we put such an ansatz correctly by trials and errors, we can fix all the coefficients uniquely.
With the similar argument for the Picard--Fuchs equation, the relation \eqref{eq:Z1p0} immediately leads to the contour integral relation,
\begin{equation}
\begin{aligned}
\Omega_1(E)=D_1 \Omega_0(E).
\end{aligned}
\end{equation}
Since we know $\Omega_0(E)$ exactly, we obtain an analytic form of $\Omega_1(E)$ without any integration!

This nice property seems to exist in the higher corrections. We observe that there exist second order differential operators such as
\begin{equation}
\begin{aligned}
\Omega_k(E)=D_k \Omega_0(E).
\end{aligned}
\end{equation}
We confirmed this statement up to $k=6$.

Now we can evaluate the Bohr--Sommerfeld condition with the quantum corrections.
We want to solve
\begin{equation}
\begin{aligned}
\sum_{k=0}^\infty \hbar^{2k} \Omega_k(E) \simeq 2\pi \hbar \( n+\frac{1}{2} \).
\end{aligned}
\label{eq:qBS-QNM}
\end{equation}
Of course, in the practical computation, we have to truncate the sum on the left hand side at a certain finite order.
We set $(s,\ell)=(2,2)$, for example. We solve the 12th order ($k=6$) quantization condition for $n=0$, and obtain the numerical eigenvalue,
\begin{equation}
\begin{aligned}
M\omega_{s=2,\ell=2,n=0}^{(12)}\approx 0.3736290-0.08894092i.
\end{aligned}
\end{equation}
This is indeed close to the correct value \eqref{eq:QNM-22}.
The approximation is improved by using Pad\'e approximants for the asymptotic series \eqref{eq:qBS-QNM}.
If we use the $[6/6]$ Pad\'e approximant of the left hand side, we get
\begin{equation}
\begin{aligned}
M\omega_{s=2,\ell=2,n=0}^{[6/6]}\approx 0.3736770-0.08896645i.
\end{aligned}
\end{equation}
The more detailed result is shown in Table~\ref{tab:WKB}.

\begin{table}[tp]
\caption{The numerical fundamental QNM frequency for the gravitational perturbation $s=2$ with $\ell=2$. The numerical values in the WKB analysis are determined by solving \eqref{eq:qBS-QNM} for $n=0$. The values in the perturbative analysis are obtained from \eqref{eq:omega-pert-s2-n0}. The same result as \eqref{eq:omega-pert-s2-n0} is also obtained by the uniform WKB analysis. These approximate methods get better in the eikonal limit $\ell \to \infty$, and thus $\ell=2$ is the most severe.}
\begin{center}
\begin{tabular}{ccl}
\hline
Order & WKB & Perturbation/Uniform WKB \\
\hline
$\hbar^0$ & $0.4686939-0.09646671 i$ & $0.4714045$\\
$\hbar^2$ & $0.3824606-0.08924751i$ & $0.3748360 - 0.1210154i$\\
$\hbar^4$ & $0.3745148-0.08847747i$ & $0.3723472 - 0.0896396i$\\
$\hbar^6$ & $0.3734578-0.08859956i$ & $0.3735116 - 0.0882316i$\\
$\hbar^{8}$ & $0.3734641-0.08878269i$ & $0.3737158 - 0.0887288i$ \\
$\hbar^{10}$ & $0.3735665-0.08889202i$ & $0.3736968 - 0.0889343i$\\
$\hbar^{12}$ & $0.3736290-0.08894092i$ & $0.3736751 - 0.0889682i$ \\
\hline
$[6/6]$ Pad\'e & $0.3736770-0.08896645i$ & $0.3736670 - 0.0889626i$ \\
\hline
Wronskian & $0.3736717-0.08896232i$ & $0.3736717-0.08896232i$\\
\hline
\end{tabular}
\end{center}
\label{tab:WKB}
\end{table}%

\subsubsection{Perturbative method}
Next, we use perturbation theory reviewed in subsection~\ref{subsec:perturbation}.
Fortunately we have a natural quantum parameter $\hbar$ in terms of the multipole index $\ell$.
We derive the perturbative expansions of the QNM frequencies in this quantum parameter. 
Compared with the WKB method, this method has an advantage that directly gives the small-$\hbar$ expansion of the QNM frequency.
As we have already mentioned, the same result is also obtained by the uniform WKB method in subsection~\ref{subsec:WKB}. 

For this purpose, the tortoise variable \eqref{eq:tortoise} is useful. We rewrite \eqref{eq:RW-2} as
\begin{equation}
\begin{aligned}
\biggl[ -\hbar^2 \frac{d^2}{dx^2}+V(x) \biggr] \phi(x)=E \phi(x),
\end{aligned}
\label{eq:Sch-RW-2}
\end{equation}
where
\begin{equation}
\begin{aligned}
V(x)=V_0(x)+\hbar^2 V_1(x)=f(z)\( \frac{2}{z^2}+\hbar^2 \frac{1-s^2}{z^3} \).
\end{aligned}
\end{equation}
The potential now depends on $\hbar$, but we can use the results in the previous section.
$V_0(x)$ has the maximum at $z=3/2$. We fix the integration constant $c$ in \eqref{eq:tortoise} so that $z=3/2$ is mapped to $x=0$. Hence we have $c=-3/2+\log 2$. We then expand the potential $V_0(x)$ and $V_1(x)$ around $x=0$.
\begin{equation}
\begin{aligned}
V_0(x)&=\frac{8}{27}-\frac{32 x^2}{729}+\frac{128 x^3}{19683}+\frac{256 x^4}{59049}-\frac{256 x^5}{177147}-\frac{2048 x^6}{7971615}+\cO(x^7), \\
V_1(x)&=(1-s^2) \biggl( \frac{8}{81}-\frac{16x}{729}-\frac{32x^2}{2187}+\frac{128x^3}{19683}+\frac{448 x^4}{531441} -\frac{8192 x^5}{7971615}+\frac{17408 x^6}{215233605}+\cO(x^7)\biggr).
\end{aligned}
\end{equation}
Recall that we are considering the resonant problem for the QNMs. Strictly speaking, to use perturbation theory in subsection~\ref{subsec:perturbation}, we need to analytically continue the Planck parameter $\hbar \to -i\hbar$ in order to invert the potential. See \cite{hatsuda2020} for detail. However, the result in subsection \ref{subsec:perturbation} can be formally used if we instead consider the analytic continuation of $\Omega$. Let us see it in detail.

The equation \eqref{eq:Sch-RW-2} can be written as
\begin{equation}
\begin{aligned}
-\frac{1}{2} \psi''(q)+\frac{v(gq)}{g^2} \psi(q)=\epsilon \psi(q).
\end{aligned}
\label{eq:BW-eq-2}
\end{equation}
where $g=\sqrt{\hbar}$ and $x=gq$, 
\begin{equation}
\begin{aligned}
\frac{v(gq)}{g^2}&=\frac{v_0(gq)}{g^2}+g^2 v_4(gq), \\
v_0(x)&=\frac{V_0(x)-V_0(0)}{2},\qquad v_4(x)=\frac{V_1(x)-V_1(0)}{2},
\end{aligned}
\end{equation}
and
\begin{equation}
\begin{aligned}
\epsilon=\frac{E-V(0)}{2\hbar}.
\end{aligned}
\end{equation}
The perturbative series of $\epsilon$ can be computed by the method reviewed in subsection \ref{subsec:perturbation}.
The angular frequency $\Omega$ is given by
\begin{equation}
\begin{aligned}
\Omega^2=2v_0''(0)=-\frac{32}{729}.
\end{aligned}
\end{equation}
which is now purely imaginary due to the resonant problem. However we can regard $\Omega$ as a formal parameter during the computation.

Here we focus on the lowest overtone $n=0$ for simplicity. The perturbative corrections are then computed as follows.
For $l=1,2,\dots$, we start with the recurrence relation (see \eqref{eq:A-recur-1})
\begin{equation}
\begin{aligned}
A_{0,l}^k=\frac{1}{2\Omega k}\biggl[(k+2)(k+1)A_{0,l}^{k+2}+\sum_{j=1}^{l-1}2\epsilon_{0,j} A_{0,l-j}^k
-2\sum_{j=1}^l v_{0,j}A_{0,l-j}^{k-j-2}\\
-2\sum_{j=1}^l v_{4,j}A_{0,l-j-2}^{k-j}\biggr]\qquad
(1 \leq k \leq 3l).
\end{aligned}
\label{eq:recur-RW-1}
\end{equation}
where we regard $A_{0,l}^k=0$ for $k>3l$ and for $k<0$. This determines $A_{0,l}^k$ for $k=3l, 3l-1, \dots, 1$ in this order.
Then the energy correction is obtained by
\begin{equation}
\begin{aligned}
\epsilon_{0,l}=-A_{0,l}^{2}.
\end{aligned}
\end{equation}
Recall that we set $A_{0,l}^0=\delta_{l0}$.

For $l=1$, the recurrence relation \eqref{eq:recur-RW-1} gives
\begin{equation}
\begin{aligned}
A_{0,1}^3=-\frac{64}{59049\Omega},\qquad A_{0,1}^2=0,\qquad A_{0,1}^1=-\frac{64}{19683\Omega^2},\qquad A_{0,1}^0=0.
\end{aligned}
\end{equation}
Then we find $\epsilon_{0,1}=0$.

For $l=2$, we have the following non-zero components:
\begin{equation}
\begin{aligned}
A_{0,2}^6&=\frac{2048}{3486784401\Omega^2},\qquad
A_{0,2}^4=-\frac{32}{59049 \Omega }+\frac{5632}{1162261467 \Omega ^3},\\
A_{0,2}^2&=-\frac{32}{19683 \Omega ^2}+\frac{5632}{387420489 \Omega ^4},
\end{aligned}
\end{equation}
and
\begin{equation}
\begin{aligned}
\epsilon_{0,2}=\frac{32}{19683 \Omega ^2}-\frac{5632}{387420489 \Omega ^4}.
\end{aligned}
\end{equation}
Up to this order, the results do not depend on $s$, but the spin-dependence appears for $l \geq 3$.
In this way, we can easily fix $\epsilon_{0,l}$ up to quite high orders.

The perturbative series of the fundamental QNM frequency is given by
\begin{equation}
\begin{aligned}
(M\omega_{n=0}^\text{pert})^2=\frac{\ell(\ell+1)}{8}\biggl[ V(0)+\hbar \Omega+2\hbar \sum_{l=1}^\infty \hbar^{l}\epsilon_{0,2l}\biggr].
\end{aligned}
\end{equation}
Plugging $\Omega=- 4\sqrt{2}i/27$ and $s=2$, we finally get\footnote{For $\Omega=4\sqrt{2}i/27$, we get the complex conjugate to the right hand side in \eqref{eq:omega-pert-s2-n0}. Because of the physical requirement of the QNMs, we have to choose a branch of the square root of $(M\omega_{n})^2$ so that the imaginary part of $M\omega_{n}$ must be negative. Therefore the two possible values of $\Omega$ finally lead to two branches of $M\omega_{n}$ whose real parts have opposite signs to each other.}
\begin{equation}
\begin{aligned}
(M\omega_{s=2,n=0}^\text{pert})^2=\frac{\ell(\ell+1)}{8}\biggl[\frac{8}{27}-\frac{4\sqrt{2} i\hbar}{27}-\frac{281 \hbar ^2}{729}
+\frac{6163 i \hbar ^3}{26244 \sqrt{2}}+\frac{969319 \hbar ^4}{17006112}\\
+\frac{60687295 i \hbar ^5}{2448880128 \sqrt{2}}+\frac{21306926297 \hbar ^6}{528958107648}
-\frac{4107036251635 i \hbar ^7}{114254951251968 \sqrt{2}}\\
+\frac{172267964131073 \hbar ^8}{24679069470425088}+\cO(\hbar^9)
\biggr].
\end{aligned}
\label{eq:omega-pert-s2-n0}
\end{equation}
Recall that $\hbar=\sqrt{2/[\ell(\ell+1)]}$. The higher order corrections are easily and quickly computed by using \textit{Mathematica}.

Note that this result is closely related to the uniform WKB method in \cite{mashhoon1983, schutz1985, iyer1987, konoplya2003}.
Also, the result can be compared with an expansion in \cite{dolan2009}\footnote{In fact, the idea in \cite{dolan2009} is very similar to the uniform WKB approach. The continuity condition in \cite{dolan2009} just corresponds to the regularity condition in the uniform WKB method.} by re-expanding \eqref{eq:omega-pert-s2-n0} in terms of $L^{-1}:=(\ell+1/2)^{-1}$.
As we have mentioned many times, the semiclassical perturbative series \eqref{eq:omega-pert-s2-n0} is in general an asymptotic divergent series, and we have to truncate the sum at the optimal order or to resum it by the Borel(-Pad\'e) summation \cite{hatsuda2020, eniceicu2020} or the Pad\'e approximants \cite{matyjasek2017, konoplya2019, matyjasek2019}. 

For $\ell=2$, we show the numerical values of $M\omega_{s=2,\ell=2,n=0}^\text{pert}$ for lower orders in Table~\ref{tab:WKB}.
Of course, as in the WKB analysis, the perturbative approximation is better for larger $\ell$.

An advantage of this method is that it is widely applicable to many potentials. In fact, to compute the QNM frequency we need only the information near the potential peak (i.e., the Taylor series around it). The global information is encoded in high-order corrections in principle.
Moreover, the overtone number $n$ of the QNMs naturally appears as the quantum number of the bound states.

The application range of perturbation theory is quite wide. One can consider other types of perturbative expansions of the QNM spectra.
For instance, some parameters of black holes can be regarded as deformation parameters from simpler geometries (e.g., the Schwarzschild spacetime).  
The Reissner-N\"ordstrom spacetime is regarded as a deformation of the Schwarzschild one by the electric charge, and the Kerr case is also regarded as a deformation by the angular momentum.
Then one can consider the perturbative expansions of the QNM spectra for such deformation parameters \cite{cardoso2019, mcmanus2019, kimura2020, hatsuda2020b, hatsuda2021}.

\subsubsection{Comments on even parity perturbation}
So far, we have looked into the analysis in the odd parity sector as well as in the scalar and the electromagnetic perturbations.
As discussed in Appendix~\ref{app:BH-pert}, the even parity gravitational perturbation of the Schwarzschild spacetime leads to the Zerilli equation \eqref{eq:V-Zerilli}. 
Since the QNM spectra of the Zerilli equation is exactly the same as those of the Regge--Wheeler equation, it is sufficient to study either of them. However we should comment that our formulation is also applicable to the Zerilli equation.
We briefly see it.

In our convention in this section, the Zerilli equation is written as
\begin{equation}
\begin{aligned}
\( f(z)\frac{d}{dz} f(z)\frac{d}{dz} +(2M\omega)^2-V_+(z) \) \phi(z)=0,
\end{aligned}
\label{eq:Zerilli}
\end{equation}
where
\begin{equation}
\begin{aligned}
f(z)=1-\frac{1}{z},\qquad
V_+(z)=f(z)\frac{\lambda^2(\lambda+2)z^3+3\lambda^2 z^2+9\lambda z+9}{z^3(\lambda z+3)^2},
\end{aligned}
\label{eq:Zerilli-potential}
\end{equation}
with $\lambda=\ell^2+\ell-2$.
This ODE has regular singular points at $z=0, 1, -3/\lambda$ and an irregular singular point at $z=\infty$ with Poincar\'e rank $1$.
Then, the local series solutions near $z=1$ and $z=\infty$ can be constructed straightforwardly. This means that the Wronskian method works in this case.

The semiclassical analyses are also applied. To see them, we rewrite the Zerilli potential as
\begin{equation}
\begin{aligned}
V_+(z)=f(z)\biggl( \frac{\ell(\ell+1)}{z^2}+\delta V_+(z) \biggr),\qquad
\delta V_+(z)=-\frac{3(\lambda(\lambda+4)z^2+6z-3)}{z^3(\lambda z+3)^2}.
\end{aligned}
\end{equation}
In the eikonal limit $\ell \to \infty$, we can see $\delta V_+(z) \sim \cO(1)$, and then this part can be regarded as the ``quantum'' correction, as in the Regge--Wheeler potential.
We can play the same game. In particular, the perturbative correction is computed by expanding the potential as
\begin{equation}
\begin{aligned}
V_+(z)=\frac{f(z)}{\hbar^2}\biggl[ \frac{2}{z^2}-\hbar^2\frac{3}{z^3}-\hbar^4 \frac{3(2z-3)}{z^4}-\hbar^6\frac{3(4z^2-15z+12)}{2z^5}+\cO(\hbar^8)\biggr].
\end{aligned}
\label{eq:V_Z-pert}
\end{equation}
with $\hbar=\sqrt{2/[\ell(\ell+1)]}$. We can use the general formulation in subsection~\ref{subsec:perturbation}.
Note that to get a perturbative correction to the eigenvalue, we can truncate the sum in the potential \eqref{eq:V_Z-pert} to a certain order.
It is a good exercise to check that the perturbative correction to the QNM frequency for this potential agrees with \eqref{eq:omega-pert-s2-n0} for the Regge--Wheeler potential. This is nothing but the isospectrality of the Regge--Wheeler/Zerilli potential.
See Appendix~\ref{app:susytr}.

\subsection{Mode stability analysis: Gregory--Laflamme instability for black strings}
In this subsection, we discuss the mode stability problem in black hole perturbation theory.
If the linear perturbation of a spacetime geometry has an exponentially growing mode, then it is unstable.
In the language of master equations, there exists a bound state solution with a negative energy ($E=\omega^2<0$) that corresponds to an unstable mode.
As we reviewed in subsection~\ref{subsec:numerical}, the number of bound states of the Schr\"odinger type equation is captured by counting functions. We can use them to judge whether the linear perturbation of a given geometry is stable or not.

Here we use Milne's counting function \eqref{eq:N-Milne} for this purpose.
The number of bound states below an energy $E$ are computed by $\lceil N^\text{Milne}(E)\rceil$.
Therefore the mode stability condition is simply presented by
\begin{equation}
\begin{aligned}
N^\text{Milne}(0) < 0.
\end{aligned}
\end{equation}
If this condition is satisfied, then there is no bound state for $E<0$.

Let us consider a concrete example. We discuss the mode (in)stability of black strings in five-dimension.
It is well-known that this black string shows the so-called Gregory--Laflamme instability.
We show that Milne's counting function actually explains the critical value of this transition.
The metric of the black string solution in five-dimension is given by
\begin{equation}
\begin{aligned}
ds^2=-f(r)dt^2+\frac{dr^2}{f(r)}+r^2(d\theta^2+\sin^2\theta d\phi^2)+dz^2,
\end{aligned}
\end{equation}
where $f(r)=1-r_H/r$, and the $z$-direction is compactified. The master equation reduces to the Schr\"odinger type equation,
\begin{equation}
\begin{aligned}
\( \frac{d^2}{dx^2}+\omega^2-V(x) \)\psi(x)=0,
\end{aligned}
\end{equation}
where $x$ is the tortoise variable, defined by
\begin{equation}
\begin{aligned}
x=r+r_H \log \left( \frac{r}{r_H}-1 \right)+c.
\end{aligned}
\end{equation}
The integration constant $c$ is fixed as follows.
 
\begin{figure}[tb]
\begin{center}
  \begin{minipage}[b]{0.45\linewidth}
    \centering
    \includegraphics[width=0.95\linewidth]{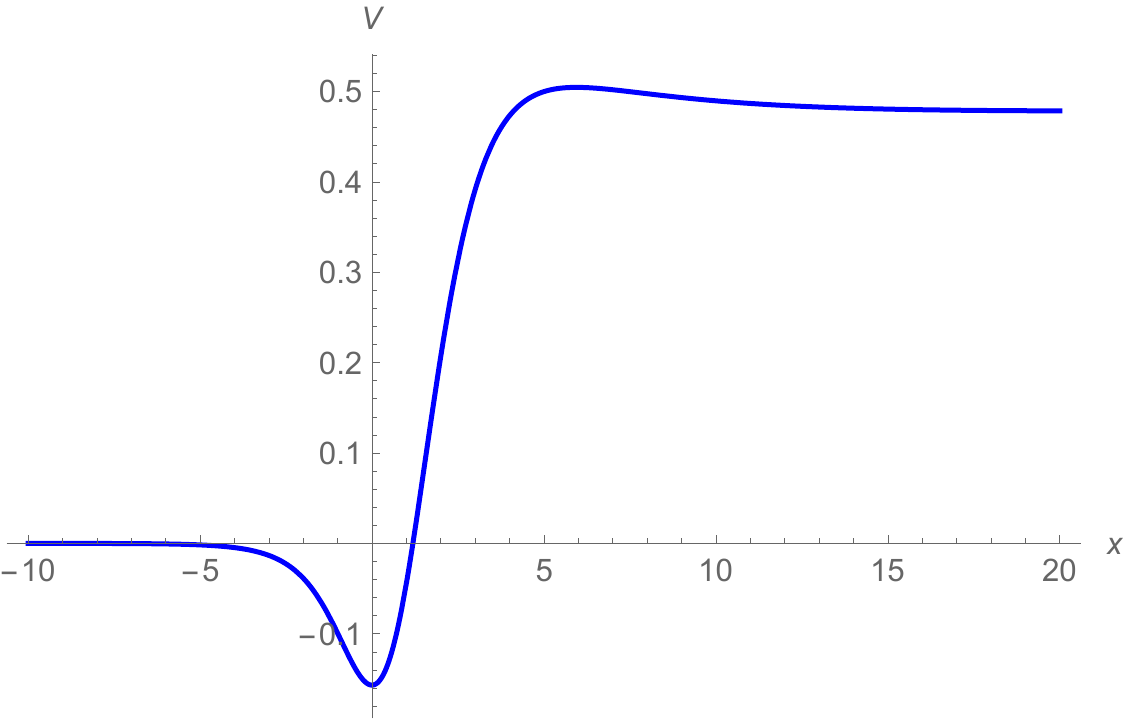}
  \end{minipage}
  \begin{minipage}[b]{0.45\linewidth}
    \centering
    \includegraphics[width=0.95\linewidth]{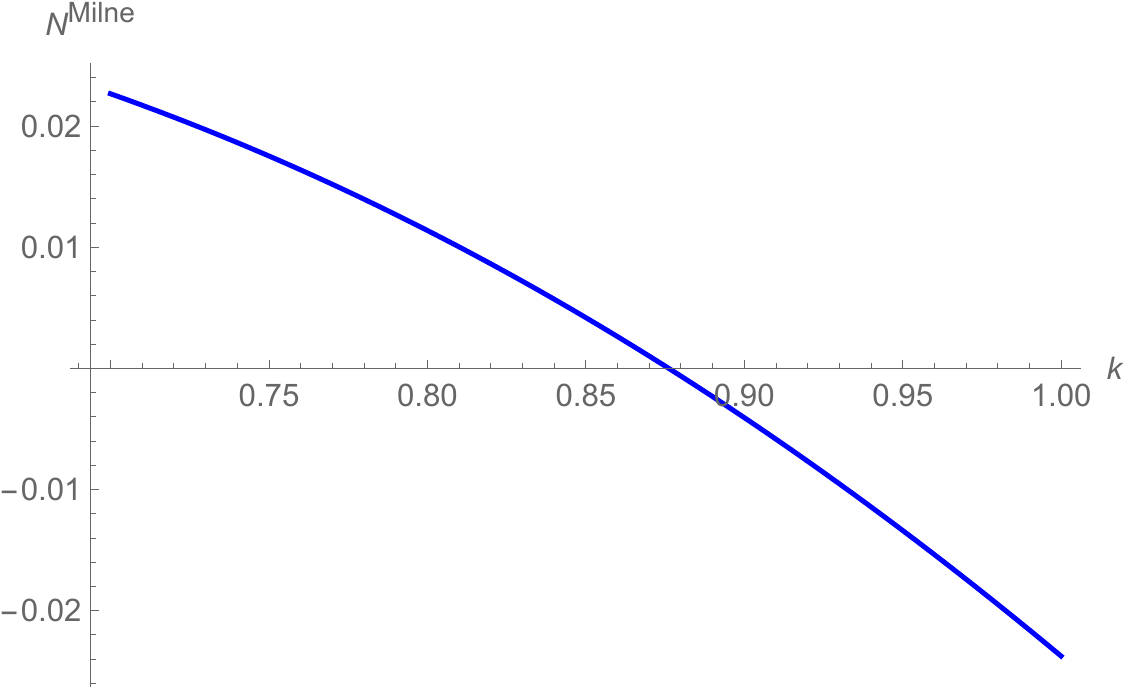}
  \end{minipage} 
\end{center}
  \caption{Left: The shape of the potential \eqref{eq:V-BS} is shown. The parameter is fixed as $k=0.7$. Our interest is to know whether this potential admits a bound state or not. Right: This figure show Milne's counting function for the black string potential \eqref{eq:V-BS}. There is the critical value $k_c \approx 0.876$ at which the counting function changes its sign.}
  \label{fig:GL}
\end{figure}

The potential is explicitly given by
\begin{equation}
\begin{aligned}
V(x)=\frac{f(r)}{r^3}\frac{k^6r^9+6k^4r^7-9k^4r^6r_H -12k^2r^4r_H+9k^2r^3r_H^2+r_H^3}{(k^2r^3+r_H)^2},
\end{aligned}
\label{eq:V-BS}
\end{equation}
where $k$ is the wave number along the $z$-direction.
For $k^2<2+\sqrt{3}$, the potential has a global minimum in $r>r_H$.
Also $V \to 0$ in $x \to -\infty$, and $V \to k^2$ in $x \to \infty$.
We focus on the case $0<k^2<2+\sqrt{3}$, and fix the integration constant $c$ so that the potential $V(x)$ has a minimum at $x=0$.
The typical shape of $V(x)$ is shown in the left panel of Figure~\ref{fig:GL}.

We first solve the differential equation
\begin{equation}
\begin{aligned}
\( \frac{d^2}{dx^2}+E-V(x) \)w_{E}(x)=\frac{1}{w_E(x)^3},\qquad w_E(0)=1,\qquad w_E'(1)=0,
\end{aligned}
\label{eq:w_E-eq-BS}
\end{equation}
where $E=\omega^2$. Milne's counting function is then obtained by
\begin{equation}
\begin{aligned}
N^\text{Milne}(E)=\frac{1}{\pi}\int_{-\infty}^\infty \frac{dx}{w_E(x)^2}-1.
\end{aligned}
\end{equation}
Of course, the non-linear differential equation \eqref{eq:w_E-eq-BS} cannot be solved analytically. We numerically solve it for a finite segment $[x_\text{min}, x_\text{max}]$.
We finally want to evaluate $N^\text{Milne}(0)$.
We have to care about its numerical evaluation.
In the region $x>x_\text{max}$, the potential is greater than zero-energy. See the left of Figure~\ref{fig:GL}. Since $1/w_0(x)^2$ very rapidly decays in $x>x_\text{max}$, we can ignore the contribution to $N^\text{Milne}(0)$ from this region as long as $V(\infty)=k^2>0$.
On the other hand, in $x<x_\text{min}$, the potential asymptotically goes to zero.
The contribution from this region is not exponentially small for $E=0$. We need to evaluate this contribution approximately.
If $|x_\text{min}|$ is large, we can ignore the potential. Therefore the approximate equation at $E=0$ in $x<x_\text{min}$ is
\begin{equation}
\begin{aligned}
\bar{w}_0''(x)=\frac{1}{\bar{w}_0(x)^3}.
\end{aligned}
\end{equation}
This equation can be solved analytically,
\begin{equation}
\begin{aligned}
\bar{w}_0(x)^2=C_1+\frac{(x+C_2)^2}{C_1},
\end{aligned}
\end{equation}
where $C_1$ and $C_2$ are integration constants.
We have to connect it, at $x=x_\text{min}$, with the (numerical) solution in $x_\text{min}\leq x \leq x_\text{max}$ smoothly. This determines the integration constants.
The result is the following:
\begin{equation}
\begin{aligned}
C_1=\frac{w_0(x_\text{min})^2}{1+w_0(x_\text{min})^2w_0'(x_\text{min})^2},\qquad
C_2=-x_\text{min}+\frac{w_0(x_\text{min})^3w_0'(x_\text{min})}{1+w_0(x_\text{min})^2w_0'(x_\text{min})^2}.
\end{aligned}
\end{equation}
where $w_0(x_\text{min})$ and $w_0'(x_\text{min})$ are evaluated by solving \eqref{eq:w_E-eq-BS} at $E=0$ numerically.
Therefore the counting function at $E=0$ is approximately evaluated by
\begin{equation}
\begin{aligned}
N^\text{Milne}(0)=\frac{1}{\pi}\int_{-\infty}^\infty \frac{dx}{w_0(x)^2}-1
\approx \frac{1}{\pi}\int_{x_\text{min}}^{x_\text{max}} \frac{dx}{w_0(x)^2}+\frac{1}{\pi}\int_{-\infty}^{x_\text{min}} \frac{dx}{\bar{w}_0(x)^2}-1,
\end{aligned}
\end{equation}
where the integration over $(-\infty, x_\text{min})$ can be computed exactly,
\begin{equation}
\begin{aligned}
\int_{-\infty}^{x_\text{min}} \frac{dx}{\bar{w}_0(x)^2} =
\int_{-\infty}^{x_\text{min}} dx \frac{C_1}{C_1^2+(x+C_2)^2}
=\arctan[ w_0(x_\text{min}) w_0'(x_\text{min}) ]+\frac{\pi}{2}.
\end{aligned}
\end{equation}
Our final expression is thus given by
\begin{equation}
\begin{aligned}
N^\text{Milne}(0) \approx \frac{1}{\pi}\int_{x_\text{min}}^{x_\text{max}} \frac{dx}{w_0(x)^2}
+\frac{1}{\pi}\arctan[ w_0(x_\text{min}) w_0'(x_\text{min}) ]-\frac{1}{2}.
\end{aligned}
\end{equation}
The remaining integral is evaluated by solving \eqref{eq:w_E-eq-BS} with $E=0$ numerically for $x_\text{min} \leq x \leq x_\text{max}$.

Now we apply this formalism to the black string potential \eqref{eq:V-BS}. Technically it is not hard to solve \eqref{eq:w_E-eq-BS} numerically for a given potential because it is an initial value problem. In the right panel of Figure~\ref{fig:GL}, we show $N^\text{Milne}(0)$ against the parameter $k$.
We set $x_\text{min}=-100$ and $x_\text{max}=100$.
It is clear to see that there is a value at which the sign of $N^\text{Milne}(0)$ changes.
This is nothing but the Gregory--Laflamme instability. Our estimation of the critical value $k_c$ is
\begin{equation}
\begin{aligned}
k_c \approx 0.8762.
\end{aligned}
\end{equation}
For $k<k_c$, there exists a negative bound state mode, and we conclude that the black string solution is unstable.

We note that there is another efficient way to discuss the mode stability of black holes \cite{kodama2003, ishibashi2003, kimura2017, kimura2018}.
The approach in this article is effectively useful to see the (in)stability quickly. On the other hand, the method in \cite{kodama2003, ishibashi2003, kimura2017, kimura2018} is useful to show the (in)stability more rigorously.

\section{Concluding remarks}
In this article, we reviewed various techniques in quantum mechanics, and explained how these techniques are applied to spectral problems appearing in black hole perturbation theory. Each of them has its own advantage, and it would be important to see the same problem from various perspectives. We hope that the techniques in this article will be useful in future developments on black hole perturbation theory.

Our main interest is QNM frequencies which is particularly important in recent gravitational wave observations. As spectral problems, they are understood as resonant eigenvalues for Schr\"odinger-type wave equations. We showed various methods to compute the QNM frequencies with high precision both numerically and analytically.

We also proposed a simple criteria to see the mode (in)stability of the linear perturbation of a given geometry. The problem is mapped to an existence condition of negative energy bound states. We can easily write it down by using Milne's very old idea in \cite{milne1930}. As an application, we explicitly showed the Gregory--Laflamme instability of 5d black string solutions.

In theories beyond general relativity, one often encounters coupled master equations. There are some approaches to analyze such systems \cite{amann1995, pani2013, mcmanus2019, kimura2018, kimura2018a, pierini2021, wagle2021, srivastava2021, cano2021a}. It would be interesting to develop the techniques in this article for the coupled master equations.

Although we did not review in this article, there are also new perspetives to study the black hole spectral problems from four-dimensional supersymmetric gauge theories \cite{aminov2020, hatsuda2021, bianchi2021} and from two-dimensional conformal filed theories \cite{novaes2014, dacunha2015, carneirodacunha2020, bonelli2021a, carneirodacunha2021}.\footnote{These two approaches turn out to be interrelated due to the Alday--Gaiotto--Tachikawa relation \cite{alday2010}. However, the relation is quite non-trivial. See the introductory section in \cite{aminov2020}.} The ideas are based on the common mathematical structure in the spectral problems. These new perspectives provide us powerful analytic treatments on the spectral problems. However, a nice physical interpretation of the relation is still obscure.

~\\
%%%%%%%%%%%%%%%%%%%%%%%%%%%%%%%%%%%%%%%%%%
\textbf{Acknowledgments:} This work is supported by JSPS KAKENHI Grant Numbers JP18K03657 (YH) and JP20H04746 (MK). We are grateful to Hiroyuki Nakano for reading the draft carefully.

%% optional
\appendixtitles{yes} % Leave argument "no" if all appendix headings stay EMPTY (then no dot is printed after "Appendix A"). If the appendix sections contain a heading then change the argument to "yes".
\appendix

\section{Derivation of the master equations around the four-dimensional Schwarzschild spacetime}\label{app:BH-pert}

In this appendix, we review a derivation of the master equations around the Schwarzschild spacetime~\cite{regge1957, zerilli1970}.
For simplicity, we assume that the spacetime is four dimensional and it has the global spherical symmetry.
The basic technique used in this section can also be applied for the various cases such as modified gravity theories.
See also reviews and textbooks~\cite{nakamura1987, nollert1999, chandrasekhar1998, maggiore2018, andersson2020}.
For further general cases, which treat higher dimensions and the maximally symmetric angular part, 
are discussed, {\it e.g.,} in~\cite{ishibashi2011, jansen2019a}.

\subsection{Four-dimensional Schwarzschild spacetime and Killing vectors}
The metric of the Schwarzschild spacetime is given by
\begin{align}
ds^2 = g^{\rm (Sch)}_{\mu \nu}dx^\mu dx^\nu = - f dt^2 + \frac{dr^2}{f} + r^2 (d\theta^2 + \sin^2\theta d\phi^2),
\end{align}
with $f = 1- r_H /r$.
This spacetime admits four linearly independent Killing vectors
\begin{align}
\xi_{(t)}^\mu \frac{\partial }{\partial x^\mu} &= \frac{\partial}{\partial t},
\\
\xi_{(1)}^\mu \frac{\partial }{\partial x^\mu} &= 
-\sin\phi\frac{\partial}{\partial \theta}
-
\frac{\cos\theta \cos \phi}{\sin \theta}
\frac{\partial}{\partial \phi},
\\
\xi_{(2)}^\mu \frac{\partial }{\partial x^\mu} &= 
\cos\phi\frac{\partial}{\partial \theta}
-
\frac{\cos\theta \sin \phi}{\sin \theta}
\frac{\partial}{\partial \phi},
\\
\xi_{(3)}^\mu \frac{\partial }{\partial x^\mu} &= 
\frac{\partial}{\partial \phi},
\end{align}
where those vectors satisfy the Killing equations $\nabla_\mu \xi_\mu + \nabla_\nu \xi_\mu = 0$.
The generator of the spherical symmetry 
$\xi_{(I)}^\mu$ with $I = 1,2,3$ form the $SO(3)$ algebraic relations
\begin{align}
[{\cal L}_{\xi_{(I)}}, {\cal L}_{\xi_{(J)}}] = \epsilon_{IJK} {\cal L}_{\xi_{(K)}},
\end{align}
where $I, J, K$ run over $1,2,3$, 
the operator
${\cal L}_{\xi_{(I)}}$ denotes the Lie derivative along $\xi_{(I)}^\mu$,
and $\epsilon_{IJK}$ is the totally anti-symmetric symbol with $\epsilon_{123} = 1$.
Defining the Casimir operator $L^2$ as
\begin{align}
L^2 := \sum_{I = 1}^3 {\cal L}_{\xi_{(I)}} {\cal L}_{\xi_{(I)}},
\end{align}
the relations
\begin{align}
[L^2, {\cal L}_{\xi_{(I)}}] = 0,
\end{align}
hold for $I = 1,2,3$.

\subsection{Massless Klein--Gordon equation}
First, we consider the massless Klein--Gordon equation on the Schwarzschild spacetime
\begin{align}
\square \Phi = 0,
\end{align}
where $\square = \nabla^\mu \nabla_\mu$.
Before showing the explicit calculation, we discuss the 
general relation among the separation of variables and Killing vectors.

\subsubsection{Killing vector and separation of variable}
When a spacetime admits a Killing vector $\xi^\mu$, we can choose the coordinate system 
so that $\xi^\mu$ becomes one of the coordinate basis.
We denote the corresponding coordinate by $\lambda$, then
\begin{align}
\xi^\mu \frac{\partial}{\partial x^\mu} = \frac{\partial}{\partial \lambda},
\end{align}
holds.
In particular, the Killing equations ${\cal L}_{\xi} g_{\mu \nu} = 0$ become
\begin{align}
\frac{\partial g_{\mu \nu}}{\partial \lambda} = 0,
\end{align}
in this coordinate system and the relation $[\partial/\partial \lambda, \square] = 0$ holds
because the d'Alembertian, which acts on scalar fields, $\square = (1/\sqrt{|g|})\partial_\mu (\sqrt{|g|} g^{\mu \nu}\partial_\nu)$ does not 
depend on $\lambda$.
If we assume the form of the Klein--Gordon field $\Phi$ as
\begin{align}
\Phi = e^{i L \lambda} \tilde{\Phi}(x^i),
\end{align}
where $L$ is the separation constant and $x^i$ denote the coordinates except for $\lambda$, we obtain a relation
\begin{align}
\frac{\partial}{\partial \lambda} \square \Phi &=
\square \frac{\partial \Phi}{\partial \lambda}
\notag\\&= 
i L \square e^{i L \lambda} \tilde{\Phi}(x^i)
\notag\\&= 
i L \square \Phi.
\end{align}
The solution of this equation can be written by
\begin{align}
\square \Phi = e^{i L \lambda} F(x^i),
\end{align}
where $F$ is a function of $x^i$.
Thus, we conclude that the Klein--Gordon equation reduces to the equation $F(x^i) = 0$ which only contains $x^i$,
while the explicit form of the function $F(x^i)$ is not known.

If the spacetime admits two or more commutative Killing vectors, 
we can choose coordinate system so that these Killing vectors are coordinate bases,
and the same discussion as single Killing vector case holds
by using 
the method of separation of variables with respect to the corresponding coordinates.

\subsubsection{Separation of variable using $SO(3)$ symmetry}
While the $SO(3)$ Killing vectors $\xi_{(I)}^\mu$ are not commutative, 
we can consider the simultaneous eigenfunction of one of the Killing vectors and $L^2$.
We denote the simultaneous eigenfunction with respect to ${\cal L}_{\xi_{(t)}}, {\cal L}_{\xi_{(3)}}$ and the Casimir operator $L^2$ by
$e^{- i\omega t} e^{im\phi} S(\theta)$ where $\omega$ and $m$ are constants,
and we assume the eigenvalue of $L^2$ is $-\ell (\ell +1)$. 
The operator $L^2$ satisfies 
\begin{align}
[L^2, \square] = 0,
\end{align}
because $L^2 = \sum_{I = 1}^3 {\cal L}_{\xi_{(I)}} {\cal L}_{\xi_{(I)}}$ and $[{\cal L}_{\xi_{(I)}}, \square] = 0$.
Assuming the form of the Klein--Gordon field $\Phi$ as
\begin{align}
\Phi = e^{- i\omega t} e^{im\phi} S(\theta) \tilde{\Phi}(r),
\end{align}
the relations
\begin{align}
{\cal L}_{\xi_{(t)}}\Phi &= - i\omega \Phi,
\\
{\cal L}_{\xi_{(3)}}\Phi &=  im \Phi,
\\
L^2 \Phi &= - \ell(\ell + 1) \Phi,
\end{align}
hold.
Because the operator $L^2$ is the Laplacian on $S^2$, 
$L^2 \Phi = \partial_\theta^2 \Phi + \cot \theta \partial_\theta\Phi +\csc^2 \theta\partial_\phi^2\Phi$,
the regular solutions to the above equations 
exist only when $\ell = 0,1,2,\cdots$ and $m = 0, \pm1, \pm2, \cdots, \pm\ell$ from the Sturm-Liouville theory,
then $e^{im\phi} S(\theta)$ is the spherical harmonics $Y_{\ell m}$
\begin{align}
e^{im\phi} S(\theta) = Y_{\ell m}.
\end{align}
Defining ${\cal L}_\pm = -i {\cal L}_{\xi_{(1)}} \pm  {\cal L}_{\xi_{(2)}}$, 
the relations
\begin{align}
{\cal L}_\pm Y_{\ell m} = \sqrt{(\ell \mp m)(\ell \pm m + 1)} Y_{\ell m\pm1},
\label{eq:ylmladder}
\end{align}
hold for different modes $m$.
Using the commutation relations 
$[{\cal L}_{\xi_{(t)}}, \square] = [{\cal L}_{\xi_{(3)}}, \square] = [L^2, \square] = 0$,
we can also show that $\square \Phi$ satisfies
\begin{align}
{\cal L}_{\xi_{(t)}}\square \Phi &= - i\omega \square\Phi,
\\
{\cal L}_{\xi_{(3)}}\square\Phi &=  im \square\Phi,
\\
L^2 \square\Phi &= - \ell(\ell + 1) \square\Phi.
\end{align}
The regular solution of the above equations can be written by
\begin{align}
\square\Phi =  e^{- i\omega t} Y_{\ell m} F(r),
\label{eq:squarephi1}
\end{align}
where $F(r)$ is a function of $r$.
This implies that the Klein--Gordon equation reduces to an
ordinary differential equation $F(r) = 0$ which only depends on $r$.\footnote{
This conclusion holds even if we do not use the global spherical symmetry of the spacetime.
Because the only non-trivial operator which commutes with all $SO(3)$ Killing vectors is the Casimir operator $L^2$,
the derivatives with respect to the angular coordinates in $\square$ should be proportional to $L^2$.
Thus, assuming $\Phi = e^{-i\omega t}\tilde{\Phi}(r) Y(\theta,\phi)$ where the function $Y$ 
satisfies $L^2 Y = -\lambda Y$ with a separation constant $\lambda$, 
$\square \Phi$ should take the form of $e^{-i\omega t}Y F(r)$.}

\subsubsection{Explicit calculation}
We write the form of the Klein--Gordon field $\Phi$ as
\begin{align}
\Phi = \sum_{\ell, m} \Phi^{\ell m},
\end{align}
with
\begin{align}
\Phi^{\ell m} = e^{-i\omega t}Y_{\ell m}(\theta,\phi) \frac{\Phi^{(0)}(r)}{r},
\end{align}
where we assumed the time dependence as $ e^{-i\omega t}$ 
and omitted to write indices $\ell, m$ in $\Phi^{(0)}$.
Because Eq.~\eqref{eq:squarephi1} holds for each mode $\ell, m$,
and the spherical harmonics $Y_{\ell m}$ with different indices are linearly independent,
we can separately discuss different $\ell, m$ modes.
The Klein--Gordon equation for each mode reduces to\footnote{
We can easily calculate the explicit form of $\square \Phi^{\ell m}$ by using \textit{Mathematica}.} 
\begin{align}
\square \Phi^{\ell m} = \frac{e^{-i\omega t}Y_{\ell m}}{rf}
\left(
f \frac{d}{dr}\left(
f\frac{d\Phi^{(0)}}{dr}
\right)
+
\left(\omega^2  - V^{(0)}\right)\Phi^{(0)}
\right) = 0,
\end{align}
where the effective potential $V^{(0)}$ is given by
\begin{align}
V^{(0)} &= f\left(\frac{\ell (\ell + 1)}{r^2} + \frac{r_H}{r^3}\right).
\label{eq:effVkgeq}
\end{align}
Thus, we obtain the master equation
\begin{align}
f \frac{d}{dr}\left(
f\frac{d\Phi^{(0)}}{dr}
\right)
+
\left(\omega^2  - V^{(0)}\right)\Phi^{(0)} =0.
\label{eq:mastereqkgeq}
\end{align}
We comment that if we consider the massive Klein--Gordon equation $(\square - m^2)\Phi =0$,
the effective potential is changed as $V^{(0)} \to V^{(0)} + f m^2$.

\subsection{Vector harmonics on $S^2$}
Before discussing the case of the Einstein equations, we introduce 
some mathematical arguments.

\subsubsection{Helmholtz--Hodge decomposition}
We introduce the
Helmholtz--Hodge decomposition of vector and symmetric two tensor fields on $S^2$.\footnote{Generalization to
compact Riemannian (Einstein) manifolds can be seen in~\cite{ishibashi2004}.}
Let $\gamma_{ij}$ be the metric of a unit two sphere $S^2$
\begin{align}
\gamma_{ij}dx^i dx^j = d\theta^2 + \sin^2\theta d\phi^2.
\label{eq:unittwosphere}
\end{align}
In this subsection, $i,j$ denote the tensor indices on $S^2$, and we raise or lower tensor indices by $\gamma_{ij}$.
We denote by $\hat{\nabla}_i$ the covariant derivative with respect to $\gamma_{ij}$.

Any regular vector field $v_i$ on $S^2$ can be uniquely decomposed into 
the scalar part and the vector part as
\begin{align}
v_i = \hat{\nabla}_i S + V_i ,
\label{eq:vi}
\end{align}
with 
\begin{align}
\hat{\nabla}^i V_i = 0.
\end{align}
The proof is simple.
Taking the divergence of Eq~\eqref{eq:vi}, we obtain an equation $\hat{\nabla}^i v_i = \hat{\nabla}^i \hat{\nabla}_i S$.
Solving this Poisson equation on $S^2$, $\hat{\nabla}_i S$ is uniquely determined.\footnote{
The homogeneous equation $\hat{\nabla}^i \hat{\nabla}_i S =0$ has a constant solution,
and this gives an ambiguity of the solution for $\ell = 0$ mode. However, it does not affect $\hat{\nabla}_i S$.
} 
Then, $V_i$ is given by $V_i = v^i - \hat{\nabla}_iS$ which clearly satisfies $\hat{\nabla}^i V_i = 0$.

In a similar way, 
any regular symmetric tensor field $t_{ij}$ on $S^2$ can be uniquely decomposed as\footnote{
While one may think that there might exitst a tensor part which satisfies $t_{ij}^t$
with trace free and divergence free conditions
$t^t_{i}{}^i = \hat{\nabla}^it_{ij}^t = 0$, such a term does not exist for $S^2$.
If we assume that a tensor part $t_{ij}^t$ exists, we can consider a perturbed metric
$\bar{\gamma}_{ij} = \gamma_{ij} + \epsilon t_{ij}^t$, where $\gamma_{ij}$ is the metric of a unit two sphere.
Using the conditions $t^t_{i}{}^i = \hat{\nabla}^it_{ij}^t = 0$,
we can show that the Ricci scalar of $\bar{\gamma}_{ij}$ becomes $2 + {\cal O}(\epsilon^2)$.
This implies that $\bar{\gamma}_{ij}$ is the metric of a unit two sphere at this order,
and the metric can be transformed into the same form as Eq.~\eqref{eq:unittwosphere} by 
a gauge transformation at ${\cal O}(\epsilon)$.
Thus, $t_{ij}^t$ should be $t_{ij}^t = 2 \hat{\nabla}_{(i} \zeta_{j)}$ with some vector field $\zeta_j$.
If we write $\zeta_i = \hat{\nabla}_iS + V_i$ with $\hat{\nabla}^iV_i = 0$,
the conditions $t^T_{i}{}^i = \hat{\nabla}^it_{ij}^T = 0$ implies $\hat{\nabla}_iS = V_i = 0$,
and then we conclude $t_{ij}^t = 0$.
}
\begin{align}
t_{ij} = \left(\hat{\nabla}_i \hat{\nabla}_j  - \frac{1}{2}\gamma_{ij}\hat{\triangle}\right)t_T
+
t_L \gamma_{ij} + 2 \hat{\nabla}_{(i}t_{j)}^v,
\label{eq:tijdecompsition}
\end{align}
with 
\begin{align}
\hat{\nabla}^it_i^v = 0.
\end{align}
To prove this, first, taking the trace of Eq.~\eqref{eq:tijdecompsition}, we obtain $t_L = t^i{}_i/2$.
Next, acting an operator $\hat{\nabla}^i\hat{\nabla}^j$ on Eq.~\eqref{eq:tijdecompsition},
after some calculation, we obtain
\begin{align}
\hat{\triangle} \hat{\triangle} t_T + 2 \hat{\triangle}t_T = 2 \hat{\nabla}^i \hat{\nabla}^j(t_{ij} - t_L\gamma_{ij}).
\label{eq:triangletriangleeq}
\end{align}
Solving Eq.~\eqref{eq:triangletriangleeq} with respect to $\hat{\triangle} t_T$, 
an equation $\hat{\triangle} t_T = A$ with a regular function $A$ is obtained, 
and then we obtain $t_T$ by solving this equation.
Then, the form of
$(\hat{\nabla}_i \hat{\nabla}_j  - (1/2)\gamma_{ij}\hat{\triangle})t_T$
is uniquely determined.\footnote{
While there are homogeneous solutions in $t_T$ for $\ell = 0$ and $1$ modes, these do not affect 
$(\hat{\nabla}_i \hat{\nabla}_j  - (1/2)\gamma_{ij}\hat{\triangle})t_T$.
}
Finally, taking a divergence of Eq.~\eqref{eq:tijdecompsition},
we obtain an equation
\begin{align}
\hat{\nabla}^i\left(
t_{ij} - \left(\hat{\nabla}_i \hat{\nabla}_j  - \frac{1}{2}\gamma_{ij}\hat{\triangle}\right)t_T
-
t_L \gamma_{ij} 
\right)
=
\hat{\triangle}t_j^v + t_j^v.
\label{eq:lapviv}
\end{align}
The solution of Eq.~\eqref{eq:lapviv} with $\hat{\nabla}^it_i^v = 0$ is uniquely determined except for the homogenous solutions. Because the homogenous solutions correspond to the Killing vectors on $S^2$, they do not affect
the form of $\hat{\nabla}_{(i}t_{j)}^v$.

\subsubsection{Vector harmonics}
It is well known that any regular function on $S^2$ can be written by the spherical harmonics $Y_{\ell m}$.
Here, we introduce the vector spherical harmonics $V^{\ell m}_i$ which can 
express any divergence free regular vector field on $S^2$.\footnote{
The proof of the completeness can be seen in~\cite{ishibashi2011}.
}
The vector spherical harmonics $V^{\ell m}_i$ satisfies the equations\footnote{
The second equation can also be written as $\hat{\triangle}V^{\ell m}_i = (-\ell(\ell + 1) + 1)V^{\ell m}_i$.}
\begin{align}
{\cal L}_{\xi_{(3)}}V^{\ell m}_i &=  im V^{\ell m}_i,
\\
L^2 V^{\ell m}_i &= - \ell(\ell + 1) V^{\ell m}_i,
\end{align}
with 
\begin{align}
\hat{\nabla}^i V^{\ell m}_i = 0,
\end{align}
where $\ell = 1, 2, \cdots,$ and $m = 0, \pm 1, \pm2, \cdots, \pm \ell$.
We can explicitly construct $V^{\ell m}_i$ by using the spherical harmonics $Y_{\ell m}$ as
\begin{align}
V^{\ell m}_i = \hat{\epsilon}_i{}^j \hat{\nabla}_j Y_{\ell m},
\end{align}
where $\hat{\epsilon}_{ij}$ is the Levi--Civita tensor of $\gamma_{ij}$.\footnote{
If we use the differential form, $\hat{\epsilon}_{ij}$ becomes ${\bm {\hat{\epsilon}}} = \sin\theta d\theta \wedge d\phi$.
}
Using the vector spherical harmonics, any divergence free vector $V_i$ on $S^2$ can be written by
\begin{align}
V_i = \sum_{\ell m} d_{\ell m} V^{\ell m}_i,
\end{align}
where $d_{\ell m}$ are constants.
Then, a vector field $v_i$ in Eq.~\eqref{eq:vi} becomes
\begin{align}
v_i &= \hat{\nabla}_i S + V_i 
\notag\\&=
\sum_{\ell, m}
v_i^{\ell m}
\notag\\&=
\sum_{\ell, m}
\left(
c_{\ell m}\hat{\nabla}_i Y_{\ell m}
+
d_{\ell m} V^{\ell m}_i
\right).
\label{eq:viwithYV}
\end{align}
We note that each $\ell, m$ component $v_i^{\ell m}$ satisfies
\begin{align}
{\cal L}_{\xi_{(3)}}v_i^{\ell m} &=  im v_i^{\ell m},
\label{eq:L3vi}
\\
L^2 v_i^{\ell m} &= - \ell(\ell + 1) v_i^{\ell m}.
\label{eq:L2vi}
\end{align}
On the other hand, if $v_i$ satisfies Eqs~\eqref{eq:L3vi} and \eqref{eq:L2vi},
$v_i$ should take the form of
\begin{align}
v_i = v_i^{\ell m} = c_{\ell m}\hat{\nabla}_i Y_{\ell m}
+
d_{\ell m} V^{\ell m}_i.
\label{eq:eqvicomponent}
\end{align}
This can be shown as follows. Substituting Eq.~\eqref{eq:viwithYV} into Eqs~\eqref{eq:L3vi} and \eqref{eq:L2vi},
we obtain equations
$\sum_{\bar{\ell}, \bar{m}}
 (\bar{m}-m) v_i^{\bar{\ell} \bar{m}} 
= 0$ and
$\sum_{\bar{\ell}, \bar{m}}
 (\bar{\ell}(\bar{\ell} + 1) - \ell(\ell+1))v_i^{\bar{\ell} \bar{m}} 
= 0$. Then $v_i^{\bar{\ell} \bar{m}}$ with $\bar{\ell} \neq \ell$ and $\bar{m} \neq m$ should vanish
because of the linear independence of $Y_{\ell m}$ and $V^{\ell m}_{i}$.

\subsection{Einstein equations}
\subsubsection{Linear metric perturbation}
Next, we consider the case of the vacuum Einstein equations.\footnote{
See~\cite{martel2005} for the case with source terms.
}
We denote by $\epsilon h_{\mu \nu}$ the small deviation from the Schwarzschild metric where $\epsilon$ is a small parameter.
The total metric at ${\cal O}(\epsilon)$ becomes
\begin{align}
g_{\mu \nu} = g_{\mu \nu}^{\rm (Sch)} + \epsilon h_{\mu \nu}.
\end{align}
We need to solve the vacuum Einstein equations $G_{\mu \nu} = 0$ or equivalently $R_{\mu \nu} = 0$ at ${\cal O}(\epsilon)$
\begin{align}
G_{\mu \nu} &= \epsilon \left(
-\frac{1}{2}\nabla_\mu \nabla_\nu h^\alpha{}_\alpha
-\frac{1}{2} \nabla^\alpha \nabla_\alpha h_{\mu \nu}
+ 
\nabla^\alpha \nabla_{(\mu}h_{\nu)\alpha}
+
\frac{1}{2}g_{\mu \nu}
(\nabla^\alpha \nabla_\alpha h^\beta{}_\beta - \nabla^\alpha \nabla^\beta h_{\alpha \beta})
\right)
\notag\\
&=: 
\epsilon \hat{X} h_{\mu \nu}
=0,
\label{eq:linearizedEisteineq}
\end{align}
where $\nabla_\mu$ denotes the covariant derivative with respect to the Schwarzschild metric,
and $\hat{X}$ denotes the operator s.t. $\hat{X} h_{\mu \nu}$ gives the linearized Einstein tensor.

The perturbed metric $h_{\mu \nu}$ can be written by
\begin{align}
h_{\mu \nu}dx^\mu dx^\nu 
&= 
h_{AB} dx^A dx^B
+
h_{Ai}dx^A dx^i
+
h_{ij}dx^i dx^j,
\end{align}
where $A,B$ run $t,r$, and $i,j$ run $\theta, \phi$.
We can regard $h_{AB}, h_{Ai}, h_{ij}$ as scalar, vector and symmetric tensor on $S^2$, respectively.
Then, using the Helmholtz--Hodge decomposition, perturbed metric $h_{\mu \nu}$ 
can be expressed by $Y_{\ell m}$ and $V^{\ell m}_i$
\begin{align}
h_{\mu \nu}dx^\mu dx^\nu &= \sum_{\ell, m}h_{\mu \nu}^{\ell m}dx^\mu dx^\nu
=
\sum_{\ell, m}\left(h_{\mu \nu}^{(+)\ell m} + h_{\mu \nu}^{(-)\ell m}\right)dx^\mu dx^\nu,
\end{align}
where $h_{\mu \nu}^{(+)\ell m}$ is the even parity perturbation
\begin{align}
h_{\mu \nu}^{(+)\ell m}dx^\mu dx^\nu
&=
e^{-i\omega t} \left(
H_0 dt^2 + 2 H_1 dt dr + H_2 dr^2
\right)Y_{\ell m}
\notag\\&\quad -
2 re^{-i\omega t}
\left( H_{t\Omega}dt +  H_{r\Omega}dr\right)
\frac{\hat{\nabla}_i Y_{\ell m} dx^i}{\sqrt{\ell(\ell+1)}}
\notag\\&\quad+
2 r^2 e^{-i\omega t} \left(
K \gamma_{ij}Y_{\ell m}
+
 \frac{K_2}{\ell (\ell+1)}
\left(
\hat{\nabla_i}\hat{\nabla}_j Y_{\ell m}
-
\frac{1}{2}\gamma_{ij}\hat{\triangle}Y_{\ell m}
\right)
\right)
dx^i dx^j,
\label{eq:hplus}
\end{align}
and $h_{\mu \nu}^{(-)\ell m}$ is the odd parity perturbation
\begin{align}
h_{\mu \nu}^{(-)\ell m}dx^\mu dx^\nu
&=
2 e^{-i\omega t}
( h_0 dt +  h_1 dr)
V_i^{\ell m}dx^i
-
2 r^2 e^{-i\omega t}h_\Omega 
\frac{\hat{\nabla}_{(i}V_{j)}^{\ell m}dx^i dx^j}{\sqrt{\ell(\ell + 1) - 1}}.
\label{eq:hminus}
\end{align}
We assumed the time dependence of $h_{\mu \nu}$ as $e^{-i\omega t}$.
There are seven unknown functions $H_0, H_1, H_2, H_{t\Omega}, H_{r\Omega}, K, K_2$
for the even parity perturbation, 
and three unknown functions $h_0, h_1, h_\Omega$ for the odd parity perturbation.
Note that we omitted to write indices $\ell, m$ in these unknown functions.

Similar to the case of the Klein--Gordon equation, 
$h_{\mu \nu}^{\ell m}$ satisfies relations
\begin{align}
{\cal L}_{\xi_{(t)}}h_{\mu \nu}^{\ell m} &= - i\omega h_{\mu \nu}^{\ell m},
\\
{\cal L}_{\xi_{(3)}}h_{\mu \nu}^{\ell m} &=  im h_{\mu \nu}^{\ell m},
\\
L^2 h_{\mu \nu}^{\ell m} &= - \ell(\ell + 1) h_{\mu \nu}^{\ell m}.
\end{align}
We should note that if a symmetric tensor $h_{\mu \nu}$ satisfies the above equations, $h_{\mu \nu}$
should take the form of Eqs.~\eqref{eq:hplus} and \eqref{eq:hminus}
from a similar discussion to obtain Eq.~\eqref{eq:eqvicomponent} for the vector case.

Because the operator $\hat{X}$ in Eq.~\eqref{eq:linearizedEisteineq}
depends only on the covariant derivative of the background metric $g_{\mu \nu}^{\rm (Sch)}$,
$\hat{X}$ and the Lie derivative with respect to the Killing vectors are commutative,
and
the commutation relations $[\hat{X},{\cal L}_{\xi_{(t)}}] = [\hat{X},{\cal L}_{\xi_{(3)}}] = [\hat{X},L^2] = 0$ hold.
Thus, the relations
\begin{align}
{\cal L}_{\xi_{(t)}}\hat{X}h_{\mu \nu}^{\ell m} &= - i\omega \hat{X}h_{\mu \nu}^{\ell m},
\\
{\cal L}_{\xi_{(3)}}\hat{X}h_{\mu \nu}^{\ell m} &=  im \hat{X}h_{\mu \nu}^{\ell m},
\\
L^2 \hat{X}h_{\mu \nu}^{\ell m} &= - \ell(\ell + 1) \hat{X}h_{\mu \nu}^{\ell m},
\end{align}
hold and
$\hat{X}h_{\mu \nu}^{\ell m}$ should take the same form of Eqs.~\eqref{eq:hplus} and \eqref{eq:hminus}
but coefficients are different functions.
This implies that we can separately discuss different $\ell, m, \omega$ modes 
and we need to solve ordinary differential equations of $r$ for each $\ell, m, \omega$ mode.

\subsubsection{Parity transformation}
In this subsection, we show that we can separately discuss the even and odd parity perturbations.
We define the parity transformation operator $P$ which is generated by the coordinate transformation
\begin{align}
(\theta, \phi) \to (\pi - \theta, \pi+\phi).
\label{eq:paritytr}
\end{align}
Then, the relations
\begin{align}
P Y_{\ell m} &= (-1)^\ell Y_{\ell m},
\\
P V_i^{\ell m} &= (-1)^{\ell +1} V_i^{\ell m},
\end{align}
hold, where we used $P \hat{\epsilon}_{ij} = - \hat{\epsilon}_{ij}$.\footnote{
We define the parity transformation of the tensor fields $PW_{\mu_1\mu_2\cdots\mu_n}$ as 
the components of the tensor after performing the coordinate transformation Eq.~\eqref{eq:paritytr} to the
whole tensor $W_{\mu_1\mu_2\cdots\mu_n}dx^{\mu_1}\otimes dx^{\mu_2}\otimes \cdots \otimes dx^{\mu_n}$ 
which includes the coordinate basis, where $\otimes$ denotes the tensor product.
}
Using these relations, $h^{(\pm)\ell m}_{\mu \nu}$ satisfy
\begin{align}
P h^{(+)\ell m}_{\mu \nu} &= (-1)^\ell h^{(+)\ell m}_{\mu \nu},
\\
P h^{(-)\ell m}_{\mu \nu} &= (-1)^{\ell+1} h^{(-)\ell m}_{\mu \nu}.
\end{align}
Because the background metric $g_{\mu \nu}^{\rm (Sch)}$ is invariant under the parity transformation
and $\hat{X}$ has the same symmetry as that of $g_{\mu \nu}^{\rm (Sch)}$,
the commutation relation $[P, \hat{X}] = 0$ holds, and then
\begin{align}
P \hat{X} h^{(+)\ell m}_{\mu \nu} = (-1)^\ell \hat{X} h^{(+)\ell m}_{\mu \nu},
\label{eq:phplus}
\end{align}
holds.
We write $ \hat{X}h^{(+)\ell m}_{\mu \nu}$ as
\begin{align}
 \hat{X} h^{(+)\ell m}_{\mu \nu} = C_+ Z^{(+)}_{\mu \nu} + C_-Z^{(-)}_{\mu \nu},
\label{eq:hatXhplus1}
\end{align}
where $C_\pm$ are constants, and $Z^{(+)}_{\mu \nu}$ and $Z^{(-)}_{\mu \nu}$ take the same form as Eqs.~\eqref{eq:hplus} and \eqref{eq:hminus}, respectively.
On the other hand, using Eq.~\eqref{eq:phplus}, we obtain
\begin{align}
(-1)^\ell \hat{X} h^{(+)\ell m}_{\mu \nu} &= P \hat{X} h^{(+)\ell m}_{\mu \nu}
\notag\\&=
(-1)^\ell  C_+ Z^{(+)}_{\mu \nu} + (-1)^{\ell +1}C_-Z^{(-)}_{\mu \nu},
\end{align}
and this implies
\begin{align}
 \hat{X} h^{(+)\ell m}_{\mu \nu} = C_+ Z^{(+)}_{\mu \nu} - C_-Z^{(-)}_{\mu \nu}
\label{eq:hatXhplus2}.
\end{align}
Eqs.~\eqref{eq:hatXhplus1} and \eqref{eq:hatXhplus2} show $C_-Z^{(-)}_{\mu \nu} = 0$,
and we conclude that $\hat{X} h^{(+)\ell m}_{\mu \nu}$ does not contain the odd parity components.
In a similar way, we can see that $\hat{X} h^{(-)\ell m}_{\mu \nu}$ does not contain the even parity components.
Thus, even and odd parity perturbations can be separately discussed because they are not coupled in the Einstein equations.

\subsubsection{Gauge transformation}
The total line element at ${\cal O}(\epsilon)$ is given by
\begin{align}
ds^2 = (g_{\mu \nu}^{\rm (Sch)} + \epsilon h_{\mu \nu})dx^\mu dx^\nu.
\end{align}
If we introduce a coordinate transformation
\begin{align}
x^\mu \to x^\mu - \epsilon \zeta^\mu,
\end{align}
the line element changes
\begin{align}
ds^2 \to g_{\mu \nu}^{\rm (Sch)}dx^\mu dx^\nu + \epsilon \left(
2 \nabla_{(\mu} \zeta_{\nu)}
+
h_{\mu \nu}\right) dx^\mu dx^\nu,
\end{align}
where $\nabla_\mu$ is the covariant derivative with respect to the Schwarzschild metric.
Thus, there is an gauge ambiguity due to the term $2 \nabla_{(\mu} \zeta_{\nu)}$ at ${\cal O}(\epsilon)$.
Writing $\zeta_\mu$ as
\begin{align}
\zeta_\mu dx^\mu &= \sum_{\ell,m}\left( e^{-i\omega t}(\zeta_t dt + \zeta_rdr)Y_{\ell m}
-
 e^{-i\omega t}\zeta_S \frac{\hat{\nabla}_i Y_{\ell m}dx^i}{\sqrt{\ell(\ell+1)}}
+
 e^{-i\omega t} \zeta_V V_i^{\ell m}dx^i\right),
\end{align}
the perturbed metric transforms 
\begin{align}
H_ 0 &\to H_0 - 2i\omega \zeta_t - f f^\prime \zeta_r,
\label{eq:metgaugetr1}
\\
H_1 &\to H_1 - i\omega \zeta_r - \frac{f^\prime}{f}\zeta_t + \zeta_t^\prime,
\\
H_2 &\to H_2 + \frac{f^\prime}{f}\zeta_r + 2 \zeta_r^\prime,
\\
H_{t\Omega} &\to H_{t\Omega} - \frac{\sqrt{\ell(\ell + 1)}}{r}\zeta_t -\frac{i\omega}{r}\zeta_S,
\\
H_{r\Omega} &\to H_{r\Omega} - \frac{\sqrt{\ell(\ell+1)}}{r}\zeta_r - \frac{2}{r^2}\zeta_S + \frac{1}{r}\zeta_S^\prime,
\\
K &\to K + \frac{f}{r}\zeta_r + \frac{\sqrt{\ell (\ell + 1)}}{2 r^2} \zeta_S,
\\
K_2 &\to K_2 - \frac{\sqrt{\ell(\ell + 1)}}{r^2} \zeta_S,
\label{eq:metgaugetr7}
\end{align}
for the even parity modes, and
\begin{align}
h_0 &\to h_0 -i\omega \zeta_V,
\\
h_1 &\to h_1 - \frac{2}{r}\zeta_V + \zeta_V^\prime,
\\
h_\Omega &\to h_\Omega - \frac{\sqrt{\ell(\ell+1)-1}}{2r^2}\zeta_V,
\label{eq:metgaugetr8}
\end{align}
for the odd parity modes, where $\prime$ denotes the radial derivative.

Hereafter, we focus on $\ell \ge 2$ modes because $\ell = 0, 1$ modes do not contain dynamical degrees of freedom.
Then, we can choose $H_{t\Omega} = H_{r\Omega} = K_2 = h_\Omega = 0$ gauge, which is called the Regge--Wheeler gauge,
by choosing $\zeta_\mu$ as
\begin{align}
\zeta_t &=\frac{rH_{t\Omega}}{\sqrt{\ell(\ell+1)}} - \frac{ir^2\omega K_2}{\ell(\ell+1)},
\label{eq:RWgauge1}
\\
\zeta_r &=\frac{rH_{r\Omega}}{\sqrt{\ell(\ell+1)}} + \frac{r^2 K_2^\prime}{\ell(\ell+1)},
\label{eq:RWgauge2}
\\
\zeta_S &= \frac{r^2 K_2}{\sqrt{\ell(\ell+1)}},
\label{eq:RWgauge3}
\\
\zeta_V &= \frac{2r^2h_\Omega}{\sqrt{\ell(\ell +1)-1}}.
\label{eq:RWgauge4}
\end{align}
Due to the spherical symmetry of the background spacetime, 
the master equations for each mode only depend on $\ell$ but not $m$.\footnote{
Because the ladder operators ${\cal L}_\pm = -i {\cal L}_{\xi_{(1)}} \pm  {\cal L}_{\xi_{(2)}}$
are constructed from the Killing vector on $S^2$,
Eq.~\eqref{eq:ylmladder} shows that 
different $m$ modes can be generated by the infinitesimal coordinate transformation 
associated with the Killing vectors on $S^2$.
}
Thus, we only need to consider $m=0$ cases.
If we need the explicit metric form of $m\neq 0$, it can be obtained by acting the ladder operators of 
spherical harmonics on $m=0$ modes.

\subsubsection{Remarks on complete gauge fixing and gauge invariant quantities}
In the gauge choice $H_{t\Omega} = H_{r\Omega} = K_2 = h_\Omega = 0$,
the gauge functions $\zeta^\mu$ are algebraically determined in Eqs.~\eqref{eq:RWgauge1}-\eqref{eq:RWgauge4}.
In general, this type of gauge choice is called the complete gauge fixing.\footnote{
We assume that the Einstein equations can be decouposed into some modes 
which can be separately duscussed by each mode, {\it e.g.,} even and odd parity modes 
for the Schwarzschild case, and we focus on one of the modes where 
the generator of the gauge transformation $\zeta_\mu$ contains $n$ functional degrees of freedom.
At each mode, we can completely fix the gauge iff $n$ independent components of 
$h_{\mu \nu} + 2\nabla_{(\mu}\zeta_{\nu)}$ can be set as zero for any $h_{\mu \nu}$ by algebraically choosing $\zeta_\mu$.
Then, $\zeta_\mu$ is expressed as linear combinations of $n$ components of $h_{\mu \nu}$, 
which will be set to be zero after the gauge choice, and their derivatives.}
We should note that the perturbed metric with the complete gauge fixing does not admit a pure gauge.\footnote{
If a pure gauge is possible, the perturbed metric becomes
$h_{\mu \nu} = -2 \nabla_{(\mu} \bar{\zeta}_{\nu)}$ with some non-trivial vector field $\bar{\zeta}_\mu$.
From the assumption, we already completely fixed the gauge
and $n$ independent components of $h_{\mu \nu}$ vanish,
where $n$ is a functional degrees of freedom of the gauge transformation in the mode which we focus on.
If we focus on such $n$ components of the equation $h_{\mu \nu} = -2 \nabla_{(\mu} \bar{\zeta}_{\nu)}$,
we can solve them algebraically with respect to $\bar{\zeta}_{\mu}$.
The vector $\bar{\zeta}_\mu$ is expressed as linear combinations of four components of $h_{\mu \nu}$, which are already set to be zero, and their derivatives.
This implies $\bar{\zeta}_\mu = 0$.}
This implies that non-trivial $h_{\mu \nu}$ with the complete gauge fixing always corresponds to physically meaningful perturbations.

Another remark is that we can construct gauge invariant quantities.
Let $h_{\mu \nu}$ be an arbitrary perturbed metric 
whose form is $h_{\mu \nu} = \bar{h}_{\mu \nu} + \nabla_{(\mu}\zeta^{(1)}_{\nu)}$ with some vector field $\zeta_\mu^{(1)}$.
We consider a gauge transformation $h_{\mu \nu} \to h_{\mu \nu} +  2 \nabla_{(\mu} \zeta^{(2)}_{\nu)}$
with a vector field $\zeta_\mu^{(2)}$. We assume that we can completely fix the gauge by choosing $\zeta_\mu^{(2)}$.
If we substitute the form of $\zeta_\mu^{(2)}$ for the complete gauge fixing, which are Eqs.~\eqref{eq:RWgauge1}-\eqref{eq:RWgauge4} for the Schwarzschild case, to the perturbed metric $h_{\mu \nu}$, 
all components of the perturbed metric do not contain $\zeta^{(1)}_\mu$. This is because
if $\zeta^{(1)}_\mu$ is contained, we can construct the perturbed metric $h_{\mu \nu}$ with pure gauge by setting $\bar{h}_{\mu \nu} = 0$,
and this contradicts with the complete gauge fixing. This implies that all components of 
the perturbed metric with the complete gauge fixing 
corresponds to the gauge invariant quantities.
For the Schwarzschild case, the non-trivial components of the
right hand side of Eqs~\eqref{eq:metgaugetr1}-\eqref{eq:metgaugetr8} with 
Eqs.~\eqref{eq:RWgauge1}-\eqref{eq:RWgauge4} are gauge invariant quantities.

\subsubsection{Odd parity perturbation}
For the odd parity perturbation, we need to consider the perturbed metric with
\begin{align}
h_{\mu \nu}^{(-)\ell 0} = 2 e^{-i\omega t} {\sin}\theta \frac{\partial Y_{\ell 0}}{\partial \theta} 
(h_0 dt + h_1 dr).
\end{align}
The $(\theta,\phi)$ component of Einstein equations gives\footnote{
Similar to the case of the Klein-Gordon equation, we can compute the components of the linearized Einstein equations by using \textit{Mathematica}.}
\begin{align}
h_0 = \frac{if}{\omega}(f h_1)^\prime.
\end{align}
Also, the $(r,\phi)$ component gives the second order differential equation of $h_1$
\begin{align}
f^2h_1^{\prime \prime} - f\frac{(2r-5r_H)}{r^2}h_1^\prime
+
\frac{2r^2 - 6 r r_H + 5 r_H^2 - \ell(\ell+1)f r^2}{r^4}h_1 +\omega^2 h_1= 0.
\end{align}
Introducing the master variable
\begin{align}
\Phi_-^{(2)} = \frac{r-r_H}{r^2}h_1,
\end{align}
we obtain the master equation for the odd parity perturbation
\begin{align}
f\frac{d}{dr}\left(f\frac{d\Phi_-^{(2)}}{dr}\right)
+
(\omega^2
-
V_{-}^{(2)})\Phi_-^{(2)}
=
0,
\label{eq:RWequation}
\end{align}
where $V_-^{(2)}$ is the effective potential
\begin{align}
V_-^{(2)} = f\left( \frac{\ell (\ell + 1)}{r^2} - \frac{3r_H}{r^3}\right).
\label{eq:V-Regge--Wheeler}
\end{align}
Eq.~\eqref{eq:RWequation} is the Regge--Wheeler equation~\cite{regge1957}.\footnote{See~\cite{lenzi2021} for the other possible choices of the master variables and the effective potentials.}

\subsubsection{Even parity perturbation}
For the even parity perturbation, we need to consider the perturbed metric with
\begin{align}
h_{\mu \nu}^{(+)\ell 0} = e^{-i\omega t} Y_{\ell 0} (H_0 dt^2 + 2 H_1 dt dr + H_2 dr^2 + 2 K r^2(d\theta^2 +{\sin}^2 d\phi^2)).
\end{align}
The traceless part of the angular components of the Einstein equations shows the relation
\begin{align}
H_0 = f^2 H_2.
\end{align}
The $(t,\theta)$ component of the Einstein equations gives
\begin{align}
H_2 = -\frac{2K}{f} + \frac{i}{\omega f}(f H_1)^\prime.
\end{align}
Combining $(t,r), (r,r)$ and $(r,\theta)$ components of the Einstein equations,
we obtain two equations
\begin{align}
H_1^\prime &= \left(
-\frac{r_H(6r_H + r(2 + 3\lambda))}{2 r^2 (3 r_H + r\lambda)f}
+
\frac{2r^2\omega^2}{(3 r_H + r \lambda)f}
\right)H_1
\notag\\&\quad +
\left(
\frac{9 r_H^2 + 4 r r_H (-2+\lambda) - 4 r^2 \lambda}{r(3 r_H + r \lambda)f^2}
+
\frac{4 r^3 \omega^2}{(3 r_H + r\lambda)f^2}
\right) (i\omega K),
\\
(i\omega K)^\prime &= 
\left(-\frac{(2+\lambda)(2 r_H + r\lambda)}{4r^2(3 r_H + r\lambda)}
-
\frac{r\omega^2}{3 r_H + r \lambda}
\right)H_1
\notag\\&\quad +
\left(
\frac{r_H(6 r_H + r(-4+\lambda))}{2 r^2(3 r_H + r\lambda)f}
-
\frac{2r^2\omega^2}{(3r_H + r\lambda)f}
\right)(i\omega K),
\end{align}
with $\lambda = \ell^2 + \ell - 2$.
Introducing new variables $\Phi_+^{(2)}$ and $D_+$
\begin{align}
H_1 &= a_1 \Phi_+^{(2)} + a_2 D_+,
\\
\omega K &= a_3 \Phi_+^{(2)} + a_4 D_+.
\end{align}
with functions $a_1, a_2, a_3, a_4$
\begin{align}
a_1 &= \frac{i}{f} - \frac{2ir\lambda}{r\lambda  + 3 r_H},
\\
a_2 &= -\frac{2i}{f},
\\
a_3 &= \frac{\lambda(\lambda+2)r^2+3 r r_H \lambda + 6 r_H^2}{2 r^2(r\lambda+3 r_H)},
\\
a_4 &= 1,
\end{align}
$\Phi_+^{(2)}$ and $D_+$ satisfy
\begin{align}
f\frac{d\Phi_+^{(2)}}{dr} &= D_+,
\\
f\frac{dD_+}{dr} &= (V_+^{(2)} - \omega^2)\Phi_+^{(2)},
\end{align}
where $V_+^{(2)}$ is the effective potential
\begin{align}
V_+^{(2)} = f\frac{ \lambda^2 (\lambda+2)r^3+ 3 \lambda^2 r_H r^2+9 \lambda r_H^2 r  + 9r_H^3}{r^3(\lambda r + 3 r_H)^2}.
\label{eq:V-Zerilli}
\end{align}
Thus, we obtain the master equation for the even parity perturbation
\begin{align}
f\frac{d}{dr}\left(f\frac{d\Phi_+^{(2)}}{dr}\right)
+
(\omega^2 -V_+^{(2)})
\Phi_+^{(2)}
=
0,
\label{eq:Zerillieq}
\end{align}
which is called the Zerilli equation~\cite{zerilli1970}.
\subsubsection{Relation between the Regge--Wheeler and Zerilli equations}
\label{app:susytr}

While the effective potentials for the Regge--Wheeler and Zerilli equations in Eqs.~\eqref{eq:V-Regge--Wheeler},\eqref{eq:RWequation}, \eqref{eq:V-Zerilli} and \eqref{eq:Zerillieq} are different,
it is known that these two equations have same spectra (see e.g., \cite{berti2009} for further discussion).
In fact, the solutions of these two equations are connected by transformations
\begin{align}
\Phi_\pm^{(2)}
=
\frac{1}{\beta - \omega^2}
\left(
\mp W \Phi_\mp^{(2)}
+
f \frac{d\Phi_\mp^{(2)}}{dr}
\right),
\end{align}
where
\begin{align}
\beta &= - \frac{4\lambda^2(\lambda+1)^2}{9r_H^2},
\\
W &= -f\frac{3 r_H}{r (3 r_H + 2 r \lambda)} - \frac{2\lambda(\lambda+1)}{3 r_H}.
\end{align}
Thus, we only need to analyze the Regge--Wheeler equation in Eqs.~\eqref{eq:V-Regge--Wheeler} and \eqref{eq:RWequation} to discuss the spectra of the system.

\subsection{Maxwell equations}
In a similar way, we can also derive the master equations for the Maxwell equations
\begin{align}
\nabla^\mu F_{\mu \nu} = 0,
\end{align}
where $F_{\mu \nu} = \partial_\mu A_\nu - \partial_\nu A_\mu$.
We expand the gauge field $A_{\mu}$ as
\begin{align}
A_\mu dx^\mu = \sum_{\ell, m} \left(A^{(+)\ell m}_\mu  + A^{(-)\ell m}_\mu\right)dx^\mu,
\end{align}
with
\begin{align}
A^{(+)\ell m}_\mu dx^\mu &=
e^{-i\omega t}(a_t dt + a_r dr)Y_{\ell m}
-
e^{-i\omega t} a_S \frac{\hat{\nabla}_i Y_{\ell m}dx^i}{\sqrt{\ell(\ell + 1)}},
\\
A^{(-)\ell m}_\mu dx^\mu &= e^{-i\omega t} a_V V_i^{\ell m}dx^i.
\end{align}
Note that $\ell = 0$ mode does not contain dynamical degrees of freedom, and then we focus on $\ell \ge 1$.
Because the Maxwell equations are invariant under the gauge transformation $A_\mu \to A_\mu + \nabla_\mu \Psi$ with a scalar field $\Psi$, we can choose one of components of $A^{(+)\ell m}_\mu$ as zero.
Expanding $\Psi$ as 
\begin{align}
\Psi = \sum_{\ell,m} e^{-i\omega t}\psi Y_{\ell m},
\end{align}
the gauge field $A_\mu$ transforms as
\begin{align}
a_t &\to a_t -i\omega \psi,
\\
a_r &\to a_r + \partial_r \psi,
\\
a_S &\to a_S + \sqrt{\ell(\ell + 1)}\psi.
\end{align}
We can choose $a_S = 0$ gauge and this completely fixes the gauge.
We note that $a_V$ is gauge invariant.
Defining two master variables
\begin{align}
\Phi_+^{(1)} &= \frac{f a_r}{i\omega},
\\
\Phi_-^{(1)} &= a_V,
\end{align}
after some calculations, we obtain the master equations
\begin{align}
f\frac{d}{dr}\left(f\frac{d\Phi_\pm^{(1)}}{dr}\right)
+
(\omega^2 -V^{(1)})
\Phi_\pm^{(1)}
=
0,
\label{eq:mastereqmaxwell}
\end{align}
with the effective potential
\begin{align}
V^{(1)} = f \frac{\ell (\ell + 1)}{r^2}.
\label{eq:effVmaxwell}
\end{align}

\subsection{Master equations for different spins}
The master equations for the Klein-Gordon, Maxwell and Einstein equations 
are given by Eqs.~\eqref{eq:effVkgeq}, \eqref{eq:mastereqkgeq}, \eqref{eq:RWequation}, \eqref{eq:V-Regge--Wheeler}, \eqref{eq:V-Zerilli}, \eqref{eq:Zerillieq}, \eqref{eq:mastereqmaxwell} and \eqref{eq:effVmaxwell}.
Because the Regge--Wheeler and Zerilli equations have same spectra
as discussed in Appendix.~\ref{app:susytr}, 
we can discuss the spectra of the master equations with different spins 
by studying a single equation in the form
\begin{align}
f\frac{d}{dr}\left(f\frac{d\phi}{dr}\right)
+
(\omega^2
-
V_s)\phi
=
0,
\end{align}
where $V_s$ is the effective potential
\begin{align}
V_s = f\left( \frac{\ell (\ell + 1)}{r^2} + (1-s^2)\frac{r_H}{r^3}\right),
\end{align}
and $s$ denotes the spin of the fields, i.e., $s = 0, 1, 2$ for 
the Klein-Gordon, Maxwell and Einstein equations, respectively.
The range of $\ell$ is given by $\ell \ge s$.

\section{Some basics on second order ordinary differential equations}\label{app:ODE}
In this appendix, we review fundamental results on second order linear ordinary differential equations in the complex domain.
We do not show any proofs on these results. We just refer the reader to \cite{coddington1955, wasow2018, delabaere2016} for rigorous treatments.

Let us consider a second order ordinary differential equation:
\begin{equation}
\begin{aligned}
y''(z)+p_1(z)y'(z)+p_2(z)y(z)=0.
\end{aligned}
\label{eq:diff-eq}
\end{equation}
In this article, it is sufficient to assume that $p_1(z)$ and $p_2(z)$ are rational functions.
For the Morse potential \eqref{eq:Morse}, 
by setting $z=e^{\beta x}$, the differential equations are transformed into this form. See \eqref{eq:Laguerre}.
The master equation \eqref{eq:RW} is of this form from the beginning.
Moreover most master equations including the Teukolsky equation \cite{teukolsky1972} in black hole perturbation theory belong to this class.

The first step to analyze the differential equation \eqref{eq:diff-eq} is to see singularities.
We refer to singular points of $p_1(z)$ and $p_2(z)$ as singular points of the differential equation \eqref{eq:diff-eq}.
In our assumption, all these singular points of \eqref{eq:diff-eq} are poles. 
In linear differential equations, we have the following nice property:
\begin{equation}
\begin{aligned}
\text{singular points of solutions to \eqref{eq:diff-eq}} \quad \Longrightarrow \quad \text{singular points of \eqref{eq:diff-eq}}.
\end{aligned}
\end{equation}
Note that the converse is not true in general. Also this statement is not true for non-linear differential equations because there appear movable singular points that depend on integration constants.

If both $p_1(z)$ and $p_2(z)$ are analytic at $z=z_0$, this point is an ordinary point of \eqref{eq:diff-eq}.
If $p_1(z)$ has at most a simple pole at $z=z_0$ and $p_2(z)$ has at most a double pole at the same point, then $z=z_0$ is a regular singular point of \eqref{eq:diff-eq}. 
Otherwise, singular points are called irregular singular points.
The singularity at $z=\infty$ are discussed by the variable transformation $z'=1/z$.
It is particularly important to construct formal series solutions in the neighborhood of a singular point.

\subsection{Regular singular points}\label{app:regular}
Let us see series solutions at a regular singular point. The method is well-known as the Frobenius method.
Without loss of generality, a regular singular point is located at $z=0$.
From the condition of regular singularity, $p_1(z)$ and $p_2(z)$ have the following Laurent expansions
\begin{equation}
\begin{aligned}
p_1(z)=\frac{p_{1,-1}}{z}+\cO(1),\qquad p_2(z)=\frac{p_{2,-2}}{z^2}+\cO(z^{-1}),\qquad z \to 0.
\end{aligned}
\end{equation}
The indicial equation is then given by
\begin{equation}
\begin{aligned}
\rho^2+(p_{1,-1}-1)\rho+p_{2,-2}=0.
\end{aligned}
\end{equation}
Let $\rho_i$ ($i=1,2$) be two roots of the indicial equation. These roots are called characteristic exponents.

If $\rho_1-\rho_2$ is not an integer, two independent series solutions are simply constructed by
\begin{equation}
\begin{aligned}
y_{i}(z)=z^{\rho_i} \sum_{k=0}^\infty a_{ik} z^k,\qquad i=1,2.
\end{aligned}
\label{eq:local-1}
\end{equation}
Let $\gamma$ be a counterclockwise closed path encircling $z=0$ but not encircling any other singular points.
Then the analytic continuation of the two solutions \eqref{eq:local-1} along $\gamma$ generates two solutions $y_i^\gamma(z)$ and
they are represented by
\begin{equation}
\begin{aligned}
(y_1^\gamma(z), y_2^\gamma(z))=(y_1(z), y_2(z)) \begin{pmatrix} e^{2\pi i \rho_1} & 0 \\ 0 & e^{2\pi i \rho_2} \end{pmatrix}.
\end{aligned}
\end{equation}
The matrix on the right hand side is referred to as a monodromy matrix.

If $\rho_1-\rho_2$ is an integer, the classification is slightly intricate. In this case, one of the solutions can be constructed in the same way above.
If $\rho_1-\rho_2=m$ where $m$ is a non-negative integer, the solution corresponding to the exponent $\rho_1$ is given by
\begin{equation}
\begin{aligned}
y_1(z)=z^{\rho_1} \sum_{k=0}^\infty a_{1k} z^k.
\end{aligned}
\end{equation}
The second solution in general takes the form
\begin{equation}
\begin{aligned}
y_2(z)=\alpha y_1(z) \log z+z^{\rho_2}\sum_{k=0}^\infty a_{2k} z^k,
\end{aligned}
\end{equation}
where $\alpha$ is a constant. The monodromy matrix of these solutions is given by
\begin{equation}
\begin{aligned}
(y_1^\gamma(z), y_2^\gamma(z))=(y_1(z), y_2(z))\; e^{2\pi i \rho_1}\!\! \begin{pmatrix} 1 & 2\pi i \alpha \\ 0 & 1 \end{pmatrix},
\end{aligned}
\end{equation}
where we have used $e^{2\pi i \rho_2}=e^{2\pi i \rho_1}$. 
In most cases, $\alpha$ is non-zero, and the logarithmic term exists in the second solution.\footnote{
We note that two solutions with the form of Eq.~\eqref{eq:local-1} are linearly dependent when $\alpha$ is non-zero.}
In particular, $\alpha$ is never zero for $m=0$ (or $\rho_1=\rho_2$). However in exceptional cases, it is accidentally vanishing, and neither solutions have logarithmic singularities. 
In this exceptional case, the monodromy matrix is proportional to the unit matrix:
$e^{2\pi i \rho_1} \diag(1,1)$. Furthermore, if $\rho_1$ is an integer, then the monodromy matrix becomes trivial. In this very special case, the singular point is referred to as an apparent singular point.\footnote{Note that since we can shift the exponents $\rho_i$ to $\rho_i+s$ by a transformation $\widehat{y}(z)=z^s y(z)$,
the monodromy matrix $e^{2\pi i \rho_1} \diag(1,1)$ can be always reduced to $\diag(1,1)$ by $\widehat{y}(z)=z^{-\rho_1} y(z)$. In the new function $\widehat{y}(z)$, the singular point $z=0$ is now apparent.} 

The important fact is that all the infinite sums appearing formal series solutions at a regular singular point are necessarily convergent.
Thus the formal solutions naturally lead to analytic solutions.
This is not the case for series solutions at an irregular singular point.

\subsection{Irregular singular points}\label{app:irregular}
Constructions of local series solutions at an irregular singular point are much more involved.
Discussing them with full generality is beyond the scope of this article. 
We just see a few examples.
Let $z=\infty$ be an irregular singular point of \eqref{eq:diff-eq}.
We need asymptotic expansions to construct local solutions at the irregular singular point $z=\infty$.
For all the examples in this article, formal local solutions at $z=\infty$ have the following asymptotic expansion:
\begin{equation}
\begin{aligned}
y^\text{formal}(z) = e^{a z^r}z^{b} \sum_{k=0}^\infty \frac{c_k}{z^{rk}}\qquad (c_0=1),
\end{aligned}
\label{eq:irreg-sol}
\end{equation}
where $r$ is referred to as the Poincar\'e rank.\footnote{The definition of the Poincar\'e rank seems different in the literature. In this article, we follow \cite{ronveaux1995}.}
This asymptotic expansion is generally divergent, and the solution \eqref{eq:irreg-sol} is purely formal.
Note that the Poincar\'e rank $r$ is easily computed from the differential equation \eqref{eq:diff-eq} \cite{ronveaux1995}.
If $p_1(z)$ and $p_2(z)$ in \eqref{eq:diff-eq} behaves as
\begin{equation}
\begin{aligned}
p_1(z)=\cO(z^{K_1}),\qquad p_2(z)=\cO(z^{K_2}), \qquad z \to \infty,
\end{aligned}
\end{equation}
then the rank $r$ is given by
\begin{equation}
\begin{aligned}
r=1+\max \(K_1, \frac{K_2}{2} \).
\end{aligned}
\end{equation}
If $p_1(z)$ is identically zero, we regard $K_1=-\infty$, and then we have $r=1+K_2/2$.

We first revisit the generalized Laguerre equation \eqref{eq:Laguerre}.
Using the above prescription, we immediately get $r=1$.
Then we plug the ansatz \eqref{eq:irreg-sol} into \eqref{eq:Laguerre}.
The leading behavior in $z \to \infty$ also determines the parameters $a$ and $b$ as
\begin{equation}
\begin{aligned}
(a,b)=(0,\nu), (1, -\alpha-\nu-1).
\end{aligned}
\end{equation}
These provides two independent solutions.
For each of them, one can fix the coefficients $c_k$ uniquely, and finally obtains the formal series solutions \eqref{eq:local-sols-inf}.

Next let us see Airy's differential equation in detail,
\begin{equation}
\begin{aligned}
y''(z)-zy(z)=0.
\end{aligned}
\end{equation}
It is easy to check that it has the irregular singular point at $z=\infty$.
The Poincar\'e rank is $3/2$ in this case.
Plugging the ansatz \eqref{eq:irreg-sol} into the Airy equation, one obtains
\begin{equation}
\begin{aligned}
(a,b)=\( -\frac{2}{3}, -\frac{1}{4} \), \( \frac{2}{3}, -\frac{1}{4} \).
\end{aligned}
\end{equation}
For $(a,b)=(-2/3, -1/4)$, the coefficients $c_k$ satisfy the following recurrence relation:
\begin{equation}
\begin{aligned}
48k c_k+(6k-1)(6k-5)c_{k-1}=0,\qquad c_0=1.
\end{aligned}
\end{equation}
It is solved by
\begin{equation}
\begin{aligned}
c_k=\frac{1}{2\pi k!}\( -\frac{3}{4} \)^k \Gamma\(k+\frac{1}{6} \) \Gamma \(k+\frac{5}{6} \).
\end{aligned}
\end{equation}
Obviously the other case has almost the same structure.
Hence the two formal solutions are given by
\begin{equation}
\begin{aligned}
y_{\infty 1}^\text{formal}(z)&= e^{-\frac{2}{3} z^{3/2}} z^{-1/4} \sum_{k=0}^\infty \frac{1}{2\pi k!}\( -\frac{3}{4} \)^k \Gamma\(k+\frac{1}{6} \) \Gamma \(k+\frac{5}{6} \) \frac{1}{z^{3k/2}}, \\
y_{\infty 2}^\text{formal}(z)&= e^{\frac{2}{3} z^{3/2}} z^{-1/4} \sum_{k=0}^\infty \frac{1}{2\pi k!}\( \frac{3}{4} \)^k \Gamma\(k+\frac{1}{6} \) \Gamma \(k+\frac{5}{6} \) \frac{1}{z^{3k/2}}.
\end{aligned}
\label{eq:formal-sols-Airy}
\end{equation}
Since the coefficients factorially divergent $c_k \sim k!$, the radius of convergence of these series is zero.
In the next subsection, we discuss how to treat these formal solutions.

\subsection{Borel summations and Stokes phenomena}\label{app:Borel}

We want to construct analytic solutions from formal series solutions in the previous subsection. To do so, we use the Borel summation method.
Let us consider a formal power series
\begin{equation}
\begin{aligned}
F^\text{formal}(z) =\sum_{k=0}^\infty \frac{f_k}{z^k}.
\end{aligned}
\end{equation}
We often encounter that the coefficient $f_k$ diverges factorially, $f_k \sim k!$ ($k \to \infty$).
The (generalized) Borel summation is defined by
\begin{equation}
\begin{aligned}
F_p^\text{Borel}(z):=\int_0^\infty d\zeta\,  e^{-\zeta} \zeta^{p-1} B_p\biggl(\frac{\zeta}{z} \biggr), \qquad p>0,
\end{aligned}
\label{eq:Borel}
\end{equation}
where
\begin{equation}
\begin{aligned}
B_p(\zeta):=\sum_{k=0}^\infty \frac{f_k}{\Gamma(k+p)} \zeta^k.
\end{aligned}
\label{eq:Borel-trans}
\end{equation}
Usually we set $p$ unity, but it is convenient to keep it freely.
If $f_k$ diverges factorially, the Borel transformed series \eqref{eq:Borel-trans} has a finite radius of convergence.  We can analytically continue it to the whole complex domain except for its singular points. 
Therefore the Borel summation method provides us a definite value from a formal divergent series.
One can see the Borel summation \eqref{eq:Borel} is regarded as the Laplace transform.
Then $B_p(\zeta)$ is obtained by the inverse Laplace transform.

We apply the Borel summation to the formal series \eqref{eq:formal-sols-Airy}.
By setting $p=5/6$, we obtain a good analytic expression of the first solution:
\begin{equation}
\begin{aligned}
y_{\infty 1}^\text{Borel}(z)=\frac{\Gamma(1/6)}{2^{2/3}\pi} e^{-\frac{2}{3}z^{3/2}}z^{-1/4} \int_0^\infty d\zeta \frac{e^{-\zeta}}{\zeta^{1/6}(4+3\zeta/z^{3/2})^{1/6}}\;.
\end{aligned}
\label{eq:first-Airy}
\end{equation}
To derive this result, we have implicitly assumed that $z$ is real and positive.
In this slice, the integral in \eqref{eq:first-Airy} is always convergent.
It turns out that as long as $z>0$, this solution is exactly related to the well-known Airy function:
\begin{equation}
\begin{aligned}
y_{\infty 1}^\text{Borel}(z)=2\sqrt{\pi} \Ai(z),\qquad z>0.
\end{aligned}
\label{eq:y1-Ai}
\end{equation}
In other words, the asymptotic expansion of the Airy function $\Ai(z)$ is given by the first equation in \eqref{eq:formal-sols-Airy}.
If one is interested in the solution in the complex domain, then the Stokes phenomenon appears.
We explain it briefly.
It is clear that the integrand of \eqref{eq:first-Airy} has two singularities at $\zeta=0$ and at
\begin{equation}
\begin{aligned}
\zeta=-\frac{4}{3}z^{3/2}.
\end{aligned}
\end{equation}
The latter causes the Stokes phenomenon.
When $\arg z=\pm 2\pi/3,  \pm 2\pi, \pm 10\pi/3, \dots$, the singularity is located on the positive real axis in the $\zeta$-plane, and
the integral is not defined as a result.
Moreover, along these rays, the integral \eqref{eq:first-Airy} has discontinuities.
Therefore, $y_{\infty 1}^\text{Borel}(z)$ is analytic only in each domain of
\begin{equation}
\begin{aligned}
\frac{2\pi}{3}(2m-1) < \arg z < \frac{2\pi}{3}(2m+1),\qquad m \in \mathbb{Z}.
\end{aligned}
\end{equation}
Let us recall that Airy's differential equation has only the singular point at $z=\infty$.
Hence its solutions must be analytic in the whole complex plane.
In fact, the two Airy functions $\Ai(z)$ and $\Bi(z)$ are both entire functions.
The Borel resummed solution \eqref{eq:first-Airy} is understood as a building block to construct globally analytic solutions because it is not analytic in the complex plane.
From these facts, we can extend the equality \eqref{eq:y1-Ai} to the sector $\left|\arg z\right|<2\pi/3$.
Beyond this sectorial domain, the asymptotic expansion of $\Ai(z)$ discontinuously changes, and it is well-known as the Stokes phenomenon.
Clearly this phenomenon is associated with the discontinuity of $y_{\infty 1}^\text{Borel}(z)$. The rays $\left|\arg z\right|=2\pi/3, 2\pi, 10\pi/3, \dots$ are Stokes lines, and the sectorial domains separated by these lines are Stokes sectors.
Outside the Stokes sector, the relation \eqref{eq:y1-Ai} must be modified due to the Stokes phenomenon.
$\Ai(z)$ receives an additional contribution.
We should emphasize that $\Ai(z)$ itself does not have any discontinuities along the Stokes lines.
The Borel resums of the asymptotic series solutions have them.
This point is somewhat confusing.
The additional contribution outside the Stokes sector is related to the Borel summation of the second solution $y_{\infty 2}^\text{formal}(z)$.
As the result, the asymptotic series expansion of $\Ai(z)$ changes discontinuously.

\begin{figure}[t]
\begin{center}
\includegraphics[width=0.7\linewidth]{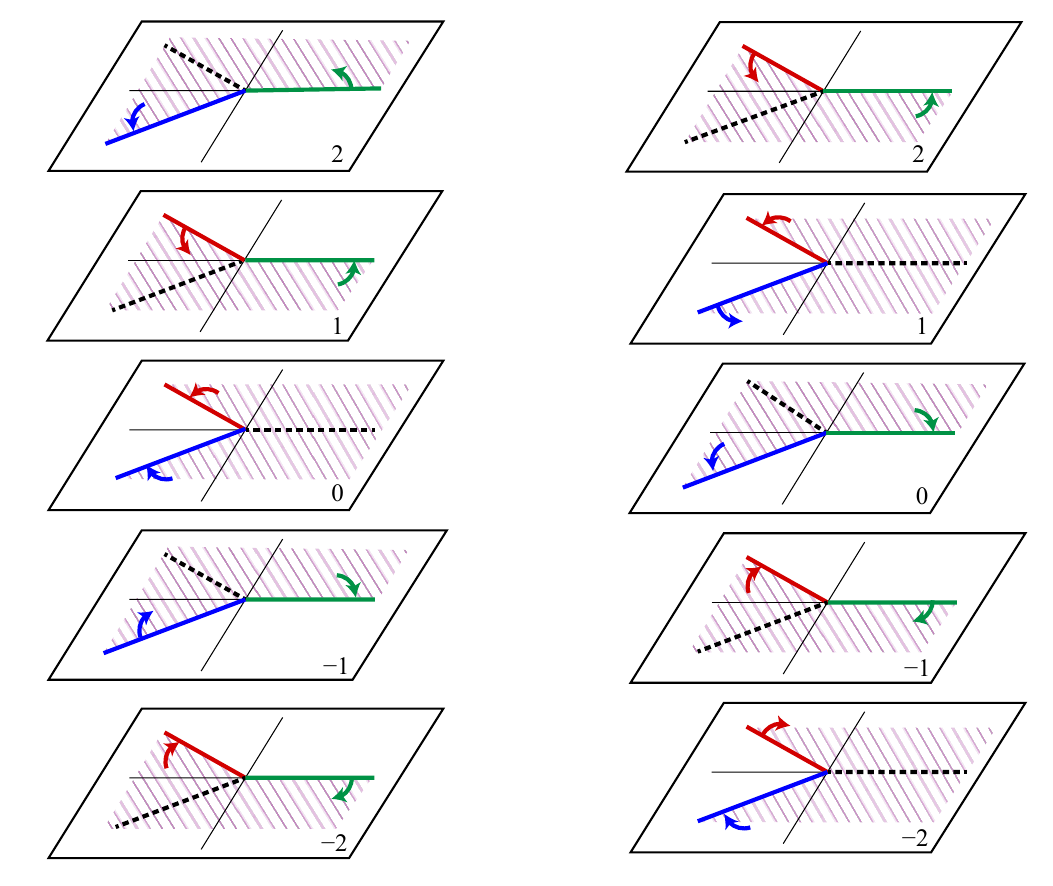}
\caption{Branch cuts of $y_{\infty 1}^\text{Borel}(z)$ (Left) and $y_{\infty 2}^\text{Borel}(z)$ (Right) in the Airy equation. These correspond to the Stokes lines of $\Ai(z)$ and $\Bi(z)$ respectively. We start with the sheet $0$, and cross the cuts in two opposite ways. The colored shaded domains are the Stokes sectors, and the dashed lines represent virtual cuts on other sheets.}
\label{fig:Stokes-Airy}
\end{center}
\end{figure}

The similar thing happens for the second solution.
The Borel resum of the second solution is given by
\begin{equation}
\begin{aligned}
y_{\infty 2}^\text{Borel}(z)=\frac{\Gamma(1/6)}{2^{2/3}\pi} e^{\frac{2}{3}z^{3/2}}z^{-1/4} \int_0^\infty d\zeta \frac{e^{-\zeta}}{\zeta^{1/6}(4-3\zeta/z^{3/2})^{1/6}}.
\end{aligned}
\end{equation}
However, this is not defined for positive $z$.
To evaluate it, we have to displace $z$ in imaginary directions slightly.
The positive real axis is just a Stokes line.
There is a discontinuity along this line.  
It is not hard to find the following exact discontinuity:
\begin{equation}
\begin{aligned}
y_{\infty 2}^\text{Borel}(z+i0)-y_{\infty 2}^\text{Borel}(z-i0)=-iy_{\infty 1}^\text{Borel}(z),\qquad z>0.
\end{aligned}
\end{equation}
Note that $y_{\infty 1}^\text{Borel}(z)$ does not have a discontinuity along the positive real axis.
Interestingly the combinations $y_{\infty 2}^\text{Borel}(z\pm i0)\pm \frac{i}{2} y_{\infty 1}^\text{Borel}(z)$ have no discontinuity along the positive real axis.
It turns out that we have the following identities: 
\begin{equation}
\begin{aligned}
y_{\infty 2}^\text{Borel}(z+i0)+\frac{i}{2} y_{\infty 1}^\text{Borel}(z)=y_{\infty 2}^\text{Borel}(z-i0)-\frac{i}{2} y_{\infty 1}^\text{Borel}(z)=\sqrt{\pi} \Bi(z),\qquad z>0.
\end{aligned}
\label{eq:Bi-Stokes}
\end{equation}
The Airy function $\Bi(z)$ has the Stokes lines along $\left|\arg z\right|=0, 4\pi/3, 8\pi/3,\dots $.
The relation \eqref{eq:Bi-Stokes} is understood as the Stoke phenomenon of $\Bi(z)$ along the Stokes line $\arg z=0$.
We show the branch cut structure of $y_{\infty 1}^\text{Borel}(z)$ and $y_{\infty 2}^\text{Borel}(z)$ (or equivalently the Stokes lines of $\Ai(z)$ and $\Bi(z)$) in Figure~\ref{fig:Stokes-Airy}.

\section{Pad\' e approximants}\label{app:Pade}
Pad\'e approximants are rational functions constructed from a given power series.
They are a very powerful and practical tool in numerics because they are not merely approximate functions but also extrapolating functions of the original power series beyond its radius of convergence. They often work even if the radius of convergence is zero.
Moreover they have important analytic information on the original function.
Pad\'e approximants are discussed in great detail in \cite{bakerjr.1996}.
We introduce very basic aspects on Pad\'e approximants.

Let us define them more precisely.
For a given formal power series
\begin{equation}
\begin{aligned}
F^\text{formal}(z)=\sum_{k=0}^\infty f_k z^k,
\end{aligned}
\end{equation}
its Pad\'e approximants of degree $[M/N]$ are defined by
\begin{equation}
\begin{aligned}
F^{[M/N]}(z):=\frac{a_0+a_1 z+\cdots+a_M z^M}{1+b_1 z+\cdots +b_N z^N},
\end{aligned}
\label{eq:Pade}
\end{equation}
where the $M+N+1$ coefficients $\{ a_m; b_n\}$ are determined uniquely by the condition
\begin{equation}
\begin{aligned}
F^\text{formal}(z)-F^{[M/N]}(z)=\cO(z^{M+N+1}),\qquad z\to 0.
\end{aligned}
\end{equation}
Therefore to obtain $F^{[M/N]}(z)$, we need the formal power series up to $z^{M+N}$.
We do not discuss an algorithm to determine these coefficients. It is easily done by recent symbolic computational systems in practice.
We refer to \eqref{eq:Pade} as the $[M/N]$ Pad\'e approximant.
Also Pad\'e approximants with $M=N$ are called diagonal Pad\'e approximants.
Obviously, for $N=0$, the $[M/0]$ Pad\'e approximant is just the (truncated) original formal power series itself.

Theoretical aspects of Pad\'e approximants seem complicated \cite{bakerjr.1996}.
For instance, it is unclear for us what is the best choice of $M$ and $N$ in general when $M+N$ is fixed.
Empirically, the diagonal Pad\'e approximant seems to be the best.
In this appendix, we rather focus on its practical aspects by seeing a few examples.

\subsection{Convergent series}
\begin{figure}[tbp]
\begin{center}
\includegraphics[width=0.5\linewidth]{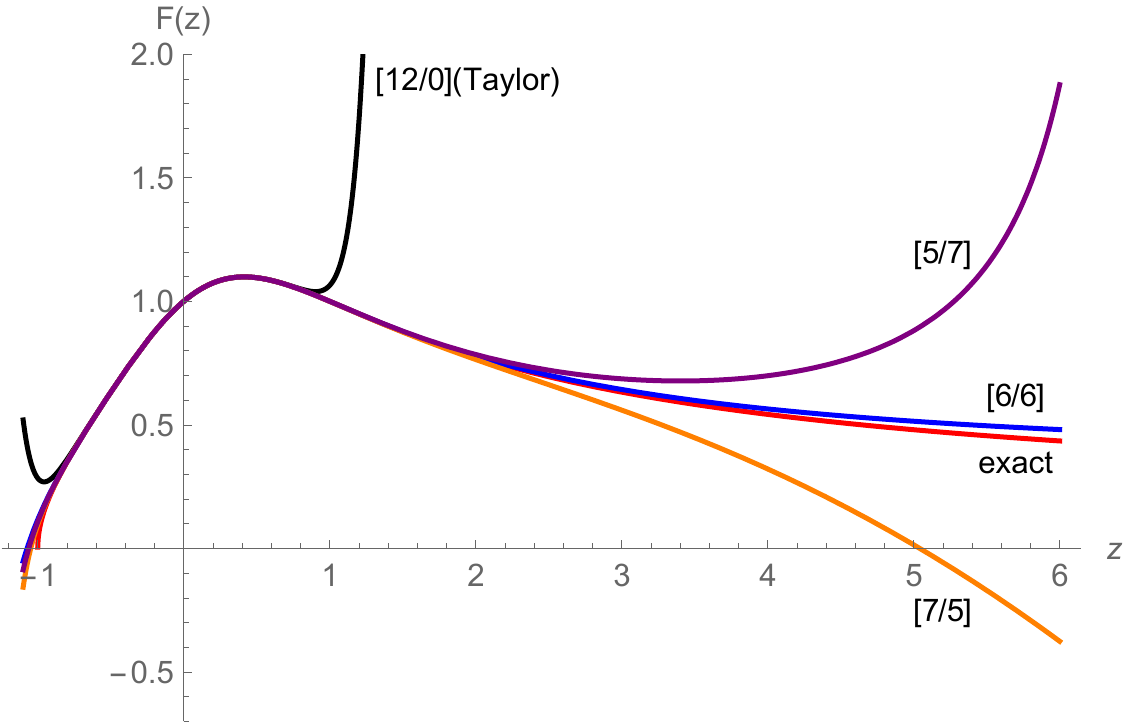}
\caption{The Pad\'e approximants of \eqref{eq:Pade-ex1} have good extrapolations to $z>1$ where the Taylor series approximation necessarily breaks down.}
\label{fig:Pade-1}
\end{center}
\end{figure}

Let us consider a simple function
\begin{equation}
\begin{aligned}
F(z)=\sqrt{\frac{z+1}{z^2+1}}=1+\frac{z}{2}-\frac{5 z^2}{8}-\frac{3 z^3}{16}+\frac{51 z^4}{128}+\frac{47 z^5}{256}-\frac{369 z^6}{1024}+\cO(z^7).
\end{aligned}
\label{eq:Pade-ex1}
\end{equation}
We compare this exact function with its Pad\'e approximants as well as its Taylor series expansion.
Clearly the complex function $F(z)$ has singularities at $z=-1, \pm i$, and the radius of convergence of its Taylor series is unity. 
Therefore the Taylor series approximation does not work for $|z|>1$.
Even in this region its Pad\'e approximants still work.
Figure~\ref{fig:Pade-1} shows graphs of $F(z)$, $F^{[12/0]}(z)$, $F^{[6/6]}(z)$, $F^{[7/5]}(z)$ and $F^{[5/7]}(z)$, obtained by the same truncated sum up to $z^{12}$.
In this example, the diagonal $[6/6]$ Pad\'e approximant looks the ``best'' rational approximate function.
Of course, the situations may change depending on $M+N$.
It is not obvious for us whether the diagonal Pad\'e approximant is always ``best'' or not.
However, we can definitely say that the Pad\' e approximants are better than the truncated power series because the latter cannot be used as an approximation beyond its radius of convergence while the former can.

\begin{figure}[tb]
\begin{center}
  \begin{minipage}[b]{0.31\linewidth}
    \centering
    \includegraphics[width=0.95\linewidth]{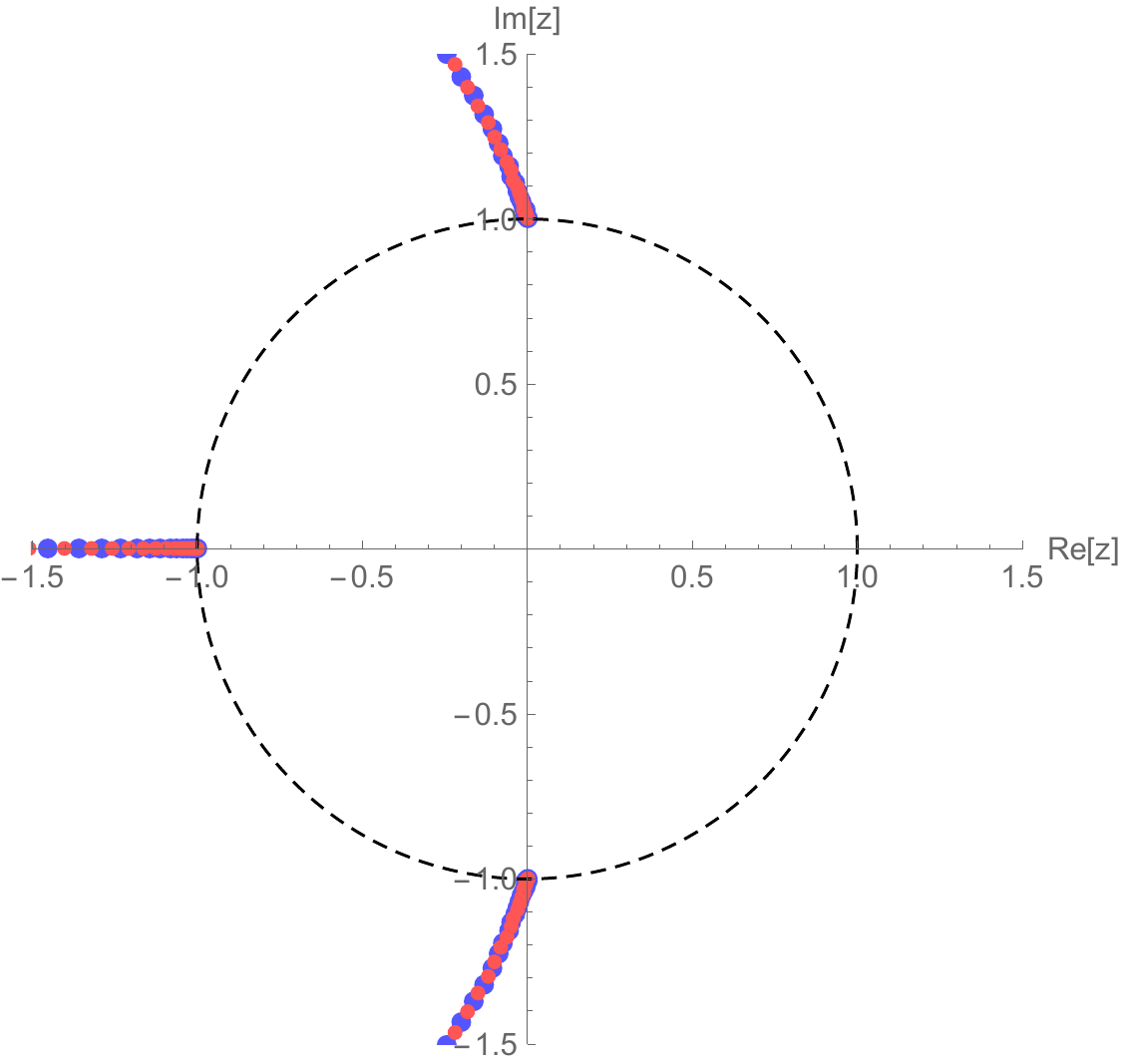}
  \end{minipage}
  \begin{minipage}[b]{0.31\linewidth}
    \centering
    \includegraphics[width=0.95\linewidth]{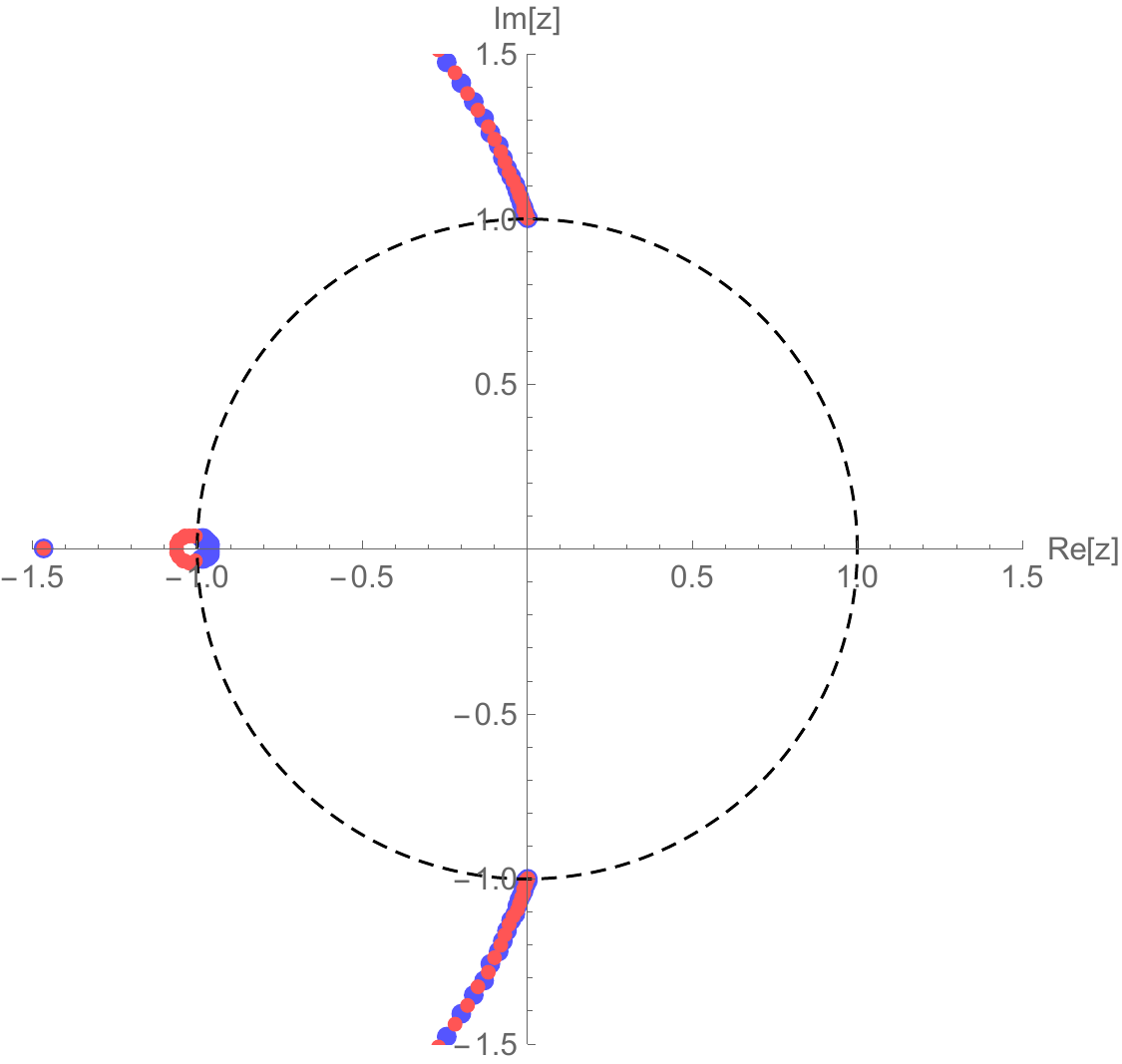}
  \end{minipage} 
    \begin{minipage}[b]{0.31\linewidth}
    \centering
    \includegraphics[width=0.95\linewidth]{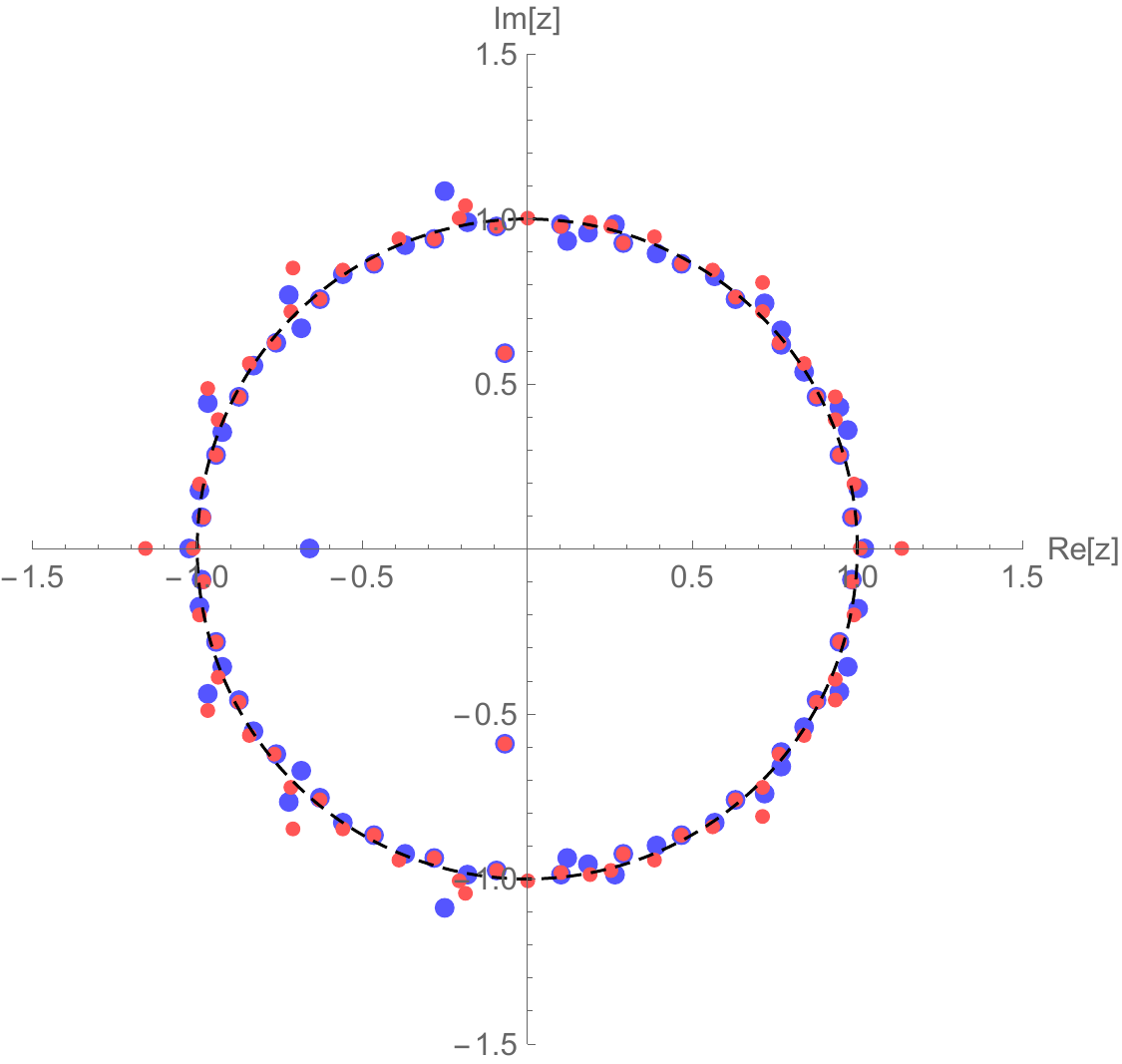}
  \end{minipage} 
\end{center}
  \caption{Pad\'e approximants tell us singularity structures of complex functions. We show the distribution of zeros (light blue) and poles (light red) for $F^{[80/80]}(z)$ (left), $G^{[80/80]}(z)$ (middle) and $H^{[80/80]}(z)$ (right). The black dashed line is the convergence circle.}
  \label{fig:Pade-singularities}
\end{figure}

Pad\'e approximants also have information on the singularity structure of the original function \cite{bakerjr.1996}.
At first glance, since all the singularities of Pad\'e approximants are poles, it seems that they cannot describe other types of singularities such as branch points nor essential singularities. However, Pad\'e approximants tell us even about these singularities.
Let us see it for the example \eqref{eq:Pade-ex1}.
In the left panel of Figure~\ref{fig:Pade-singularities}, we show zeros and poles of the $[80/80]$ Pad\'e approximant of \eqref{eq:Pade-ex1}.
Clearly we can see the branch cut structure as a cluster of them.
Interestingly the Pad\'e approximant automatically chooses the direction of the branch cuts. They are stretched in the opposite direction to the origin.

We illustrate other examples:
\begin{equation}
\begin{aligned}
G(z)&=\frac{1}{\sqrt{z^2+1}}\exp \left( \frac{z}{z+1} \)=1+z-z^2-\frac{z^3}{3}+\frac{2 z^4}{3}+\frac{2 z^5}{15}-\frac{14 z^6}{45}+\cO (z^7), \\
H(z)&=\frac{1}{\sqrt{z^2+1}}\sum_{k=0}^\infty z^{2^k-1}=1+z-\frac{z^2}{2}+\frac{z^3}{2}+\frac{3 z^4}{8}-\frac{z^5}{8}-\frac{5 z^6}{16}+\cO(z^7).
\end{aligned}
\end{equation}
The function $G(z)$ has an essential singularity at $z=-1$, and $H(z)$ has a natural boundary on $|z|=1$.
These singularities are not obvious from the power series expansions at all. We can decipher them by the Pad\'e approximants.
We show the zeros and poles of $G^{[80/80]}(z)$ and $H^{[80/80]}(z)$ in the middle and right panels in Figure~\ref{fig:Pade-singularities}, respectively.
It is interesting to observe that the branch cut structure in $H(z)$ outside the natural boundary disappears in the Pad\'e approximant. 
For more detail on singularities, see subsection 2.2 in \cite{bakerjr.1996}.

\subsection{Divergent series}\label{app:Pade-div}
\begin{figure}[tb]
\begin{center}
  \begin{minipage}[b]{0.35\linewidth}
    \centering
    \includegraphics[width=0.95\linewidth]{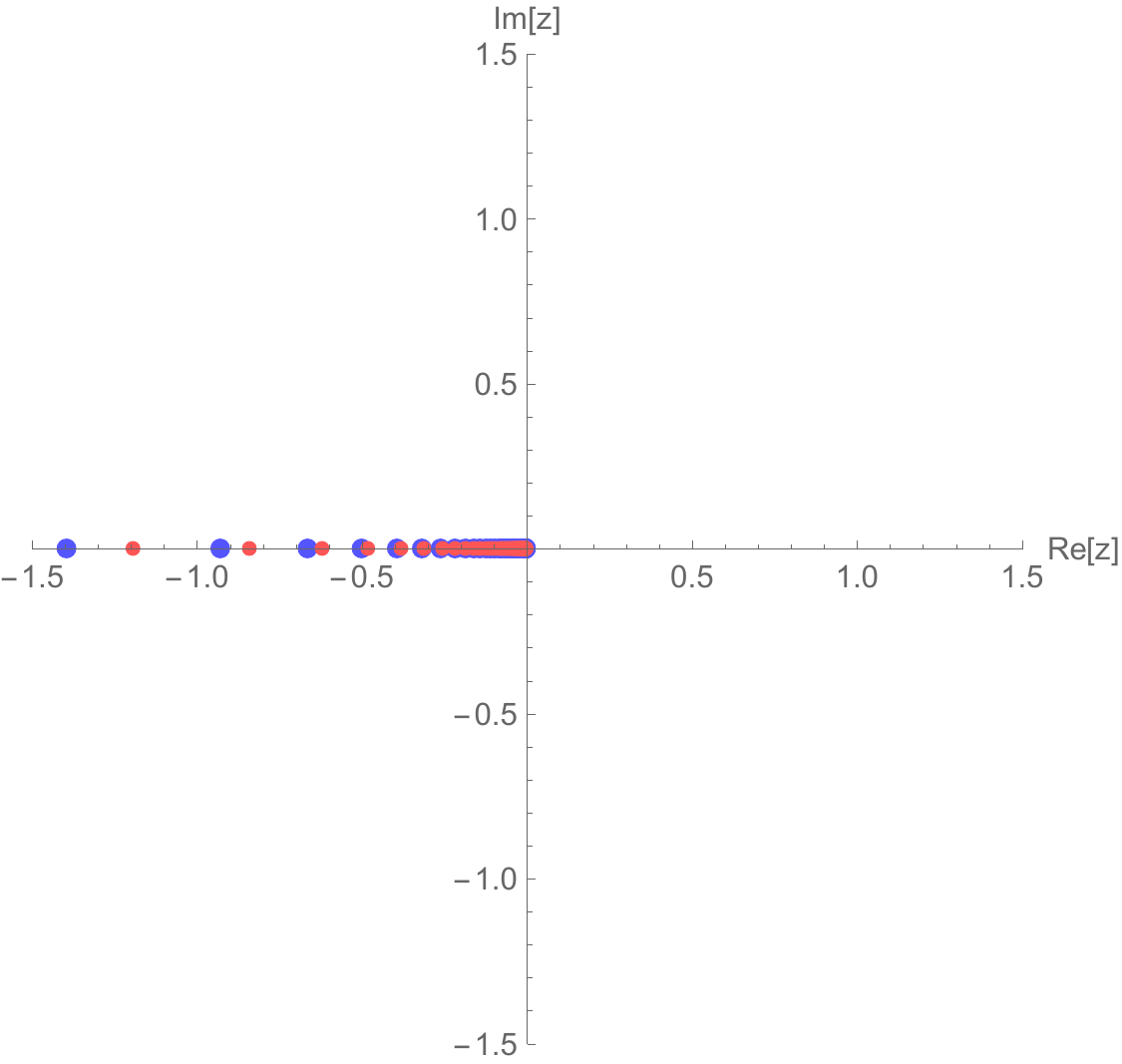}
  \end{minipage}\hspace{1truecm}
  \begin{minipage}[b]{0.45\linewidth}
    \centering
    \includegraphics[width=0.95\linewidth]{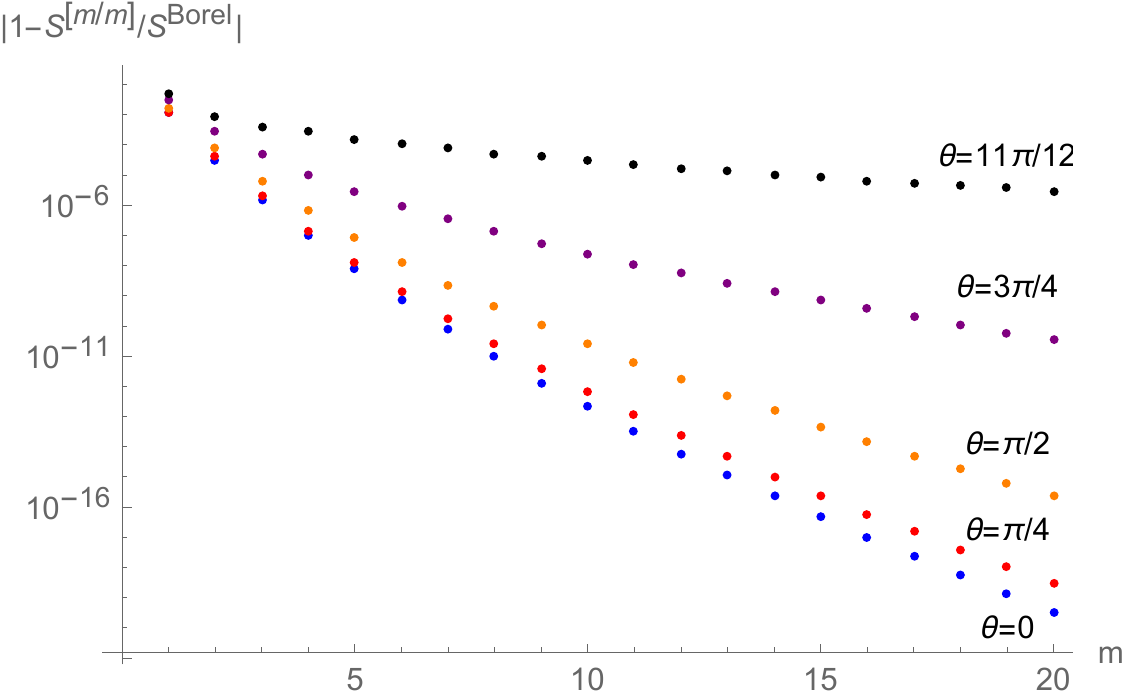}
  \end{minipage} 
\end{center}
  \caption{Left: Zeros and poles of the Pad\'e approximant $S^{[80/80]}(z)$ of the Stieltjes series \eqref{eq:St}. It implies the branch cut on the negative real axis. This is consistent with the discontinuity of the Borel sum \eqref{eq:Borel-St}. This branch cut is related to the Stokes phenomenon of the Borel summation. Right: Convergence of the Pad\'e approximants $S^{[m/m]}(e^{i\theta}/10)$. As $\theta$ gets close to $\pi$, the convergence gets slower.}
  \label{fig:St}
\end{figure}

Pad\'e approximants for formal asymptotic series are involved but interesting.
We show an illuminating example known as the Stieltjes series:
\begin{equation}
\begin{aligned}
S^\text{formal}(z)=\sum_{k=0}^\infty (-1)^k k! z^k.
\end{aligned}
\label{eq:St}
\end{equation}
Obviously it is a divergent series.
Its Borel summation is exactly given by
\begin{equation}
\begin{aligned}
S^\text{Borel}(z)=\int_0^\infty d\zeta \frac{e^{-\zeta}}{1+z\zeta}\;,
\end{aligned}
\label{eq:Borel-St}
\end{equation}
where we have taken $p=1$ in \eqref{eq:Borel} as usual.
We compute the Pad\'e approximant of the Stieltjes series \eqref{eq:St}.
For comparison, we also evaluate the truncated sum
\begin{equation}
\begin{aligned}
S_m(z)=\sum_{k=0}^m (-1)^k k! z^k.
\end{aligned}
\end{equation}
We show numerical values of $S_{2m}(1/10)$, $S^{[m/m]}(1/10)$ and $S^{[m/m]}(e^{11\pi i/12}/10)$ for $1\leq m \leq 10$ in Table~\ref{tab:Stieltjes}.
The Pad\'e approximant rapidly converges to the Borel sum for $z>0$ while the truncated sum has an optimal order.
However, if $z$ is close to the negative real axis, the convergence gets quite worse.
This is because the Borel sum \eqref{eq:Borel-St} has a discontinuity along $z<0$.
In fact, the Pad\'e approximant implies the cut structure as shown in the left panel of Figure~\ref{fig:St}.
This is perfectly consistent with the expectation from the Borel analysis.
Finally, we show the convergence behavior of the Pad\'e approximants in the right panel of Figure~\ref{fig:St}.

\begin{table}[tp]
\caption{Truncated sums and Pad\'e approximants of the Stieltjes series are compared with the Borel sum. For $z>0$, the Pad\'e approximant quickly converges to the numerical value of the Borel sum, but it is not the case if $z$ is near the branch cut $z<0$ of the Borel sum.}
\begin{center}
\begin{tabular}{c|ll|c}
\hline
$m$  &  \multicolumn{1}{c}{$S_{2m}(1/10)$} & \multicolumn{1}{c|}{$S^{[m/m]}(1/10)$} & \multicolumn{1}{c}{$S^{[m/m]}(e^{11\pi i/12}/10)$} \\
\hline
$1$  & ${\bf 0.9}20000000000000$  & ${\bf 0.91}6666666666667$ & ${\bf 1.11}71803947 - {\bf 0.0}395971996 i$\\
$2$  & ${\bf 0.91}6400000000000$  & ${\bf 0.9156}62650602410$ & ${\bf 1.11}94368945 - {\bf 0.044}1279793 i$ \\
$3$  & ${\bf 0.915}920000000000$  & ${\bf 0.91563}4674922601$ & ${\bf 1.118}5374181 - {\bf 0.044}9593628 i$ \\
$4$  & ${\bf 0.915}819200000000$  & ${\bf 0.915633}423180593$ & ${\bf 1.1186}500893 - {\bf 0.044}2145546 i$ \\
$5$  & ${\bf 0.915}819200000000$  & ${\bf 0.9156333}46051126$ & ${\bf 1.118}5456350 - {\bf 0.044}6695805 i$ \\
$6$  & ${\bf 0.915}899033600000$  & ${\bf 0.9156333}40032624$ & ${\bf 1.1186}923372 - {\bf 0.044}4312796 i$ \\
$7$  & ${\bf 0.91}6148114432000$  & ${\bf 0.915633339}468114$ & ${\bf 1.118}5170114 - {\bf 0.04450}36358 i$ \\
$8$  & ${\bf 0.91}6932719052800$  & ${\bf 0.915633339}406670$ & ${\bf 1.1186}359811 - {\bf 0.0445}494854 i$ \\
$9$  & ${\bf 0.91}9778218477568$  & ${\bf 0.91563333939}9102$ & ${\bf 1.1186}152844 - {\bf 0.044}4638222 i$ \\
$10$  & ${\bf 0.9}31942728518451$  & ${\bf 0.91563333939}8066$ & ${\bf 1.118}5679982 - {\bf 0.04450}73821 i$ \\
\hline
$S^\text{Borel}(z)$ & ${\bf 0.915633339397881}$ & ${\bf 0.915633339397881}$ & ${\bf 1.1186001330} -{\bf 0.0445042079} i$ \\
\hline
\end{tabular}
\end{center}
\label{tab:Stieltjes}
\end{table}%

In summary, if one uses Pad\' e approximants to divergent series, one has to see the branch cut structure of its Borel summation.
If a parameter region one is interested in is far from these cuts, one can safely use the Pad\'e approximants.
If the parameter is close to the cuts, one needs a modification. A good resolution is to combine the Borel summation and the Pad\'e approximants, which is sometimes referred to as the Borel--Pad\'e summation. In this method, we first consider the Borel transform \eqref{eq:Borel-trans}, and compute its Pad\'e approximant. A big difference is that the infinite series \eqref{eq:Borel-trans} is convergent. 
We can evaluate the Borel summation \eqref{eq:Borel} by replacing $B_p(\zeta/z)$ by its Pad\'e approximant.

%%%%%%%%%%%%%%%%%%%%%%%%%%%%%%%%%%%%%%%%%%
\reftitle{References}


\externalbibliography{yes}
\bibliography{review}

%=====================================
% References, variant B: internal bibliography
%=====================================

% The following MDPI journals use author-date citation: Arts, Econometrics, Economies, Genealogy, Humanities, IJFS, JRFM, Laws, Religions, Risks, Social Sciences. For those journals, please follow the formatting guidelines on http://www.mdpi.com/authors/references
% To cite two works by the same author: \citeauthor{ref-journal-1a} (\citeyear{ref-journal-1a}, \citeyear{ref-journal-1b}). This produces: Whittaker (1967, 1975)
% To cite two works by the same author with specific pages: \citeauthor{ref-journal-3a} (\citeyear{ref-journal-3a}, p. 328; \citeyear{ref-journal-3b}, p.475). This produces: Wong (1999, p. 328; 2000, p. 475)

%%%%%%%%%%%%%%%%%%%%%%%%%%%%%%%%%%%%%%%%%%
%% optional
%\sampleavailability{Samples of the compounds ...... are available from the authors.}

%% for journal Sci
%\reviewreports{\\
%Reviewer 1 comments and authors’ response\\
%Reviewer 2 comments and authors’ response\\
%Reviewer 3 comments and authors’ response
%}

%%%%%%%%%%%%%%%%%%%%%%%%%%%%%%%%%%%%%%%%%%
\end{document}